\documentclass[article,10pt]{IEEEtran}
\usepackage[usenames,dvipsnames,svgnames]{xcolor}
\usepackage{amsmath}
\usepackage{amsfonts}
\usepackage{amssymb}
\usepackage{setspace}
\usepackage[latin1]{inputenc}
\usepackage{graphicx}
\usepackage{amsthm}
\usepackage{cleveref}
\usepackage{color}
\usepackage{cite}
\usepackage{bm}
\usepackage{epsfig,psfrag}
\usepackage{tabularx}
\usepackage{multirow}
\usepackage{bbm}
\usepackage{mathrsfs} 
\usepackage{colortbl,pgfplotstable}
\usepackage{tikz}
\usepackage{pst-node}
\usepackage{acronym}
\usepackage[yyyymmdd,hhmmss]{datetime}
\usepackage{notation}
\usepackage{subfigure}
\usepackage[linesnumbered,ruled,vlined]{algorithm2e}
\usepackage {soul}
\usepackage[normalem]{ulem}

\usetikzlibrary{arrows,backgrounds,calc,positioning,shapes,shadows}

\providecommand{\ist}{\hspace*{.3mm}}
\providecommand{\rmv}{\hspace*{-.3mm}}
\providecommand{\iist}{\hspace*{1mm}}

\providecommand{\nn}{\nonumber}
\newcommand{\T}{\mathrm{T}}
\makeatletter
\newif\ifAC@uppercase@first%
\def\Aclp#1{\AC@uppercase@firsttrue\aclp{#1}\AC@uppercase@firstfalse}%
\def\AC@aclp#1{%
  \ifcsname fn@#1@PL\endcsname%
    \ifAC@uppercase@first%
      \expandafter\expandafter\expandafter\MakeUppercase\csname fn@#1@PL\endcsname%
    \else%
      \csname fn@#1@PL\endcsname%
    \fi%
  \else%
    \AC@acl{#1}s%
  \fi%
}%
\def\Acp#1{\AC@uppercase@firsttrue\acp{#1}\AC@uppercase@firstfalse}%
\def\AC@acp#1{%
  \ifcsname fn@#1@PL\endcsname%
    \ifAC@uppercase@first%
      \expandafter\expandafter\expandafter\MakeUppercase\csname fn@#1@PL\endcsname%
    \else%
      \csname fn@#1@PL\endcsname%
    \fi%
  \else%
    \AC@ac{#1}s%
  \fi%
}%
\def\Acfp#1{\AC@uppercase@firsttrue\acfp{#1}\AC@uppercase@firstfalse}%
\def\AC@acfp#1{%
  \ifcsname fn@#1@PL\endcsname%
    \ifAC@uppercase@first%
      \expandafter\expandafter\expandafter\MakeUppercase\csname fn@#1@PL\endcsname%
    \else%
      \csname fn@#1@PL\endcsname%
    \fi%
  \else%
    \AC@acf{#1}s%
  \fi%
}%
\def\Acsp#1{\AC@uppercase@firsttrue\acsp{#1}\AC@uppercase@firstfalse}%
\def\AC@acsp#1{%
  \ifcsname fn@#1@PL\endcsname%
    \ifAC@uppercase@first%
      \expandafter\expandafter\expandafter\MakeUppercase\csname fn@#1@PL\endcsname%
    \else%
      \csname fn@#1@PL\endcsname%
    \fi%
  \else%
    \AC@acs{#1}s%
  \fi%
}%
\edef\AC@uppercase@write{\string\ifAC@uppercase@first\string\expandafter\string\MakeUppercase\string\fi\space}%
\def\AC@acrodef#1[#2]#3{%
  \@bsphack%
  \protected@write\@auxout{}{%
    \string\newacro{#1}[#2]{\AC@uppercase@write #3}%
  }\@esphack%
}%
\def\Acl#1{\AC@uppercase@firsttrue\acl{#1}\AC@uppercase@firstfalse}
\def\Acf#1{\AC@uppercase@firsttrue\acf{#1}\AC@uppercase@firstfalse}
\def\Ac#1{\AC@uppercase@firsttrue\ac{#1}\AC@uppercase@firstfalse}
\def\Acs#1{\AC@uppercase@firsttrue\acs{#1}\AC@uppercase@firstfalse}

\makeatletter
\renewcommand*{\@algocf@post@ruled}{}
\makeatother

\acrodef{harp}[HARP]{high-frequency acoustic recording package}
\acrodef{is}[IS]{importance sampling}
\acrodef{bp}[BP]{belief propagation}
\acrodef{spa}[SPA]{sum-product algorithm}
\acrodef{da}[DA]{data association}
\acrodef{mmse}[MMSE]{minimum mean-square error}
\acrodef{po}[PO]{potential object}
\acrodef{pmf}[pmf]{probability mass function}
\acrodef{pdf}[pdf]{probability density function}
\acrodef{iid}[iid]{independent and identically distributed}
\acrodef{rmse}[RMSE]{root-mean-squared error}
\acrodef{ospa}[OSPA]{optimal sub-pattern assignment}
\acrodef{bpf}[BPF]{bootstrap particle filter}
\acrodef{upf}[UPF]{unscented particle filter}
\acrodef{pde}[PDE]{partial differential equation}
\acrodef{edh}[EDH]{exact Daum and Huang}
\acrodef{ledh}[LEDH]{localized exact Daum and Huang}
\acrodef{map}[MAP]{maximum a posteriori}
\acrodef{tdoa}[TDOA]{time-difference-of-arrival}
\acrodef{pf}[PFL]{particle flow}
\acrodef{mot}[MOT]{multiobject tracking}
\acrodef{pda}[PDA]{probabilistic data association}
\acrodef{jpda}[JPDA]{Joint \ac{pda}}
\acrodef{phd}[PHD]{probability hypothesis density}
\acrodef{cphd}[CPHD]{cardinalized \ac{phd}}
\acrodef{mht}[MHT]{multi-hypothesis tracking}
\acrodef{slam}[SLAM]{simultaneous localization and mapping}
\acrodef{iid}[iid]{independent and identically distributed}
\acrodef{rfs}[RFS]{random finite sets}
\acrodef{mospa}[MOSPA]{mean \ac{ospa}}
\acrodef{gmm}[GMM]{Gaussian mixture model}
\acrodef{ekf}[EKF]{extended Kalman filter}
\acrodef{ukf}[UKF]{unscented Kalman filter}
\acrodef{mse}[MSE]{mean squared error}
\acrodef{map}[MAP]{maximum a posteriori probability}
\acrodef{roi}[ROI]{region of interest}
\acrodef{smcmc}[SMCMC]{sequential Markov chain Monte Carlo}
\acrodef{mou}[MOU]{measurement-origin uncertainty}
\acrodef{fp}[FP]{false positive}

\definecolor{temporalgreen}{RGB}{0,128,0}
\definecolor{spatialred}{RGB}{255,0,0}
\definecolor{temporalblue}{RGB}{204,204,255}
\definecolor{temporalgray}{RGB}{204,204,204}

\newcolumntype{L}[1]{>{\raggedright\arraybackslash}p{#1}}
\newcolumntype{C}[1]{>{\centering\arraybackslash}p{#1}}
\newcolumntype{R}[1]{>{\raggedleft\arraybackslash}p{#1}}
\newcolumntype{a}[1]{>{\columncolor{temporalblue}\centering\arraybackslash}p{#1}}
\newcolumntype{b}[1]{>{\columncolor{temporalgray}\centering\arraybackslash}p{#1}}

\DeclareMathAlphabet{\mathpzc}{OT1}{pzc}{m}{it}

\allowdisplaybreaks
\providecommand{\rd}{\textcolor{black}}
\providecommand{\rdd}{\textcolor{black}}
\begin{document}

\title{Multisensor Multiobject Tracking\\ with Improved Sampling Efficiency}

\author{\normalsize Wenyu~Zhang,~\IEEEmembership{\normalsize Student Member,~IEEE} and
Florian~Meyer,\hspace{-.3mm} \IEEEmembership{\normalsize Member,~IEEE} \vspace{-8mm}

\thanks{This material is based upon work supported by the Under Secretary of Defense for Research and Engineering under Air Force Contract No. FA8702-15-D-0001 and by the National Science Foundation (NSF) under CAREER Award No. 2146261. Parts of this work have been presented at the IEEE RADAR-21, Atlanta, Georgia, May 2021.}

\thanks{Wenyu~Zhang is with the Department of Electrical and Computer Engineering, University of California San Diego, La Jolla, CA, USA (e-mail: \texttt{wez078@eng.ucsd.edu}).}

\thanks{Florian~Meyer is with the Scripps Institution of Oceanography and the Department of Electrical and Computer Engineering, University of California San Diego, La Jolla, CA, USA (e-mail: \texttt{flmeyer@ucsd.edu}). }

}

\maketitle

\begin{abstract}
Passive monitoring of acoustic or radio sources has important applications in modern convenience, public safety, and surveillance. A key task in passive monitoring is \ac{mot}. This paper presents a Bayesian method for multisensor \ac{mot} for challenging tracking problems where the object states are high-dimensional, and the measurements follow a nonlinear model. Our method is developed in the framework of factor graphs and the \ac{spa} and implemented using random samples or ``particles''. The multimodal \acp{pdf} provided by the \ac{spa} are effectively represented by a \ac{gmm}. To perform the operations of the \ac{spa} with improved sample efficiency, we make use of \ac{pf}. Here, particles are migrated towards regions of high likelihood based on the solution of a partial differential equation. This makes it possible to obtain good object detection and tracking performance even in challenging multisensor \ac{mot} scenarios with single sensor measurements that have a lower dimension than the object positions. We perform a numerical evaluation in a passive acoustic monitoring scenario where multiple sources are tracked in 3-D from 1-D \ac{tdoa} measurements provided by pairs of hydrophones. Our numerical results demonstrate favorable detection and estimation accuracy compared to state-of-the-art reference techniques. 
\end{abstract}

\begin{IEEEkeywords}
Multiobject tracking, particle flow, factor graphs, sum-product algorithm.
\vspace{-2.5mm}
\end{IEEEkeywords}

\section{Introduction}\label{sec:introduction}

\ac{mot} is an important capability for a variety of applications, including surveillance, autonomy,  and marine mammal research. \ac{mot} is a high-dimensional nonlinear filtering problem complicated by \ac{mou}, i.e., the associations between measurements and objects, and an unknown number of objects to be tracked. In this paper, we develop a sequential Bayesian \ac{mot} framework for the particularly challenging scenarios where object states are high-dimensional and measurement models are nonlinear. We expect that our approach is particularly useful for the passive monitoring of acoustic \cite{Zim:B11} or radio \cite{CumMur:J09} sources in 3-D.
\vspace{-3mm}

\subsection{State-of-the-Art}\label{sec:state-of-the-art}

Traditional methods for \ac{mot} include \ac{pda} \cite{BarWilTia:B11}, \ac{mht} \cite{Rei:J79}, and methods based on \ac{rfs} \cite{Mah:B07,Wil:J15,VoVoHoa:J17,SauLiCoa:17}. Most of these traditional approaches suffer from a computational complexity that is exponential in important system parameters, including the number of measurements, objects, and sensors. \ac{mot} methods that are scalable with respect to these parameters have been recently developed in the framework of factor graphs and the \ac{spa} \cite{MeyBraWilHla:J17,MeyKroWilLauHlaBraWin:J18,SolMeyBraHla:J19,TesMeyBee:20,MeyWil:J21}. Factor graphs represent statistical independencies of random variables. The \ac{spa} is known to provide accurate solutions to high-dimensional Bayesian estimation problems efficiently. In particular, by performing local operations (``messages'') on the factor graph, accurate approximations (``beliefs") of the marginal posterior \acp{pdf} of unknown states \cite{KscFreLoe:01} are computed. \ac{spa}-based methods are versatile and have been successfully applied to a variety of applications, including cooperative localization\cite{WymLieWin:J09, MeyHliHla:J16, WinMeyLiu:J18, TeaLiuMeyConWin:J22}, \ac{slam} \cite{LeiMey:19,LiLeiVenTuf:J22,LeiVenTeaMey:A22}, and focalization for underwater localization \cite{MeyGem:J21}.

\begin{figure*}
        \centering
        \subfigure[][]{\includegraphics[scale=0.24]{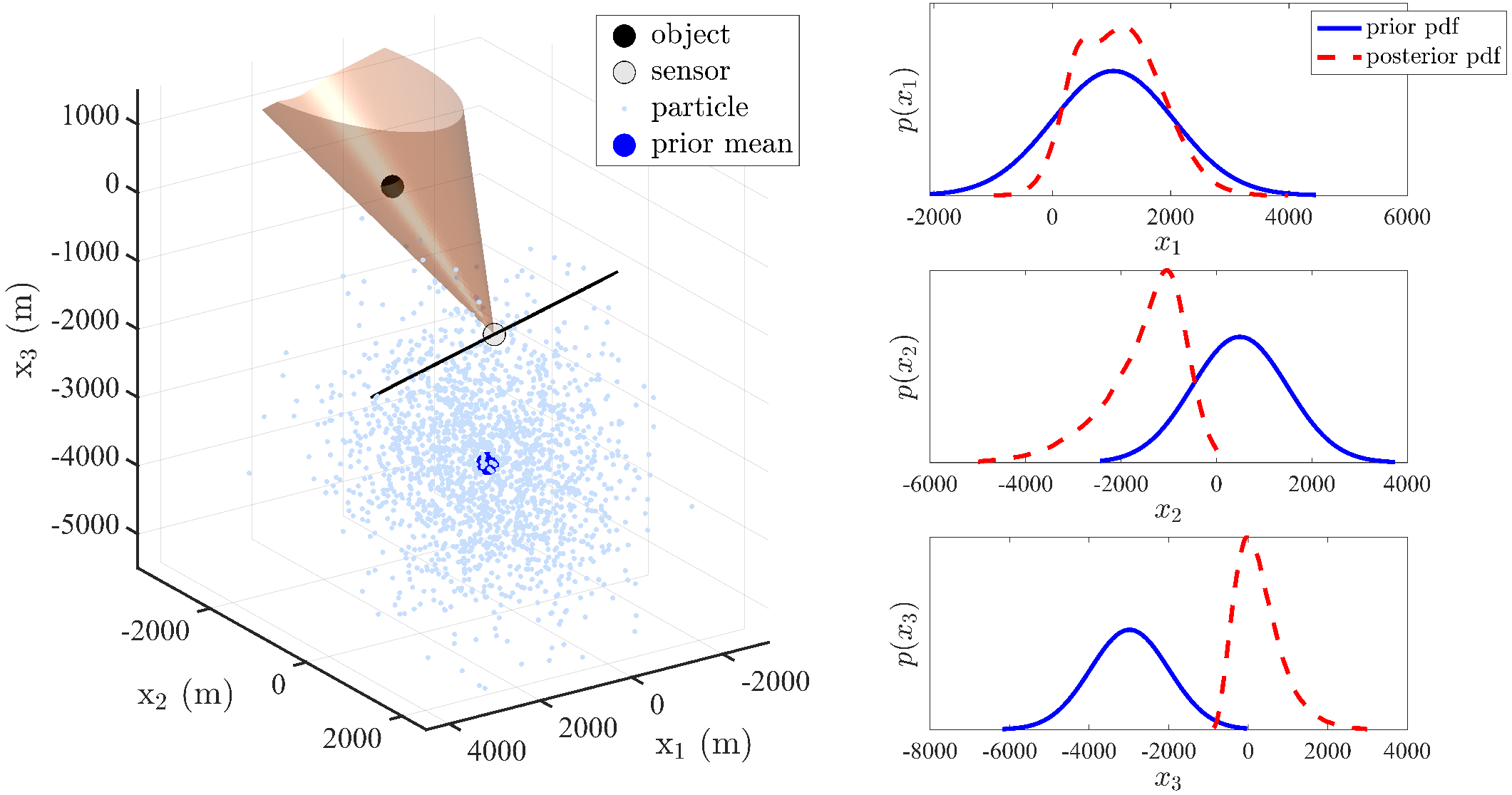}\label{fig:AISprior}}
        \hspace{1.8mm} \subfigure[][]{\includegraphics[scale=0.24]{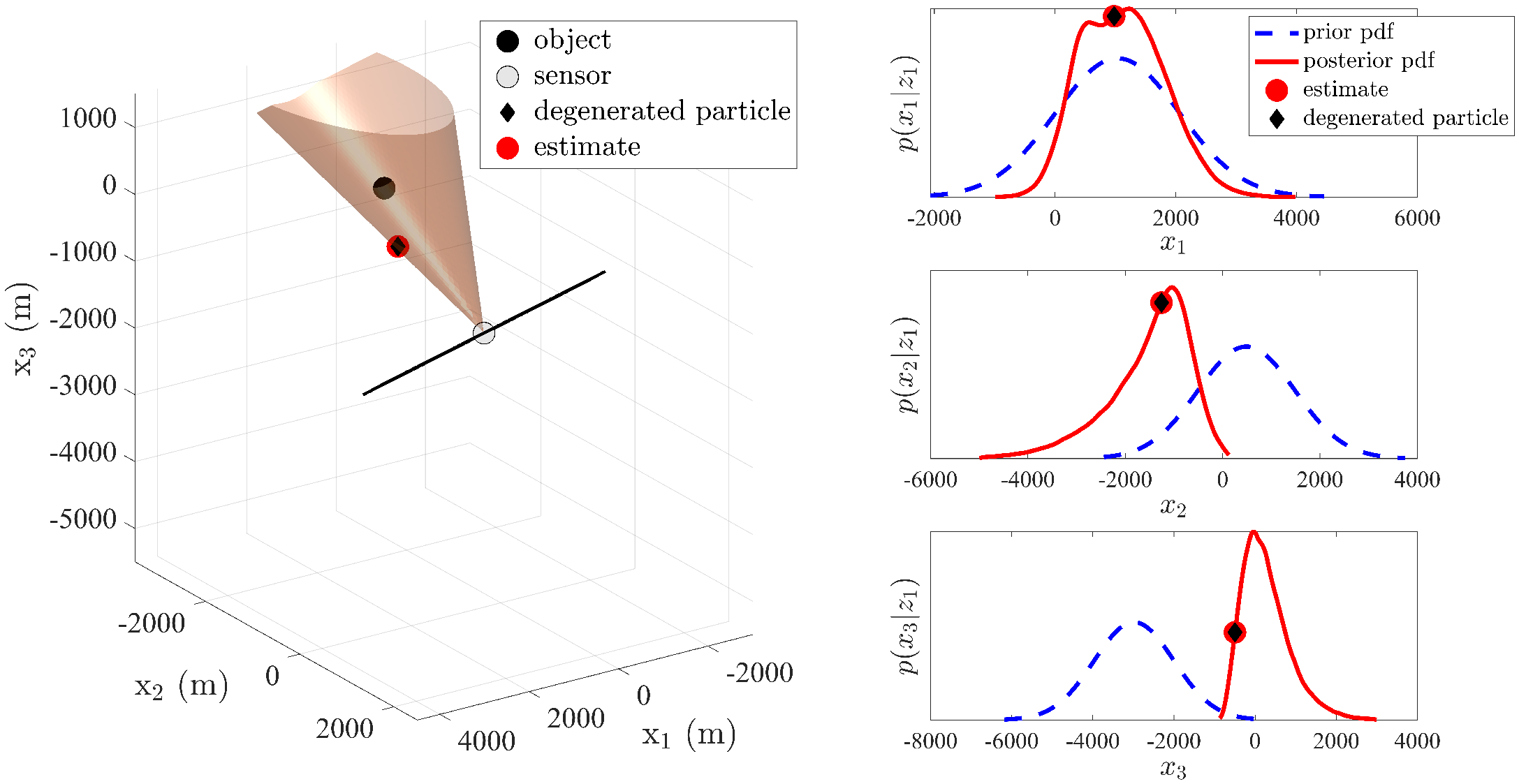}\label{fig:AISposterior}}
        \vspace{-1.5mm}
        \caption{\small \rd{Particle degeneracy in a tracking scenario with 3-D object state, $\V{x} = [x_1 \ist\ist  x_2 \ist\ist x_3]^{\mathrm{T}}\rmv\rmv$, and a single 1-D \ac{tdoa} measurement $z_1$. A single time step in considered. The 1-D \ac{tdoa} measurement is generated by the sensor shown as gray circle. Assuming no measurement noise, the 1-D \ac{tdoa} measurement describes potential 3-D object locations on the hyperboloid shown in red. The object, shown in black, is located on the hyperboloid. Note that any other location on the hyperboloid will lead to the same measurement in the case without noise. (a): The prior \ac{pdf}, $f(\V{x})$, is Gaussian, with the mean depicted as a big blue dot. 2000 particles, shown as small light blue dots, are drawn from the prior distribution.  On the right, the prior and posterior \acp{pdf} for the case with measurement noise are shown in three separate 2-D plots. Each of these plots is obtained by depicting the prior and posterior \acp{pdf} along the three axes of the coordinate system. (b): After importance sampling, as performed by the conventional ``bootstrap'' particle filter, only a single particle has a nonzero weight. This single particle does not accurately represent the posterior \ac{pdf}  $p(\V{x}|z_1)$ for future processing, e.g., of a measurement, $z_2$, provided by a second sensor.}}
        \vspace{-4mm}
        \label{fig:particleDegeneracy}
\end{figure*}

To calculate messages that, due to nonlinearities in the system model, cannot be evaluated in closed form, \ac{spa}-based methods for \ac{mot} typically rely on particle-based computations that closely follow the \ac{bpf} \cite{GorSalSmi:93,DouFreGor:01,AruMasGorCla:02} and rely on importance sampling. \rdd{A known drawback of this approach is that it typically fails in tracking problems where (i) the states of individual objects have dimensions higher than four, (ii) measurements are very informative compared to the predicted/prior \acp{pdf}.} In particular, tracking of objects in 3-D Cartesian coordinates or employing sensors that yield low measurement variance often leads to a failure of particle-based computations due to particle degeneracy \cite{BicLiBen:B08}. The particle degeneracy problem is related to the fact that predicted \acp{pdf} are used as proposal \acp{pdf} for sampling. Since predicted \acp{pdf} can have completely different shapes than the posterior \acp{pdf}, this sampling strategy is highly inefficient, i.e., few or none of the generated particles are suitable to represent the posterior \acp{pdf}. \rd{Fig. \ref{fig:particleDegeneracy} shows an example of particle degeneracy in a 3-D tracking scenario with a single object and a single \ac{tdoa} measurement.} Particle degeneracy is exacerbated in high dimensional problems and in problems with low measurement variance. \rd{In particular, it can lead to the unwanted behavior that filter performance degrades as measurement variance is reduced.} As the dimension of the problem increases, or measurement variance is reduced, the likelihood function becomes ``peakier,'' and it becomes more unlikely that a particle, sampled from the prior distribution, is located in a region of high likelihood.
%


Sometimes particle degeneracy can be avoided by using vast numbers of particles or by implementing regularization strategies \cite{MusOudLeg:J01,AruMasGorCla:02,MeyHliHla:J16}. A straightforward approach to improve sampling efficiency and avoid particle degeneracy is to design proposal \acp{pdf} that are similar to the posterior \acp{pdf} \cite{GorSalSmi:93,DouFreGor:01,AruMasGorCla:02}. However, finding a distribution that is easy to sample from and simultaneously similar to posterior \acp{pdf} \cite{GorSalSmi:93,DouFreGor:01,AruMasGorCla:02} is often challenging.
\rd{To improve samping efficiency, adaptive importance sampling can be employed \cite{BugElvMar:J17,ElvMarBug:J19}. In particular, auxiliary particle filters use a delayed resampling strategy to increase the number of particles with significant weights after importance sampling \cite{PitShe:99}. This approach can improve sampling efficiency but can only be applied in combination with a prediction step, which may be unavailable for newly introduced object states in  MOT scenarios. Furthermore, multiple particle filtering \cite{Pet:07}, similar to a particle-based implementation of the \ac{spa}, aims to increase sample efficiency by exploiting factorization of the underlying statistical model  \cite{BugElvMar:J17}. Since it relies on a suitable factorization of the conditional posterior, its applicability is restricted. Incorporating \ac{smcmc} methods into particle filters \cite{Ber:97,Wal:01} is another general approach for nonlinear sequential Bayesian estimation, which is known to be very computationally expensive in high-dimensional state spaces.}
An alternative approach to improve sampling efficiency in sequential estimation is to perform the update step of an unscented Kalman filter \cite{JulUhl:04} and use the resulting Gaussian \ac{pdf} as a proposal \ac{pdf} for particle filtering. The unscented particle filter \cite{MerDouFre:00,MetDouFre:R00} combines this idea with a Gaussian mixture representation of predicted and posterior \acp{pdf}. To the best of our knowledge, the unscented particle filtering approach has not yet been extended to problems with \ac{mou} and an unknown number of states to be\vspace{-1mm} estimated.

\ac{pf} \cite{DuaHua:07,DuaHua:09,DuaHua:10,DuaHua:13,BunGod:16} is a promising strategy for challenging nonlinear estimation problems that has recently received significant attention \cite{LiCoates:17}.  \rdd{It has the potential to avoid particle degeneracy due to its ability to actively move particles representing a prior or predicted \ac{pdf} to locations of high likelihood.\footnote{\rdd{Conventional strategies that rely on resampling also move particles but do so in a more passive way.  In particular, in conventional strategies, after particles are randomly drawn, only those that correspond to locations of high likelihood remain after resampling. In challenging problems, this is prone to particle degeneracy, i.e.,  the number of remaining particles can become too low to be a representative description of the underlying posterior \ac{pdf}.}}}

\begin{figure*}
        \centering
        \subfigure[][]{\includegraphics[scale=0.16]{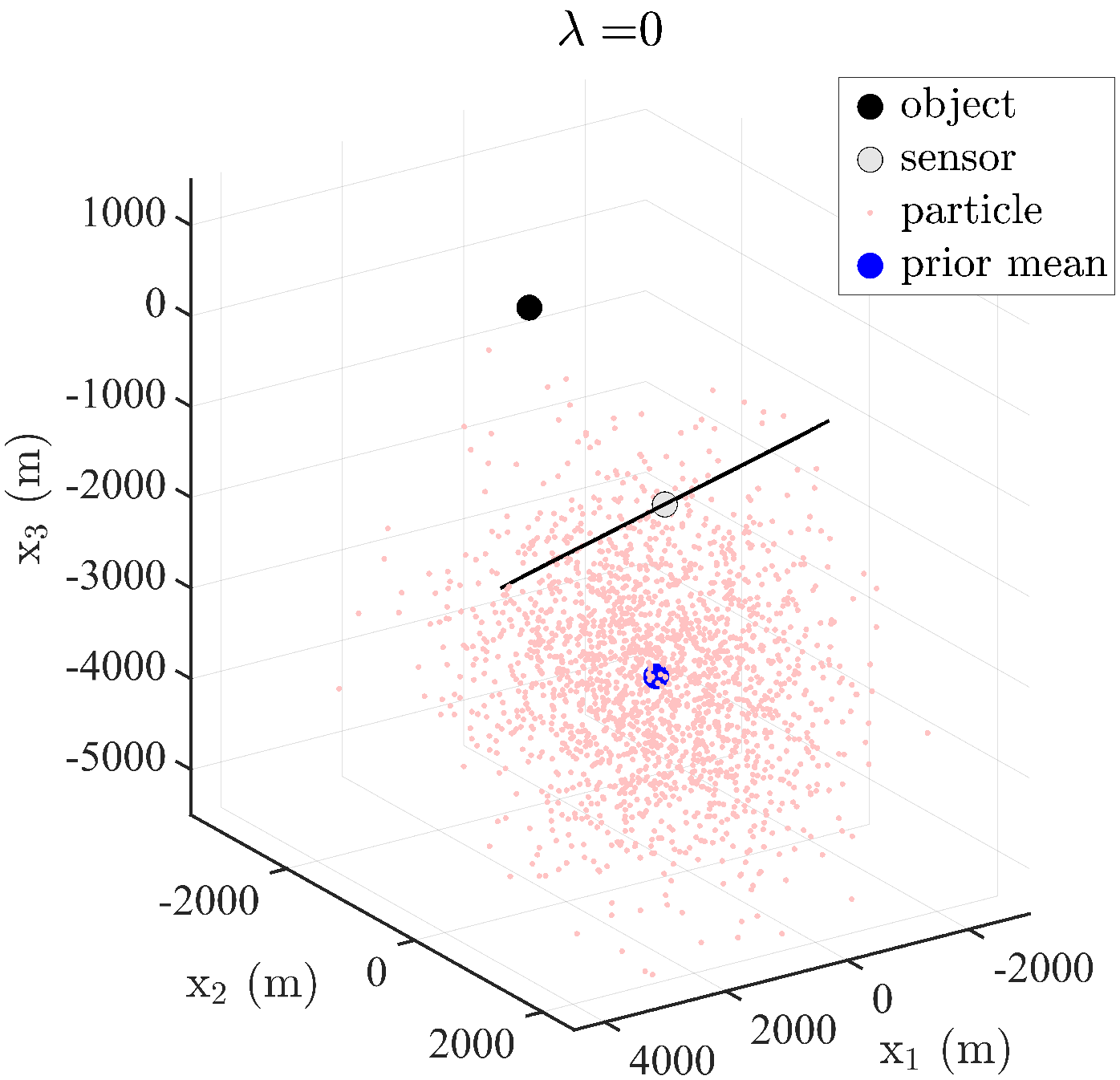}\label{fig:prior}}
        \hspace{.8mm} \subfigure[][]{\includegraphics[scale=0.16]{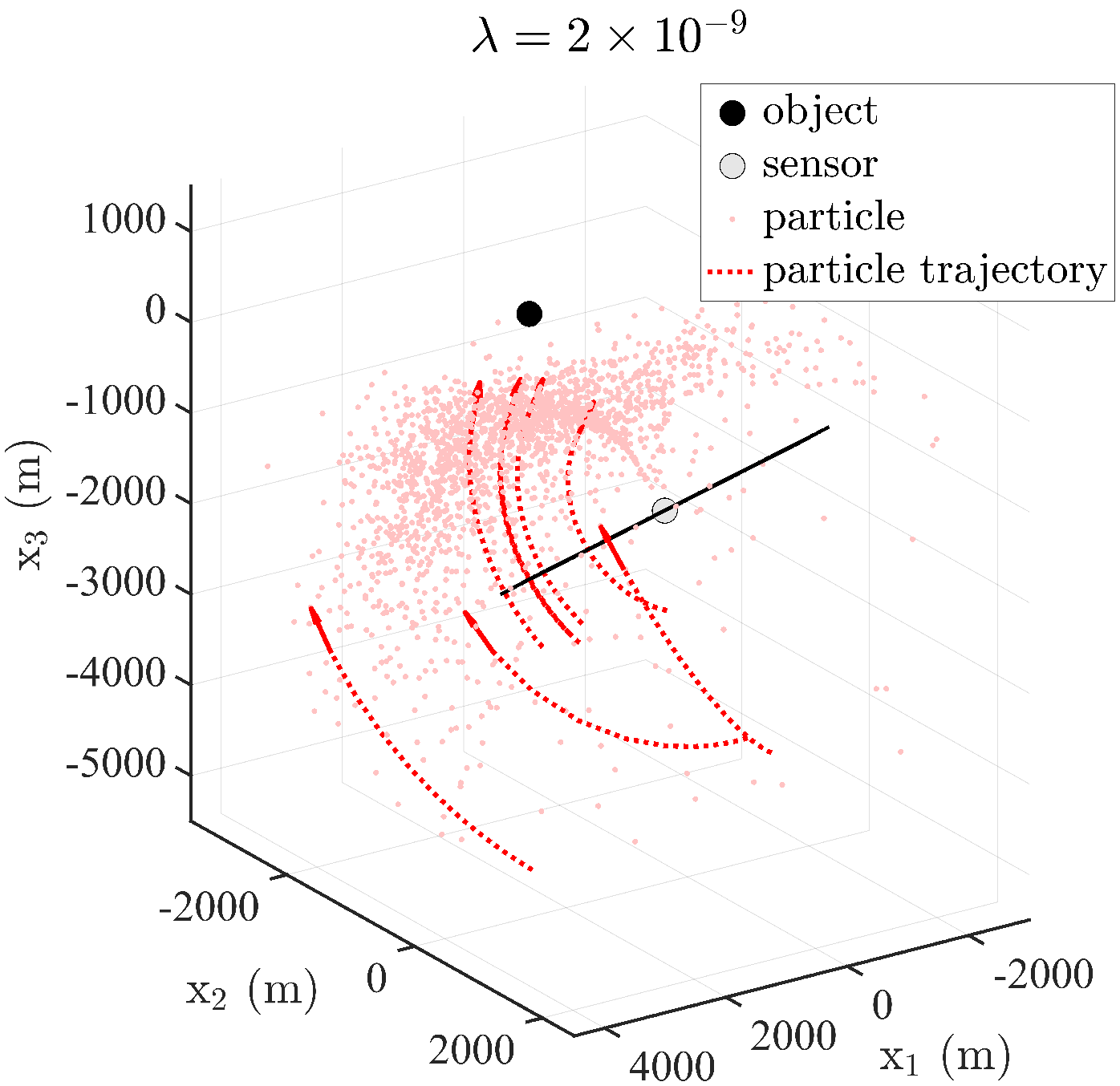}\label{fig:PF1}}
        \hspace{.8mm} \subfigure[][]{\includegraphics[scale=0.16]{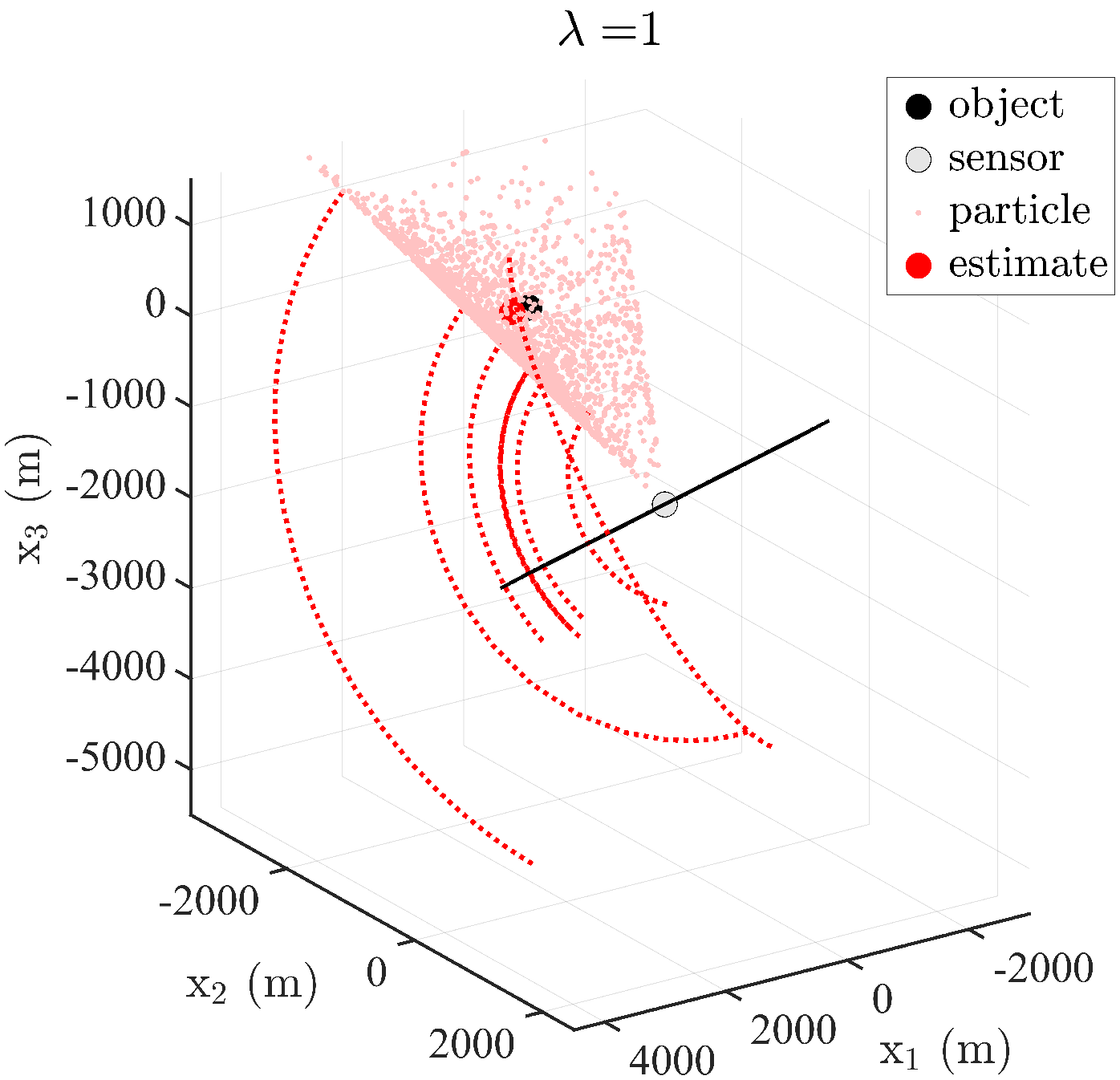}\label{fig:PF2}}
        \hspace{.8mm} \subfigure[][]{\includegraphics[scale=0.165]{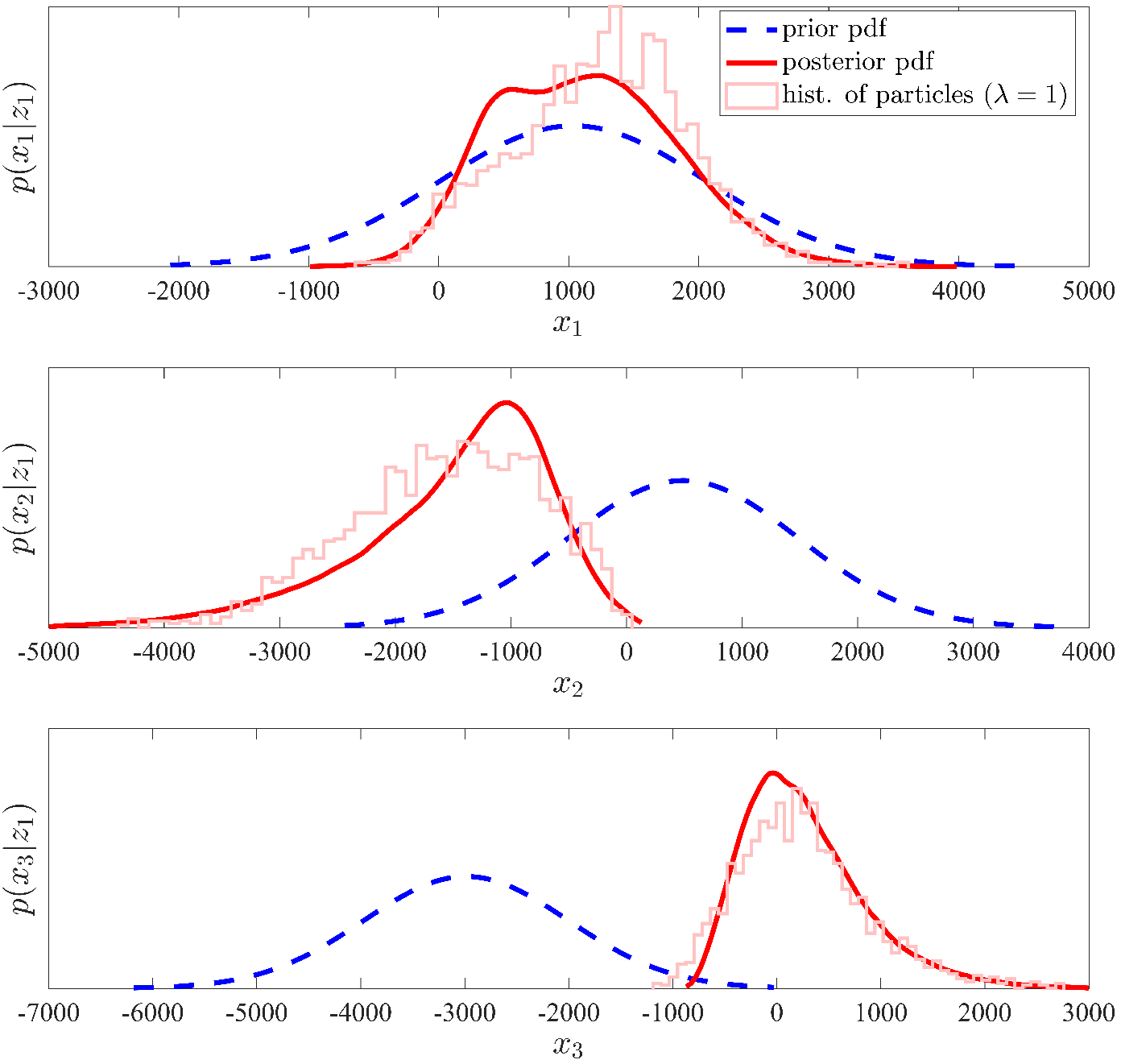}\label{fig:PFpdf}}
        \vspace{-1.5mm}
        
        \caption{\small \rd{Example of \ac{pf} in the tracking scenario with 3-D object state and a single 1-D \ac{tdoa} measurement discussed in Fig.~\ref{fig:particleDegeneracy}. (a): 2000 particles represent the prior \ac{pdf} at the onset of the flow, i.e., at pseudo time $\lambda=0$, are depicted. (b): An intermediate flow state corresponding to $\lambda=2 \times 10^{-9}$ is shown.  The tracks of 8 selected particles are indicated as red dashed line with arrows. (c): At pseudo time $\lambda=1$, particle migration is completed and the resulting particles represent the hyperboloid-shaped posterior \ac{pdf}. (d): The histogram of the flowed particles together with 1-D prior and posterior \acp{pdf} is drawn. The representation of the posterior \ac{pdf} provided by the particles after the flow is much more accurate than the single ``degenerated'' particle resulting from conventional particle filtering discussed in Fig.~\ref{fig:particleDegeneracy}. Due to approximations performed in \ac{pf}, there can be a small mismatch of particles after the flow and the true posterior \ac{pdf}. Such a mismatch can also be seen in (d) by comparing the posterior pdf with the histogram of particles at $\lambda \rmv=\rmv 1$. Invertible \ac{pf} can eliminate such mismatch and provide an asymptotical optimal representation of the posterior \ac{pdf} by making it possible to compute particle weights for importance sampling. }}
        \vspace{-5mm}
        \label{fig:particleFlowHyperboloid}
\end{figure*}

 This active motion is illustrated Fig.~\ref{fig:particleFlowHyperboloid}. For \ac{pf} a homotopy function is defined to formulate a pdf that can be smoothly deformed from the predicted pdf (or prior pdf) to the posterior pdf. \ac{pf} then makes use of the homotopy function to incrementally move a set of particles sampled from the predicted \ac{pdf}. In particular, a \ac{pde} for particle velocity is obtained by combining the homotopy function with the Fokker-Planck equation. The particle velocity solution to the \ac{pde} can be discretized and used as a transport equation for particle migration. After migration, the set of particles represents the posterior \ac{pdf}. There are two different types of \ac{pf} resulting in the \ac{edh} filter and the \ac{ledh} filter. In the \ac{edh} filter, the \ac{pf} equations are computed once for the mean of all particles. In contrast, in the computationally more demanding \ac{ledh} filter, the \ac{pf} equations are computed for each particle individually. \rd{\ac{pf} has been demonstrated to achieve a superior performance complexity tradeoff compared to existing approaches that aim at improving sampling efficiency \cite{LiCoates:17}.} \rdd{As other particle filtering approaches, \ac{pf} is highly parallelizable \cite{DuaHua:09,DuaHua:10,DuaHua:13} and thus ideal for real-time processing on graphical processing units (GPUs).}

Traditional \ac{pf} methods avoid importance sampling and can only provide an approximate representation of posterior \acp{pdf} in general nonlinear systems \cite{DuaHua:07,DuaHua:09,DuaHua:10,DuaHua:13,BunGod:16}. Nevertheless, these ``proposal-free'' methods often lead to accurate estimation results at a significantly reduced computational complexity compared to \ac{bpf} \cite{LiCoates:17}. Recently, it has been shown that \ac{pf} can be described by an invertible mapping and can thus be used as a measurement-driven proposal \ac{pdf} for importance sampling \cite{LiCoates:17}. The resulting invertible \ac{pf} filter \cite{LiCoates:17}  is an asymptotically optimal approach to nonlinear filtering that avoids particle degeneracy and can provide accurate estimation results in high-dimensional and nonlinear problems. 

 A significant limitation of the \ac{pf} filter presented in \cite{LiCoates:17} is that it assumes that the prior or predicted \acp{pdf} follow Gaussian distributions. It is thus unsuitable for problems that involve multimodal \acp{pdf}. For problems where the measurement noise follows a Gaussian mixture \ac{pdf}, \cite{SouCoa:17} introduces the Gaussian sum \ac{pf} filter. Here, the means of the Gaussian mixture components are updated by performing an update step similar to the \ac{ledh}. On the other hand, the covariance matrices of the components are updated by extended Kalman filters that also run in parallel. An extension of \cite{SouCoa:17} to the case where both driving noise and measurement noise are distributed by a Gaussian mixture \ac{pdf} is presented in \cite{PalCoa:C18}. Here, invertible flow is used for particle weight update in an importance sampling step. For problems where both driving noise and measurement noise can be multimodal, \cite{LiSouCoa:19} combines the invertible \ac{pf} with a \ac{smcmc} method that relies on the Metropolis-Hastings approach, i.e., a Metropolis-Hastings kernel is constructed using a \ac{pf} algorithm based on a \ac{gmm}. However, aforementioned \ac{pf} approaches that can represent multimodal \acp{pdf} \cite{SouCoa:17,PalCoa:C18,LiSouCoa:19} are unsuitable for \ac{mot} since neither model \ac{mou} nor an unknown number of states to be estimated.
\rd{For the cooperative localization problem, a method that relies on invertible \ac{pf} is presented in \cite{Luk:23,WieLeiMeyTeaWit:C22}. This method is not suitable for the more challenging multiobject tracking problems since it can only be applied to problems without measurement origin uncertainty, known number of states to be estimated, and posterior \acp{pdf} with simple, unimodal shapes.} A variant of the \ac{pf} filter has been proposed for \ac{mot} \cite{SauLiCoa:17}. In particular, \ac{edh} and \ac{ledh} variants of the single-sensor $\delta$-Generalized Labeled Multi-Bernoulli filter \cite{VoVoHoa:J17} with invertible flow are presented. These approaches are unsuitable for multiobject tracking problems where measurements are provided by multiple sensors.
 \vspace{-2mm}
  
\subsection{Contributions, Paper Organization, and Notation}\label{subsec:contributions}
 \vspace{-.5mm}

We develop a method for multisensor \ac{mot} with improved sample efficiency that can be used in scenarios with high-dimensional object states and informative measurements. \rd{Of particular interest are multisensor \ac{mot} problems, where inexpensive sensors are used and the tracking of objects in Cartesian coordinates is impossible based on the measurements provided by a single sensor. In this type of tracking problems, the measurement of a single sensor typically has a lower dimension than the positions of objects.} Consider a scenario where object positions are 3-D, but sensors only provide 1-D measurements, e.g., times of arrival (TOAs), time differences of arrival (TDOAs), or directions of arrival (DOAs). In this type of \ac{mot} problem, prior or predicted \acp{pdf} can have complicated multimodal shapes, e.g., spheres, hyperboloids, or cones at the initial step after the appearance of a new object. \rd{As an example, Figs.~\ref{fig:particleDegeneracy} and \ref{fig:particleFlowHyperboloid} show the hyperboloid-shaped \acp{pdf} resulting from a TDOA measurement model in a 3-D tracking scenario.}

Our approach performs \ac{spa}-based message passing on the factor graph for scalable multisensor \ac{mot} developed in \cite{MeyKroWilLauHlaBraWin:J18}. The messages of the \ac{spa} are computed sequentially across sensors. \rd{To improve sampling efficiency, we embed invertible particle flow into \ac{spa} computations. For the evaluations of particle weights, invertible particle flow relies on a Gaussian representation of the prior or predicted \ac{pdf} at the onset of the flow. To represent beliefs of object states with complicated non-Gaussian shapes, such as, e.g., hyperboloids, as in the example in Figs.~\ref{fig:particleDegeneracy} and \ref{fig:particleFlowHyperboloid}, we make use of a \ac{gmm} representation that is known to be asymptotically optimal \cite{Sil:B86}. Combining a \ac{gmm} with an efficient sampling approach to represent \acp{pdf} with complicated shapes in high dimensions is inspired by unscented particle filtering \cite{MerDouFre:00,MetDouFre:R00}.} A general proposal pdf that takes \ac{mou} into account and consists of a mixture of pdfs related to different particle flows is developed. The resulting computations are asymptotically optimal. In particular, since particles are migrated towards regions of high likelihood, an accurate approximation of \ac{spa} messages with a relatively small number of particles is obtained. 

\rd{The technical novelty of the proposed method lies in a new method for multitarget tracking that can achieve a superior runtime--estimation accuracy tradeoff in nonlinear and high-dimensional problems by improving sampling efficiency. The improved tradeoff is obtained by carefully embedding invertible \ac{pf}. In particular, to address \ac{mou}, association probabilities are computed by performing parallel flows, one for each component of the \ac{gmm} and each possible measurement-to-object association. The particles of the parallel flows are weighted based on association probabilities and combined into a mixture of flows. The mixture of flows provides samples of the proposal for importance sampling. Since all flows are invertible, it is possible to evaluate the proposal \ac{pdf} represented by the mixture of flows at each particle. Thus, the resulting \ac{spa}-based computation of beliefs is asymptotically optimal in the sense that the resulting particle representation of the beliefs provided by the \ac{spa} is arbitrarily accurate for an increasingly large number of Gaussian components and a number of particles. Our method, for the first time, performs \ac{mot} with probabilistic data association based on \ac{pf}.}

We further demonstrate that the proposed multisensor \ac{mot}  can outperform reference methods based on conventional (``bootstrap'') and unscented particle filtering in a 3-D passive source tracking scenario. \rd{In particular, in the considered realistic source tracking scenario, graph-based MOT based on conventional particle filtering \cite{MeyBraWilHla:J17} cannot provide acceptable estimation accuracy. The also considered, yet unpublished, implementation of graph-based MOT based on unscented particle filtering, has a lower estimation accuracy but a higher runtime compared to the proposed method that embeds\vspace{1mm} invertible \ac{pf}.}

Key contributions of this paper are as\vspace{.5mm} follows.
\begin{itemize}
\item We develop a graph-based MOT method based on a GMM and invertible PFL for challenging scenarios \rd{with high-dimensional object states and arbitrarily shaped posterior \acp{pdf}}.
\vspace{1.5mm}

\item We demonstrate that the proposed method can significantly outperform reference techniques in a challenging 3-D passive source multisensor \ac{mot} scenario \rd{and show tracking results using real passive acoustic data.}
\vspace{.5mm}
\end{itemize}

 This paper advances over the preliminary account of our method provided in the conference publication \cite{ZhaMey:21} by (i) introducing a \ac{gmm} for multimodal state distribution with dynamic kernel resampling; (ii) considering the multisensor \ac{mot} problem; (iii) presenting an improved proposal distribution based on \ac{pf}; (iv) performing a comprehensive numerical evaluation in a 3-D passive source tracking scenario; \rd{and (v) applying the proposed method to an underwater acoustic dataset\footnote{More details on the application of the proposed method to the problem of tracking multiple whales underwater by performing \ac{tdoa} measurements of their echolocation clicks, is presented in the companion paper \cite{JanMeySnyWigBauHil:A22}}}. Contrary to the approach presented in \cite{SauLiCoa:17}, the proposed method is suitable for multisensor scenarios. In addition, \ac{spa}-based processing makes our approach scalable with respect to relevant system parameters.

\emph{Notation:} Random variables are displayed in sans serif, upright fonts and their realizations in serif, italic fonts. 
Vectors and matrices are denoted by bold lowercase and uppercase letters, respectively. For example, a random variable and its realization are denoted by $\rv x$ and $x$, respectively, and a random vector and its realization 
by $\RV x$ and $\V x$, respectively. 
Furthermore, $\|\V{x}\|$ and ${\V{x}}^{\text T}$ denote the Euclidean norm and the transpose of vector $\V x$, respectively; and
$\propto$ indicates equality up to a normalization factor. $\Set{N}(\V{x}; \V{x}^{\ast},\M{P})$ denotes the Gaussian \ac{pdf}  (of random vector $\RV{x}$) with mean $\V{x}^{\ast}$ and covariance \vspace{0mm} matrix $\M{P}$. The trace of matrix $\M{M}$ is denoted as $\mathrm{Tr}\{\M{M}\}$. Finally, $1(a)$ denotes the indicator function of the event $a \rmv=\rmv 0$, i.e., $1(a) \rmv=\rmv 1$ if $a \rmv=\rmv 0$ and $0$\vspace{-1mm} otherwise.

\section{Review of Invertible \ac{pf}}\label{sec:reviewPF}
\vspace{0mm} 

We consider the general setting of calculating the posterior \vspace{0mm} \ac{pdf} based on Bayes' rule $f(\boldsymbol{x}|\boldsymbol{z}) \propto f(\boldsymbol{x}) \ist f(\boldsymbol{z}|\boldsymbol{x})$ with the state of interest $\V{x}$ and the observed (fixed) measurement $\V{z}$. If the prior \ac{pdf} $f(\V{x})$ follows a Gaussian distribution and the likelihood function $f(\V{z}|\V{x})$ represents a linear measurement model $\RV{z} = \M{H} \RV{x} + \RV{v}$ with Gaussian measurement noise $\RV{v}$, the posterior \ac{pdf} $f(\V{x}|\V{z})$ also follows a Gaussian distribution. In this special case, the mean and covariance of the Gaussian posterior \ac{pdf} $f(\V{x}|\V{z})$ can be calculated in closed form by the Kalman update step \cite{ShaKirLi:B02}.

If the measurement model is nonlinear, e.g., $\RV{z} = \M{h} (\RV{x}) + \RV{v}$, a popular approach is to approximate the posterior \ac{pdf} $f(\boldsymbol{x}|\boldsymbol{z})$ by a set of $N_{\mathrm{p}}$ weighted particles $\{(\boldsymbol{x}^{(i)}\rmv,w^{(i)})\}^{N_{\mathrm{p}}}_{i=1}$. Note that the weights are normalized to one, i.e., $\sum^{N_{\mathrm{p}}}_{i=1} w^{(i)} = 1$ and can be computed based on the importance sampling principle \cite{AruMasGorCla:02} as\vspace{-2mm} follows
\begin{equation}
\label{eq:importanceSampling}
  w^{(i)} \propto \frac{  f(\boldsymbol{x}^{(i)}) f(\boldsymbol{z} |\boldsymbol{x}^{(i)})}{q(\boldsymbol{x}^{(i)}|\boldsymbol{z})}.
 \vspace{-.5mm}
\end{equation}
Here, the proposal \ac{pdf} $q(\boldsymbol{x}|\boldsymbol{z})$ is used to sample the particles $\{\V{x}^{(i)}\}^{N_{\mathrm{p}}}_{i=1}$. It is an arbitrary \ac{pdf} that has the same support as $f(\boldsymbol{x}|\boldsymbol{z})$.  Importance sampling is asymptotically optimal if $q(\boldsymbol{x}|\boldsymbol{z})$ is ``heavier tailed'', i.e., less informative, than $f(\boldsymbol{x}|\boldsymbol{z})$ \cite{DouFreGor:01}. In particular, importance sampling can provide an approximation of $f(\boldsymbol{x}|\boldsymbol{z})$ that can be made arbitrarily good by choosing $N_{\mathrm{p}}$ sufficiently large \cite{AruMasGorCla:02}. For $N_{\mathrm{p}}$ fixed, if the proposal $q(\boldsymbol{x} |\boldsymbol{z})$ is ``more similar'' to the posterior $f(\boldsymbol{x}|\boldsymbol{z})$ \cite{DouFreGor:01}, importance sampling is ``more accurate''. 

A simple choice for the proposal \ac{pdf} used in the update step of the conventional ``bootstrap'' particle filter \cite{GorSalSmi:93,AruMasGorCla:02} is the prior \ac{pdf} $f(\boldsymbol{x})$. However, for a feasible number of particles $N_{\mathrm{p}}$ and most choices of the proposal \ac{pdf}, importance sampling can suffer from particle degeneracy \cite{BicLiBen:B08}. Particle degeneracy is especially severe if the state $\V{x}$ is high-dimensional and the measurement $\V{z}$ is informative (i.e., the likelihood function has narrow and sharp\vspace{-4mm} peaks). 

\subsection{Particle Flow (PFL)}
\vspace{-.5mm}

\ac{pf} \cite{DuaHua:07,DuaHua:09,DuaHua:10,DuaHua:13} is an approach that aims at avoiding particle degeneracy. Here, particles are smoothly migrated in the state space from a representation of the prior \ac{pdf} to a representation of the posterior \ac{pdf} by solving a \ac{pde}. Let us introduce the homotopy function $\pi_\lambda(\boldsymbol{x})\rmv =  f(\boldsymbol{x}) \ist l^\lambda(\boldsymbol{x})$ where $\lambda \!\in\! [0,1]$ is the pseudo time of the flow process and $l(\boldsymbol{x}) \rmv= \rmv f(\boldsymbol{z}|\boldsymbol{x})$ is the likelihood function. Note that for $\lambda \rmv=\rmv 1$, the homotopy function is equal to the unnormalized posterior \ac{pdf}, i.e., $\pi(\boldsymbol{x}) \triangleq  \pi_1(\boldsymbol{x}) = f(\boldsymbol{x}) \ist l(\boldsymbol{x})$. The log-homotopy function is then given \vspace{0mm} by \cite{DuaHua:07,DuaHua:09},
\begin{equation}
  \phi(\boldsymbol{x},\lambda)=\log f(\boldsymbol{x})+\lambda\log l(\boldsymbol{x}).
  \label{homotopyPhi}
  \vspace{0mm}
\end{equation}
The log-homotopy function is a pseudo posterior \ac{pdf} in the log domain that defines a smooth and continuous deformation from $\phi(\boldsymbol{x},0) \rmv=\rmv \log f(\boldsymbol{x})$ to $\phi(\boldsymbol{x},1) \rmv= \log \pi(\boldsymbol{x})$. This deformation describes the \ac{pf} process.

It can be shown that the stochastic process defined by homotopy function $\pi_\lambda(\boldsymbol{x})$ satisfies the Fokker-Planck equation \cite{BunGod:16,DuaHua:13,DuaHua:10}. Combining the Fokker-Planck equation for the zero-diffusion case with \eqref{homotopyPhi} results in the following \vspace{.8mm} \ac{pde} \cite{DuaHua:13,DuaHua:10}
\begin{equation}
  \label{exactFlow}
      \frac{\partial\phi(\boldsymbol{x},\lambda)}{\partial\boldsymbol{x}}\V{\zeta}(\boldsymbol{x},\lambda)+\log l(\boldsymbol{x})=-\mathrm{Tr}\Big(\frac{\partial \V{\zeta}(\boldsymbol{x},\lambda)}{\partial\boldsymbol{x}}\Big)
       \vspace{1mm}
\end{equation}
where $\V{\zeta}(\boldsymbol{x},\lambda) \rmv=\rmv \frac{\mathrm{d} \V{x}}{\mathrm{d}\lambda}$ describes particle velocity (samples of $\RV{x}$) as the pseudo time $\lambda$ increases from $0$ to $1$, i.e., as the homotopy function is deformed from the prior pdf to the posterior pdf. This migration is referred to as the\vspace{-5.5mm} \ac{pf}.

\subsection{PFL Update Step}
\vspace{-1mm}
\label{sec:numericalImplementation}

If $f(\boldsymbol{x})$ and $l(\boldsymbol{x})$ are Gaussians or in another exponential family, then an exact and closed form solution for \eqref{exactFlow} is available. The \ac{edh} filter \cite{DuaHua:10Sol,DuaHua:10} makes use of this closed-form solution in its update step.
More precisely, let $f(\V{x}) = \Set{N}(\V{x}; \V{x}_0^{\ast},\M{P})$ and $\RV{z} = \M{H} \RV{x} + \RV{v}$ be a linear measurement model with measurement noise $\RV{v} \sim \Set{N}(\V{v}; \V{0},\M{R})$. The exact flow solution \cite{DuaHua:10Sol,DuaHua:10} now \vspace{0mm} reads  $\V{\zeta}(\boldsymbol{x},\lambda) = \M{A}(\lambda)\boldsymbol{x} + \V{b}(\lambda)$ where we introduce
\vspace{1mm}
\begin{align}
  \M{A}(\lambda) &= -\frac{1}{2}\M{P}\M{H}^\T(\lambda \M{H}\M{P}\M{H}^\T+\M{R})^{-1}\M{H} \label{EDH_A}
\end{align} 
 and
 \begin{align}
  \V{b}(\lambda) &= (\M{I}+2\lambda \M{A}(\lambda))\big[(\M{I}\rmv+\rmv\lambda \M{A}(\lambda))\M{P}\M{H}^\T \M{R}^{-1}\boldsymbol{z} \rmv+\rmv\M{A}(\lambda){\boldsymbol{x}}^\ast_0\big]\rmv.  \nn \\[.5mm] \label{EDH_b}\\[-6.5mm] 
  \nn
\end{align}
Note that in \eqref{EDH_b}, $\V{z}$ is the observed and thus fixed measurement.

This solution is extended to the nonlinear measurement model $\RV{z} = \M{h} (\RV{x}) + \RV{v}$ by performing a suboptimal linearization step. In particular, in a first-order approximation, a Jacobian matrix is computed\vspace{-.5mm}, i.e. $\M{H}(\lambda) = \left.\frac{\partial \M{h}(\boldsymbol{x})}{\partial \boldsymbol{x}}\right|_{\boldsymbol{x}={\boldsymbol{x}}^\ast_{\lambda}}\rmv$ where ${\V{x}}^\ast_\lambda$ is the approximated mean of $\V{x}$ at pseudo time $\lambda$.

In a practical implementation, we calculate $\V{\zeta}(\boldsymbol{x},\lambda)$ at $N_\lambda$ discrete values of $\lambda$, i.e., $0=\lambda_0<\lambda_1<...<\lambda_{N_\lambda}=1$, to perform the \ac{pf}. Here, we first sample $N_\mathrm{p}$ particles $\big\{\V{x}_{0}^{(i)}\big\}_{i=1}^{N_\mathrm{p}}\! \triangleq\! \big\{\V{x}_{\lambda_{0}}^{(i)}\}_{i=1}^{N_\mathrm{p}}$ from $f({\boldsymbol{x}})$. Next, at each discrete pseudo time step $l \rmv\in\rmv\{1,\dots,N_{\lambda}\}$, particles are migrated according to
\begin{equation}
\boldsymbol{x}^{(i)}_{\lambda_l} = \boldsymbol{x}^{(i)}_{\lambda_{l-1}} + \tilde{\V{\zeta}}(\boldsymbol{x}^{(i)}_{\lambda_{l-1}},\lambda_{l}) (\lambda_{l} - \lambda_{l-1})\hspace{1.5mm}  \label{eq:flow}
\vspace{1mm}
\end{equation}
for all $i\rmv\in\rmv \{1,\dots,N_\mathrm{p}\}$. Here, the linearized flow solution $\tilde{\V{\zeta}}(\boldsymbol{x}^{(i)}_{\lambda_{l-1}},\lambda_{l}) = \M{A}_l \boldsymbol{x}^{(i)}_{\lambda_{l-1}} \rmv+\rmv \V{b}_l$ is computed based on $\M{A}_l$ and $\V{b}_l$ given by (cf.~\eqref{EDH_A} and \eqref{EDH_b})
\begin{equation}
  \label{EDH_A_nonlinear}
  \M{A}_l \rmv=\rmv -\frac{1}{2}\M{P}{\M{H}_l}^\T(\lambda_l \M{H}_l\M{P}{\M{H}_l}^\T+\M{R})^{-1}\M{H}_l
\end{equation}
\begin{equation}
  \label{EDH_b_nonlinear}
  \V{b}_l \rmv=\rmv (\M{I}+2\lambda_l \M{A}_l)[(\M{I}+\lambda_l \M{A}_l)\M{P}{\M{H}_l}^\T \M{R}^{-1}(\boldsymbol{z}-\boldsymbol{e}_l)+\M{A}_l{\boldsymbol{x}}^\ast_0].
  \vspace{.5mm}
\end{equation}
Note that $\boldsymbol{e}_l = \M{h}({\boldsymbol{x}}^\ast_{\lambda_{l-1}},0)-\M{H}_l{\boldsymbol{x}}^\ast_{\lambda_{l-1}}$ is the error of linear- ization and that the linearized measurement model, $\M{H}_l$, is computed based on the mean of the last step $l-1$, i.e.,
\begin{equation}
  \label{EDH_H_nonlinear}
  \M{H}_l = \left.\frac{\partial \M{h}(\boldsymbol{x})}{\partial \boldsymbol{x}}\right|_{\boldsymbol{x}={\boldsymbol{x}}^\ast_{\lambda_{l-1}}}\rmv\rmv\rmv\rmv.
  \vspace{-1mm}
\end{equation}
The mean at which the measurement model is linearized is typically propagated in parallel to the particles, i.e., ${\V{x}}^\ast_{\lambda_{l}} = {\V{x}}^\ast_{\lambda_{l-1}} + \tilde{\V{\zeta}}({\V{x}}^\ast_{\lambda_{l-1}},\lambda_{l}) (\lambda_{l} - \lambda_{l-1})$.

After the last discrete pseudo time step, $l \rmv=\rmv N_{\lambda}$, particles $\{\V{x}_1^{(i)}\}_{i=1}^{N_\mathrm{p}} \! \triangleq\! \{\V{x}_{\lambda_{N_\lambda}}^{(i)}\}_{i=1}^{N_\mathrm{p}}$ that approximately represent the unnormalized posterior \ac{pdf} $\pi(\boldsymbol{x})$ are finally obtained.
 Pseudocode for the \ac{pf} update step is provided in \textbf{Algorithm \ref{alg:PF}}. \ac{pf} based on a linearized model has no optimality guarantees. However, it has been demonstrated numerically to typically provide an accurate representation of the posterior \ac{pdf} $f(\V{x}|\V{z})$\cite{DuaHua:07,DuaHua:09,DuaHua:10,DuaHua:13,BunGod:16,LiCoates:17}. In what follows, the \ac{pf} related to the measurement $\V{z}$ as defined by \eqref{eq:flow}--\eqref{EDH_b_nonlinear}, is denoted as $\RV{x}_{0}\rmv\xrightarrow[]\rmv \V{z} \rmv\xrightarrow[]\rmv \RV{x}_{1}$ or for notational convenience in future derivations as\vspace{-2.5mm} $\RV{x}_{0}\rmv\xrightarrow[]\rmv \V{z} \rmv\xrightarrow[]\rmv \RV{x}$.
\begin{algorithm}
  \label{alg:PF}
  \scriptsize
  $\Big[\big\{{\boldsymbol{x}}^{(i)}_{1}\big\}_{i=1}^{N_\mathrm{p}}, \{ A_l \}_{l=1}^{N_\lambda} \Big] = \text{ParticleFlow}\Big(\big\{{\boldsymbol{x}}^{(i)}_{0}\big\}_{i=1}^{N_\mathrm{p}}, \boldsymbol{x}_0^{\ast}, \boldsymbol{P},\V{z} \Big) $\\
Define pseudo time steps $0=\lambda_0<\lambda_1<...<\lambda_{N_\lambda}=1$\;

  \For{$l = 1 : N_\lambda$}{
    Calculate the linearized measurement model $\boldsymbol{H}_l$ according to \eqref{EDH_H_nonlinear}\;
    Compute $\boldsymbol{A}_l$ and $\boldsymbol{b}_l$ according to \eqref{EDH_A_nonlinear} and \eqref{EDH_b_nonlinear}\;
    \For{$i = 1 : N_\mathrm{p}$}{
    $\tilde{\boldsymbol{\zeta}}(\boldsymbol{x}^{(i)}_{\lambda_{l-1}}\rmv,\lambda_l) = \boldsymbol{A}_l\boldsymbol{x}^{(i)}_{\lambda_{l-1}} + \boldsymbol{b}_l$\; 
    $\boldsymbol{x}^{(i)}_{\lambda_l} = \boldsymbol{x}^{(i)}_{\lambda_{l-1}} +  \tilde{\boldsymbol{\zeta}}(\boldsymbol{x}^{(i)}_{\lambda_{l-1}}\rmv,\lambda_l) (\lambda_{l} - \lambda_{l-1})$\;

    }
    $\tilde{\boldsymbol{\zeta}}(\boldsymbol{x}^{\ast}_{\lambda_{l-1}}\rmv,\lambda_l) = \boldsymbol{A}_l\boldsymbol{x}^{\ast}_{\lambda_{l-1}} + \boldsymbol{b}_l$\;
    $\boldsymbol{x}^{\ast}_{\lambda_l} = \boldsymbol{x}^{\ast}_{\lambda_{l-1}} +  \tilde{\boldsymbol{\zeta}}(\boldsymbol{x}^{\ast}_{\lambda_{l-1}}\rmv,\lambda_l) (\lambda_{l} - \lambda_{l-1})$\;
  }
  Output: $\big\{{\boldsymbol{x}}^{(i)}_{1}\big\}_{i=1}^{N_\mathrm{p}}\triangleq\big\{{\boldsymbol{x}}^{(i)}_{\lambda_{N_\lambda}}\big\}_{i=1}^{N_\mathrm{p}}$ and $\{ A_l \}_{l=1}^{N_\lambda}$
  \caption{\ac{pf} Update Step}
  \vspace{-4.2mm}
  \end{algorithm}
   
As an alternative to \ac{edh} \ac{pf} as discussed above, \ac{ledh} \ac{pf} has been introduced in \cite{DingCoates:12,LiCoates:17}. Here, a linearization is performed for each particle location individually instead of only at the mean particle location, i.e., individual flow parameters $\M{H}_l^{(i)},\M{A}_l^{(i)}$, and $\M{b}_l^{(i)}$ are computed to migrate the particles along their individual flows. Although the \ac{ledh} flow usually outperforms the \ac{edh} flow, it suffers from considerable computational complexity since the main computational burden is related to calculating the flow parameters.

\rd{In Fig.~\ref{fig:particleFlowHyperboloid}, it is depicted how \ac{pf} actively move particles representing a prior or predicted \acp{pdf} to locations of high likelihood. Active motion of particles leads to a significantly improved approximation of the posterior \ac{pdf} compared to conventional importance sampling shown in\vspace{-4mm} Fig.~\ref{fig:particleDegeneracy}. }

  
\subsection{Importance Sampling with Invertible Flow}
\vspace{-1mm}
\label{sec:ImportanceSamplingFlow}
\ac{pf} can be used to compute a measurement-driven proposal \ac{pdf} $q(\boldsymbol{x}|\boldsymbol{z})$ for importance sampling (cf.~\eqref{eq:importanceSampling}) to perform asymptotically optimal estimation \cite{LiCoates:17}. Here, the mapping as performed by \ac{pf}  $\RV{x}_{0}\rmv\xrightarrow[]\rmv \V{z} \rmv\xrightarrow[]\rmv \RV{x}\rmv\rmv$ is invertible, i.e., there exists an invertible mapping of the particles after the flow $\big\{\V{x}^{(i)}\big\}_{i=1}^{N_\mathrm{p}}$ to the particles $\big\{\V{x}_0^{(i)}\big\}_{i=1}^{N_\mathrm{p}}$ if certain constraints on the differences of consecutive discrete pseudo times $\lambda_l - \lambda_{l-1}$, $l \rmv\in\rmv\{1,\dots,N_{\lambda}\}$ are satisfied \cite{LiCoates:17}.

By exploiting the invertible mapping, the proposal \ac{pdf} resulting from \ac{pf} can be evaluated at the particles\vspace{1.5mm} as \cite{LiCoates:17}
\begin{equation}
  q_{\mathrm{PFL}}(\boldsymbol{x}^{{(i)}}|\boldsymbol{z}) = \frac{f(\V{x}_{0}^{(i)})} {\theta}.
  \label{eq:invertibleMapping}
  \vspace{2mm}
\end{equation}
Here, the ``mapping factor'' $\theta$ is defined\vspace{1.5mm} as
\begin{equation}
\theta = \prod_{l=1}^{N_\lambda} \big| \ist \mathrm{det}\big[\V{I}+(\lambda_{l} - \lambda_{l-1}) \ist \M{A}_l\big] \big|. 
\label{eq:mappingFactor}
\vspace{0mm}
\end{equation}
By plugging \eqref{eq:invertibleMapping} into \eqref{eq:importanceSampling} the weight of the particle $\V{x}^{(i)}$ is obtained\vspace{1mm} as
\begin{equation}
w^{(i)} \propto \frac{\theta f(\boldsymbol{z} |\boldsymbol{x}^{(i)}) \ist f(\boldsymbol{x}^{(i)} )}{f(\boldsymbol{x}_0^{(i)})}. 
\label{eq:weightUpdate}
\vspace{1.5mm}
\end{equation}
The resulting particle set $\{\V{x}^{(i)},w^{(i)}\}_{i=1}^{N_\mathrm{p}}$ is an asymptotically optimal sample representation of the posterior \ac{pdf} $f(\V{x}|\V{z})$ that can often provide accurate estimation results in nonlinear and high-dimensional estimation problems even if the number of particles is moderate \cite{LiCoates:17}. Pseudocode for importance sampling with invertible \ac{pf} is provided in \textbf{Algorithm \ref{alg:PFPF}}. Note that since the \ac{pf} used for the measurement-driven proposal \ac{pdf} is typically based on the \ac{edh} filter update step, a Gaussian prior \ac{pdf} is\vspace{-1mm} assumed.
\begin{algorithm}
  \label{alg:PFPF}
  \scriptsize
  $\Big[\big\{{\boldsymbol{x}}^{(i)}_{1}, w_1^{(i)}\big\}_{i=1}^{N_\mathrm{p}} \Big] = \text{InvertibleFlow}\big(\boldsymbol{x}^{\ast} \rmv\rmv, \boldsymbol{P}, \V{z} \big) $\\[1mm]
  \For{$i = 1 : N_\mathrm{p}$}{ Draw $\boldsymbol{x}_{0}^{(i)} \sim \mathcal{N}\big(\boldsymbol{x}; \boldsymbol{x}^{\ast} \rmv\rmv, \boldsymbol{P}\big)$\;  }
  \vspace{1mm}
  Perform\vspace{.5mm} PF according $\Big[\big\{{\boldsymbol{x}}^{(i)}_{1}\big\}_{i=1}^{N_\mathrm{p}}, \{ A_l \}_{l=1}^{N_\lambda} \Big] =\text{ParticleFlow}\Big(\big\{{\boldsymbol{x}}^{(i)}_{0}\big\}_{i=1}^{N_\mathrm{p}},\boldsymbol{x}^{\ast}, \boldsymbol{P},$ $ \V{z} \Big)\vspace{.5mm} $\; 
      \hspace*{59mm} \tcp{see \textbf{Alg.~\ref{alg:PF}}}
    \vspace{1mm}
  Compute the mapping factor $\theta$ from $\{ A_l \}_{l=1}^{N_\lambda}$ following \eqref{eq:mappingFactor}\; 
  \vspace{.3mm}
  \For{$i= 1 : N_\mathrm{p}$}{
  \vspace{.3mm}
    Perform weight update according to \eqref{eq:weightUpdate}\vspace{.3mm}, i.e., $w_1^{(i)} \rmv=\rmv \frac{\theta \ist f\big(\boldsymbol{z} | \boldsymbol{x}_{1}^{(i)} \big) \ist  \mathcal{N}\big(\boldsymbol{x}_{1}^{(i)}; \boldsymbol{x}^{\ast} \rmv\rmv, \boldsymbol{P}\big) }{\mathcal{N}\big(\boldsymbol{x}_0^{(i)}; \boldsymbol{x}^{\ast} \rmv\rmv, \boldsymbol{P}\big)}$\;
  }
   Output: $\big\{{\boldsymbol{x}}^{(i)}_{1}, w_1^{(i)}\big\}_{i=1}^{N_\mathrm{p}}$
   \caption{\mbox{Importance Sampling with Invertible Flow}}
        \vspace{-3mm}
  \end{algorithm}

An approximate Gaussian representation of this posterior distribution can be subsequently obtained by applying \textbf{Algorithm \ref{alg:GaussianApprox}} which calculates a mean ${\boldsymbol{x}}^{\ast}_{1}$ and a covariance matrix $\boldsymbol{P}_1$ from the unnormalized weighted particles $\big\{{\boldsymbol{x}}^{(i)}_{1}\rmv, w_1^{(i)}\big\}_{i=1}^{N_\mathrm{p}}$. Note that in Algorithm \ref{alg:PFPF}, the same mapping factor $\theta$ is used to calculate all particle weights. If Algorithm 3 is applied after Algorithm 2, this factor is irrelevant since all weights are normalized in Algorithm \ref{alg:GaussianApprox}. However, making use of $\theta$ is important if multiple flows are performed in parallel, as will be discussed in\vspace{-2mm} Section \ref{sec:GaussianMixture}.

  
\begin{algorithm}
  \label{alg:GaussianApprox}
  \scriptsize
  $\big[{\boldsymbol{x}}^{\ast}, \boldsymbol{P} \big] = \text{GaussianRepresentation}\Big(\big\{{\boldsymbol{x}}^{(i)}, \omega^{(i)}\big\}_{i=1}^{N_\mathrm{p}} \Big) $\\
  \vspace{.5mm}
  Normalize particles, i.e., \\
   \For{$i= 1 : N_\mathrm{p}$}{$w^{(i)} = \frac{\omega^{(i)}}{ \sum^{N_\mathrm{p}}_{i'=1} \omega^{(i')}} $ \;
   }
   \vspace{.5mm}
   Compute mean and covariance matrix from particles, i.e., \\
   $\boldsymbol{x}^{\ast} = \sum^{N_\mathrm{p}}_{i=1} w^{(i)} \boldsymbol{x}^{(i)}$ \\[.5mm]
   $\boldsymbol{P}  = \sum^{N_\mathrm{p}}_{i=1} w^{(i)} \boldsymbol{x}^{(i)} \rmv \boldsymbol{x}^{(i) \ist \mathrm{T}} - \boldsymbol{x}^{\ast} \ist \boldsymbol{x}^{\ast \ist \mathrm{T}}$ \\[.5mm]

   Output: ${\boldsymbol{x}}^{\ast}, \boldsymbol{P}$
   \caption{Computation of Gaussian Representation}
      \vspace{-4.3mm}
  \end{algorithm}
Note that instead of a particle-based covariance matrix computation as performed in Line 7 of Algorithm~\ref{alg:GaussianApprox}, an extended or unscented Kalman update step can be used \cite{LiCoates:17}. For a large number of particles $N_\mathrm{p}$, a particle-based computation is more accurate than a computation based on the extended or unscented Kalman update\vspace{0mm} step.


\section{Gaussian Mixture Representation for Nonlinear Estimation in High-Dimensions}
\vspace{-.2mm}
\label{sec:GaussianMixture}

In this section, we use \ac{gmm} representations \cite{Sil:86,McL:19,Pre:07} for nonlinear estimation in high dimensions. In particular, we developed methods for updating the parameters of \acp{gmm} based on \ac{pf}. This approach is suitable for high-dimensional \acp{pdf} that are multimodal and thus relevant for estimation problems in multiobject tracking and\vspace{-4mm} \ac{slam} \cite{LeiMey:19}. 

\subsection{\ac{gmm} Importance Sampling with Invertible Flow}
\label{sec:PFwithInvFlow}

\rd{As discussed in the previous Section \ref{sec:reviewPF}, for the evaluations of particle weights, invertible particle flow relies on a Gaussian representation of the prior \ac{pdf} at the onset of the flow. In challenging multisensor MOT problems, the complicated multimodal shapes of prior and posterior \acp{pdf} (see, e.g., the hyperboloid-shaped posterior \ac{pdf} in Fig.~\ref{fig:PF2}) can often not be approximated accurately by a single Gaussian.} A \ac{gmm} aims at representing multimodal distributions based on an additively weighted combination of multiple Gaussian components. Each Gaussian component is typically referred to as a ``kernel''. Let $N_\mathrm{k}$ be the total number of kernels and let $h \in \{1,\dots,N_\mathrm{k}\}$ be the kernel index. A multimodal prior \ac{pdf} that follows a \ac{gmm} representation can then be written as $f(\V{x}) = \frac{1}{N_\mathrm{k}}\sum_{h=1}^{N_\mathrm{k}}\Set{N}\big(\V{x}; \V{x}^{\ast(h)},\M{P}^{(h)}\big)$. The corresponding multimodal posterior \ac{pdf} $f(\V{x}|\V{z})$ can be computed by performing Algorithm \ref{alg:PFPF} and Algorithm \ref{alg:GaussianApprox} $N_\mathrm{k}$ times in parallel, i.e., one instance of both algorithms is performed for each kernel $\Set{N}\big(\V{x}; \V{x}^{\ast(h)},\M{P}^{(h)}\big)$, $h \rmv\in\rmv \{1,\dots,N_{\mathrm{k}}\}$. To obtain a \ac{gmm} representation composed of an arbitrary number $N'_\mathrm{k}$ of kernels, a resampling step is then performed, i.e., $N'_\mathrm{k}$ particles are drawn from the overall $N_\mathrm{k}N_\mathrm{p}$ particles based on their weights $w_1^{(h,i)}\rmv$. The resampled particles represent the mean of $N'_{\mathrm{k}}$ new kernels. The covariance of the new kernels is inherited from the original kernel the mean was sampled from. Pseudocode for \ac{gmm} importance sampling with invertible \ac{pf} is provided in \textbf{Algorithm \ref{alg:invertibleFlow}}. Importance sampling with invertible \ac{pf} makes use of resampling as presented in\vspace{-1.5mm} \textbf{Algorithm \ref{alg:resampling}}.

\begin{algorithm}
  \label{alg:invertibleFlow}
  \scriptsize
  \vspace{.8mm}
  $\Big[\big\{\boldsymbol{x}_{+}^{\ast(h)}\rmv\rmv, \boldsymbol{P}_{+}^{(h)}\big\}^{N_\mathrm{k}}_{h=1}\Big]= \text{InvertibleFlowGMM}\Big(\big\{\boldsymbol{x}^{\ast(h)}, \boldsymbol{P}^{(h)}\big\}^{N_\mathrm{k}}_{h=1}, \V{z} \Big)$\\[1mm]
  \For{$h = 1 : N_\mathrm{k}$}{ 
  \vspace{.3mm}
  $\Big[\big\{{\boldsymbol{x}}^{(i,h)}_{1}, w_1^{(i,h)}\big\}_{i=1}^{N_\mathrm{p}} \Big] = \text{InvertibleFlow}\Big(\boldsymbol{x}^{\ast (h)}, \boldsymbol{P}^{(h)}, \V{z}  \vspace{.75mm} \Big) $\\[1mm]
  \hspace{53mm} \tcp{see \textbf{Alg.~\ref{alg:PFPF}}}
  \vspace{2mm}

  $\Big[\sim, \boldsymbol{P}^{(h)}_1 \Big] = \text{GaussianRepresentation}\Big(\big\{{\boldsymbol{x}}^{(i,h)}_{1}, w_1^{(i,h)}\big\}_{i=1}^{N_\mathrm{p}} \Big) $\\[1mm]
  \hspace{53mm} \tcp{see \textbf{Alg.~\ref{alg:GaussianApprox}}}
  \vspace{.3mm}
   }
  
$\Big[\big\{\boldsymbol{x}_{+}^{\ast(h)}\rmv\rmv, \boldsymbol{P}_{+}^{(h)}\big\}^{N'_\mathrm{k}}_{h=1}\Big]\vspace{.3mm} \rmv=\rmv \text{Resampling}\Big(\Big\{\boldsymbol{P}_1^{(h)}\rmv\rmv\rmv, \big\{ \boldsymbol{x}_1^{(i,h)}\rmv\rmv, w_1^{(i,h)}\big\}_{i=1}^{N_\mathrm{p}}  \Big\}^{N_\mathrm{k}}_{h=1}  \Big)$\\[.8mm]
  \hspace{58.5mm} \tcp{see \textbf{Alg.~\ref{alg:resampling}}}
  \vspace{.3mm}

  Output: $\big\{ \boldsymbol{x}_{+}^{\ast(h)}\rmv\rmv, \boldsymbol{P}_{+}^{(h)}\big\}^{N'_\mathrm{k}}_{h=1}$
  \vspace{.5mm}
  \caption{\mbox{\ac{gmm} Importance Samp. with Invertible Flow}}
\end{algorithm}
  \vspace{-2mm}

For $N_{\mathrm{k}} = 1$ and $N_{\mathrm{p}} > 1$, this importance sampling approach is equivalent to invertible \ac{pf} based on the \ac{edh} update step. Furthermore, if $N_{\mathrm{p}} = 1$, $N_{\mathrm{k}} > 1$, this importance sampling approach is equivalent to invertible \ac{pf} based on the \ac{ledh} update step. Note that for $N_{\mathrm{p}} \rmv=\rmv 1$, as performed by the \ac{ledh}, an additional extended or unscented  Kalman update step needs to be used to calculate an approximate covariance matrix\vspace{-4mm} \cite{LiCoates:17}.

\begin{algorithm}
  \label{alg:resampling}
  \scriptsize
  \vspace{.4mm}
  \mbox{$\Big[\big\{\boldsymbol{x}_{+}^{\ast(h')}\rmv\rmv\rmv, \boldsymbol{P}_{+}^{(h')}\big\}^{N'_\mathrm{k}}_{h'=1}\Big]\vspace{.3mm}= \text{Resampling}\Big(\Big\{\boldsymbol{P}_1^{(h)}\rmv\rmv\rmv, \big\{ \boldsymbol{x}_1^{(i,h)}\rmv\rmv,  w_1^{(i,h)}\big\}_{i=1}^{N_\mathrm{p}}  \Big\}^{N_\mathrm{k}}_{h=1} \ist \Big)$}\\[1mm]
  
     \For{$h = 1 : N_\mathrm{k}$}{\For{$i= 1 : N_\mathrm{p}$}{$w_1^{(i,h)} = \frac{w_1^{(i,h)}}{ \sum^{N_\mathrm{k}}_{h'=1} \rmv \sum^{N_\mathrm{p}}_{i'=1} w_1^{(i',h')}} $ \;
   }}
  \vspace{1.2mm}
  
  \For{$h' = 1 : N'_\mathrm{k}$}{
  \vspace{.3mm}
  Sample index $(i',\hbar)$ using $ \Big\{ \big\{w_1^{(i,h)}\big\}^{N_\mathrm{p}}_{i=1} \Big\}^{N_\mathrm{k}}_{h=1}$\;
  \vspace{.6mm}
  Set $\boldsymbol{x}_{+}^{\ast(h')} = \boldsymbol{x}^{(i',\hbar)}_{1}$ and $\boldsymbol{P}_{+}^{(h')} = \boldsymbol{P}^{(\hbar)}_{1}$\;
  }
  
  Output: $\big\{ \boldsymbol{x}_{+}^{\ast(h')}\rmv\rmv\rmv, \boldsymbol{P}_{+}^{(h')}\big\}^{N'_\mathrm{k}}_{h'=1}$
  \caption{Resampling}
    \vspace{-5mm}
\end{algorithm}

\subsection{Measurement-Origin Uncertainty (MOU)}
\label{eq:MOU1}

In a variety of estimation problems, the measurement model suffers from a deficiency beyond measurement noise referred to as \ac{mou} \cite{BarWilTia:B11}. Here, there is a single object but multiple measurements and it is not known which measurement was generated by the object. Consider a single object with state $\RV{x}$ and measurements $\RV{z}^{(m)}\rmv$, $m \rmv\in \{1,\dots,M\}$. If $m'$ is the measurement that was generated by the object, the corresponding measurement model is given by $\RV{z}^{(m')} = \M{h} (\RV{x}) + \RV{v}$. 
Based on this model, the conditional \ac{pdf} of the object-originated measurement $\V{z}^{(m')}$ reads $f_{\rmv \mathrm{o}\rmv}\big(\V{z}^{(m')}|\V{x}\big)$. All the other measurements are \acp{fp} that follow the \ac{fp} \ac{pdf} $f_{\mathrm{fp}\rmv}\big(\V{z}^{(m')}\big)$. It is assumed that at most one measurement originates from the object. The probability that the object generates a measurement is $p_\mathrm{d}$, and the mean number of \acp{fp} is Poisson distributed with mean $\mu_{\mathrm{fp}}$. 
Since it is unknown which measurement was generated by the object, a discrete and random association variable $\rv{a} \in \{0,1,\dots,M\}$ is introduced. Here, $\rv{a} \rmv=\rmv 0$ describes the event where no measurement originated from the object and $\rv{a} \rmv=\rmv m$, $m \rmv\in \{1,\dots, M\}$ describes the event where measurement $\RV{z}^{(m)}$ was originated from the object. Let $\RV{z} = [\RV{z}^{(1) \mathrm{T}},\dots,\RV{z}^{(M) \mathrm{T}} ]^{\mathrm{T}}$ be the joint measurement vector. Following common assumptions \cite{BarWilTia:B11}, conditioned on $\RV{x}$, the joint \ac{pdf} of $\RV{z}$ and $\rv{a}$  is given\vspace{1.2mm} by
\begin{equation}
  f(\V{z},a|\boldsymbol{x}) \propto \begin{cases}
           \frac{p_{\mathrm{d}} f( \V{z}^{(m)} \rmv |\ist \V{x} )} { \mu_{\mathrm{fp}} f_{\mathrm{fp}}( \V{z}^{(m)} )} , & a \rmv=\rmv m \in \{1,\dots,M \}  \\[1mm]
          1 - p_{\mathrm{d}}\ist, & a \rmv=\rmv 0. \ist
         \end{cases}  \label{eq:sumLikelihood}
         \vspace{1mm}
\end{equation}
For $\V{z}$ fixed, one can use this conditional \ac{pdf} to directly compute the \ac{mou} likelihood function 
\begin{align}
f(\V{z} |\boldsymbol{x}) &= \sum^{M}_{a = 0} f(\V{z},a|\boldsymbol{x}) \nn\\[-1.5mm]
&\propto 1 - p_{\mathrm{d}}  + \sum^{M}_{a = 1}  \frac{p_{\mathrm{d}} f( \V{z}^{(a)} \rmv |\ist \V{x} )} { \mu_{\mathrm{fp}} f_{\mathrm{fp}}( \V{z}^{(a)} )} \label{eq:likelihoodMOU}\\[-4mm]
\nn
\end{align}
and the marginal\vspace{.3mm} \ac{pmf} 
\begin{align}
p(a|\V{z})  &\propto f(a, \V{z}) \nn\\[.5mm]
&=  \int f(\V{z},a|\boldsymbol{x})  f(\V{x}) \mathrm{d} \V{x}.
\label{eq:marginal}
\end{align}
The values of the \ac{pmf} $p(a|\V{z})$ are also referred to as marginal association probabilities \cite{BarWilTia:B11}, i.e., they represent the probability of a particular association event $a \rmv\in\rmv \{0,\dots, M\}$ conditioned on an observed\vspace{-1mm} $\V{z}$. 

\subsection{Importance Sampling with Invertible \ac{pf} for Problems with \ac{mou}}
\vspace{0mm}
\label{eq:MOU2}

In principle, the \ac{mou} likelihood function in \eqref{eq:likelihoodMOU} can be directly used for importance sampling as in \eqref{eq:importanceSampling}. However, in problems with \ac{mou}, importance sampling based on invertible \ac{pf} is complicated by the fact that there are multiple measurements, and it is thus not clear which measurement should be used to compute \ac{pf} parameters \eqref{EDH_A_nonlinear} and   \eqref{EDH_b_nonlinear}, i.e., the \ac{pf} proposal \ac{pdf} $q_{\mathrm{PFL}}(\boldsymbol{x}_{1}^{{(i)}}|\boldsymbol{z})$ in \eqref{eq:invertibleMapping} cannot be directly used. To address this problem, we propose the combined\vspace{.5mm} proposal \ac{pdf}
\begin{align}
q(\boldsymbol{x}|\boldsymbol{z}) &= p(a = 0 | \V{z})  f(\V{x}) \nn\\[1.3mm]
&\hspace{5mm}+ \sum^{M}_{m=1} p(a = m | \V{z}) \ist q_{\mathrm{PFL}}\big(\boldsymbol{x}|\boldsymbol{z}^{(m)}\big)
\label{eq:proposal1}
\end{align}
where we used the marginal association probabilities $p(a | \V{z})$ to weight the proposal \acp{pdf} $q_{\mathrm{PFL}}\big(\boldsymbol{x}|\V{z}^{(m)}\big)$ related to \ac{pf} based on measurements $\V{z}^{(m)}\rmv$, $m \rmv\in\rmv \{1,\dots,M\}$ (cf.~\eqref{eq:invertibleMapping}) and the Gaussian prior \ac{pdf} $f(\V{x}) = \Set{N}(\V{x}; \V{x}_0^{\ast},\M{P})$. In particular, recall that the proposal\vspace{-.5mm} distribution $q_{\mathrm{PFL}}\big(\boldsymbol{x}|\boldsymbol{z}^{(m)})$ related to the \ac{pf} $\RV{x}_{0}\rmv\xrightarrow[]\rmv \V{z}^{(m)} \rmv\xrightarrow[]\rmv \RV{x}^{(m)}$, can be evaluated \vspace{0mm}as \begin{equation}
q_{\mathrm{PFL}}\big(\boldsymbol{x}^{{(i,m)}}|\boldsymbol{z}^{(m)}\big) = \frac{\Set{N}(\V{x}_{0}^{(i,m)}; \V{x}_0^{\ast},\M{P})} {\theta^{(m)}}
\vspace{.5mm}
\label{eq:mapping}
\end{equation}
where $\theta^{(m)}$ is the mapping factor.

A total of $(M+1) N_{\mathrm{p}}$ particles representing the proposal \ac{pdf} in \eqref{eq:proposal1} is obtained by drawing $N_{\mathrm{p}}$ particles for each of the $M + 1$ components in \eqref{eq:proposal1} and calculating corresponding marginal association probabilities and weights. For the first component related to association event $\rv{a} \rmv=\rmv 0$, $N_{\mathrm{p}}$ particles $\big\{\V{x}^{(i,0)}\big\}^{N_{\mathrm{p}}}_{i=1}$ are directly drawn from $f(\V{x})$, the corresponding marginal association probability is obtained by using \eqref{eq:sumLikelihood} in \eqref{eq:marginal}, i.e.,  $p(a \rmv=\rmv 0|\V{z}) \rmv\propto\rmv 1 - p_{\mathrm{d}}$, and the corresponding combined proposal weights are set to $\omega'^{(i,0)} \rmv=\rmv p(a \rmv=\rmv 0 | \V{z}) \ist\ist \Set{N}(\V{x}_{0}^{(i,0)}; \V{x}_0^{\ast},\M{P})$ (cf. \eqref{eq:proposal1}). For each other component related to association event $\rv{a} \rmv=\rmv m$, $m \rmv\in\rmv \{1,\dots,M\}$, first $N_{\mathrm{p}}$ particles\vspace{.2mm} $\big\{\V{x}^{(i,m)}_0\big\}^{N_{\mathrm{p}}}_{i=1}$ are drawn from $f(\V{x})$. Next, the \ac{pf} $\RV{x}_{0}\rmv\xrightarrow[]\rmv\rmv\rmv\rmv \RV{z}^{(m)} \rmv\rmv\xrightarrow[]\rmv\rmv   \RV{x}^{(m)}$ is applied\vspace{.3mm} to\vspace{-.1mm} the particles $\big\{\V{x}^{(i,m)}_0\big\}^{N_{\mathrm{p}}}_{i=1}$ and new particles $\big\{\V{x}^{(i,m)}\big\}^{N_{\mathrm{p}}}_{i=1}$ are obtained for each $m \rmv\in\rmv \{1,\dots,M\}$. An approximation of each marginal association probability $\tilde{p}(a=m|\V{z})$, $m \rmv\in\rmv \{1,\dots,M\}$ is finally calculated from these particles by using \eqref{eq:sumLikelihood} in \eqref{eq:marginal} and performing Monte Carlo integration \cite{DouFreGor:01} based on the proposal \ac{pdf} $q_{\mathrm{PFL}}\big(\boldsymbol{x}^{{(i,m)}}|\boldsymbol{z}^{(i,m)}\big)$ in \eqref{eq:mapping},\vspace{0mm} i.e.,
\begin{equation}
\tilde{p}(a\rmv=\rmv m|\V{z}) \propto \sum^{N_{\mathrm{p}}}_{i=1} \frac{p_{\mathrm{d}} \ist \theta^{(m)} f\big( \V{z}^{(m)} \rmv |\ist \V{x}^{(i,m)} \big) \ist \Set{N}(\V{x}^{(i,m)}; \V{x}_0^{\ast},\M{P})} { N_{\mathrm{p}}  \ist \mu_{\mathrm{fp}} \ist
f_{\mathrm{fp}} \big( \V{z}^{(m)} \big) \ist \Set{N}(\V{x}_0^{(i,m)}; \V{x}_0^{\ast},\M{P})}.\vspace{2mm}
\label{eq:assocProb}
\end{equation}
The\vspace{-.4mm} corresponding combined proposal\vspace{0mm} weights are\vspace{.3mm} set according \vspace{-.5mm} to (cf.~\eqref{eq:proposal1} and\vspace{-.5mm} \eqref{eq:mapping})
\begin{equation}
\omega'^{(i,m)} \ist =\ist \tilde{p}(a \rmv=\rmv m | \V{z}) \ist \frac{\Set{N}(\V{x}_0^{(i,m)}; \V{x}_0^{\ast},\M{P})} {\theta^{(m)}}.
\label{eq:proposalWeighs}
\vspace{.4mm}
\end{equation}
Finally, we reindex the resulting particles and\vspace{0mm} weights $\big\{  \big(\ist \omega'^{(i,a)}\rmv\rmv,$ $\V{x}^{(i,a)}\big) \big\}^{N_{\mathrm{p}}}_{i=1}$, $a\rmv\in\rmv\{0,\dots,M\}$ to\vspace{0mm} obtain $\big\{ \big(\ist \omega^{(l)},$ $\V{x}^{(l)}\big) \big\}^{L}_{l=1}$ where $l = i (a+1)$ and $L  = N_{\mathrm{p}} \ist  (M + 1)$.
\vspace{.4mm} 

Following the importance sampling principle\vspace{.33mm}, we next aim to compute particles $\big\{ \big(\ist w^{(\hspace{.2mm}l)}\rmv\rmv, \V{x}^{(\hspace{.2mm} l)}\big) \big\}^{L}_{l=1}$ that represent the posterior \ac{pdf} $f(\V{x}|\V{z})$. In particular, by\vspace{.2mm} plugging \eqref{eq:likelihoodMOU} into \eqref{eq:importanceSampling} and by using $\big\{ \big(\ist \omega^{(l)}\rmv\rmv, \V{x}^{(l)}\big) \big\}$ (cf.~\eqref{eq:proposalWeighs}) to represent $q(\boldsymbol{x}|\boldsymbol{z})$ in \eqref{eq:importanceSampling}, we \vspace{.5mm}  obtain
\begin{align}
  &w^{(\hspace{.2mm}l)} \rmv\propto\rmv \frac{ \Set{N}(\boldsymbol{x}^{(l)}; \V{x}_0^{\ast},\M{P}) \Big ( 1 \rmv-\rmv p_{\mathrm{d}} \rmv+\rmv \sum^{M}_{m = 1}  \rmv \frac{p_{\mathrm{d}} f( \V{z}^{(m)} \rmv |\ist \V{x}^{( l )}  ) f( \V{x}^{(l)}  )} { \mu_{\mathrm{fp}} f_{\mathrm{fp}}( \V{z}^{(m)} )} \Big) }{\omega^{( l )} }. \nn\\[-1mm]
  \label{eq:weightsFinal} \\[-7mm]
  \nn
\end{align}
The resulting set of particles $\big\{ \big(\ist w^{( \hspace{.2mm} l )}\rmv\rmv, \V{x}^{( \hspace{.2mm}  l )}\big) \big\}^{L}_{l =1}$ is an asymptotically optimal representation of $f(\V{x}|\V{z})$ for scenarios with \vspace{-1mm} \ac{mou}.
\begin{algorithm}
  \scriptsize
  \mbox{$\Big[ \big\{\boldsymbol{x}_{+}^{\ast(h)}\rmv\rmv, \boldsymbol{P}_{+}^{(h)}\big\}^{N_\mathrm{k}}_{h=1} \Big] = \text{InvertibleFlowGMMwithDA}\Big( \big\{\boldsymbol{x}^{\ast(h)}\rmv, \boldsymbol{P}^{(h)}\big\}^{N_\mathrm{k}}_{h=1}, \V{z} \Big) $}\\[1mm]
  $\Big[ \Big \{  \Big \{ \big\{  \V{x}^{(i,h,a)}, \omega^{(i,h,a)} \big\}^{N_{\mathrm{p}}}_{i=1}, \theta^{(h,a)} \Big\}^{N_{\mathrm{k}}}_{h=1}\rmv, \beta^{(a)}  \Big\}^{M}_{a=0} \hspace*{1mm} \Big] \vspace{1mm} = \hspace*{5mm} \text{Evaluation}\Big( \big\{\boldsymbol{x}^{\ast(h)}\rmv, \boldsymbol{P}^{(h)}\big\}^{N_\mathrm{k}}_{h=1}, 1, \V{z} \Big) $\\[.5mm]
  \hspace{59.8mm} \tcp{see \textbf{Alg.~\ref{alg:Evaluation}}}
  $\Big[ \big\{\boldsymbol{x}_{+}^{\ast(h)}\rmv\rmv, \boldsymbol{P}_{+}^{(h)}\big\}^{N_\mathrm{k}}_{h=1}, \sim \Big] \vspace{.7mm} = \hspace*{5mm} \text{Update}\Big( \Big \{  \Big \{ \big\{  \V{x}^{(i,h,a)}, \omega^{(i,h,a)} \big\}^{N_{\mathrm{p}}}_{i=1}, \theta^{(h,a)} \Big\}^{N_{\mathrm{k}}}_{h=1}\rmv, \beta^{(a)}, 1  \Big\}^{M}_{a=0} \hspace{.2mm} \Big) $\\[1.7mm]
  \hspace{59.8mm} \tcp{see \textbf{Alg.~\ref{alg:Update}}}
Output: $\big\{\boldsymbol{x}_{+}^{\ast(h)}\rmv\rmv, \boldsymbol{P}_{+}^{(h)}\big\}^{N_\mathrm{k}}_{h=1}$
  \vspace{.3mm}
  \caption{\ac{gmm} Importance Sampling with Invertible Flow and DA}
   \label{alg:PFDA}
   \vspace{-6mm}
\end{algorithm}

\subsection{\ac{gmm} Importance Sampling with Invertible Flow for Problems with \ac{mou}}
\vspace{-1mm}

For problems where the prior distribution is non-Gaussian and potentially multimodal, \ac{gmm} \ac{pf} with invertible flow discussed in Section \ref{sec:PFwithInvFlow} can be directly applied to problems with \ac{mou}. Pseudocode for \ac{gmm} importance sampling with the invertible flow for \ac{mou} problems is provided in \textbf{Algorithm \ref{alg:PFDA}}. Algorithm \ref{alg:PFDA} relies on the measurement evaluation presented in \textbf{Algorithm \ref{alg:Evaluation}} and the measurement update presented in  \textbf{Algorithm \ref{alg:Update}}.  Note that in Algorithm \ref{alg:Update}, for future reference, we have also introduced extrinsic data association information denoted as $\kappa^{(a)}$, $a \in \{0,\dots,M\}$. In the single object tracking considered here, we have $\kappa^{(a)} = 1$. 
\begin{algorithm}
  \label{alg:Evaluation}
  
  \scriptsize
  $\Big[ \Big \{  \Big \{ \big\{  \V{x}^{(i,h,a)}, \omega^{(i,h,a)} \big\}^{N_{\mathrm{p}}}_{i=1}, \theta^{(h,a)} \Big\}^{N_{\mathrm{k}}}_{h=1}, \beta^{(a)}  \Big\}^{M}_{a=0} \hspace{.2mm} \Big] \vspace{1mm} = \hspace*{0mm} \text{Evaluation}\Big( \big\{\boldsymbol{x}^{\ast(h)}, \boldsymbol{P}^{(h)}\big\}^{N_\mathrm{k}}_{h=1}, p, \V{z} \Big) $\\[1mm]
  Initialize association variables as $\beta^{(0)} \rmv=\rmv (1-p_\text{d}) \ist p + (1 - p )\vspace{.3mm}$\vspace{.8mm}\; 
  
  \For{$h = 1 : N_\mathrm{k}$}{
  \For{$i = 1 : N_\mathrm{p}$}{ Draw $\boldsymbol{x}_{0}^{(i,h)} \sim \mathcal{N}\big(\boldsymbol{x}^{(h)} ; \boldsymbol{x}^{\ast(h)} \rmv\rmv, \boldsymbol{P}^{(h)}\big)$\; }
  Initialize particles as $\big\{{\boldsymbol{x}}^{(i,h,0)}\big\}_{i=1}^{N_\mathrm{p}} = \big\{{\boldsymbol{x}}^{(i,h)}_{0}\big\}_{i=1}^{N_\mathrm{p}}\vspace{.7mm}$\; 
  Initialize proposal weights according to \\[.5mm]
  $\omega^{(i,h,0)} \rmv\rmv=\rmv \mathcal{N}\big(\V{x}^{(i,h,0)}; \boldsymbol{x}^{\ast(h)} \rmv\rmv, \boldsymbol{P}^{(h)} \big), i \rmv\rmv=\rmv\rmv 1,\dots,N_\mathrm{p}$\;
  }
  \vspace{.8mm}
 Initialized  mapping factor $\theta^{(h,0)} \rmv=\rmv 1$\;
 \vspace{.8mm}
  \For{$m = 1 : M$}{
  \vspace{.6mm}
  \For{$h = 1 : N_\mathrm{k}$}{
  \vspace{.6mm}
  
  Perform \ac{pf}, i.e.,
   \vspace{.8mm}
  
    $\Big[\big\{{\boldsymbol{x}}^{(i,h,m)}\rmv\big\}_{i=1}^{N_\mathrm{p}}, \big\{ A^{(h,m)}_l \big\}_{l=1}^{N_\lambda} \Big]$ \\[.2mm] 
    $\hspace{5mm} = \text{ParticleFlow}\Big(\big\{{\boldsymbol{x}}^{(i,h)}_{0}\rmv\big\}_{i=1}^{N_\mathrm{p}}, \boldsymbol{x}^{\ast(h)} \rmv\rmv, \boldsymbol{P}^{(h)}, \V{z}^{(m)} \Big)$\\[.2mm]
    \hspace*{49mm} \tcp{see \textbf{Alg.~\ref{alg:PF}}}
    \vspace{1mm}
    
    Compute mapping factor $\theta^{(h,m)}$ from $\big\{ A^{(h,m)}_l \big\}_{l=1}^{N_\lambda}$ as in \eqref{eq:mappingFactor}\;}
    \vspace{1mm}
    
     Precompute weights (cf.~\eqref{eq:proposalWeighs} and \eqref{eq:weightsFinal}), i.e., \\[1mm]
     $\omega^{(i,h,m)} \rmv=\rmv \frac{\mathcal{N}\big(\V{x}^{(i,h,0)}\rmv; \boldsymbol{x}^{\ast(h)} \rmv\rmv, \boldsymbol{P}^{(h)}\big)}{\mathcal{N}\big(\V{x}^{(i,h,m)}\rmv; \boldsymbol{x}^{\ast(h)} \rmv\rmv, \boldsymbol{P}^{(h)}\big) \theta^{(h,m)}}$,
      $i \rmv\rmv=\rmv\rmv 1,\dots,N_\mathrm{p}$\vspace{0mm}\;
      \vspace{1.5mm}
    
    Compute approximate association variables following \eqref{eq:assocProb},\vspace{1mm} i.e.,
    $\beta^{(m)} = \frac{p \ist p_{\mathrm{d}}}{N_{\mathrm{p}} N_{\mathrm{k}} \mu_{\mathrm{fp}}} \sum^{N_{\mathrm{k}}}_{h=1} \sum^{N_{\mathrm{p}}}_{i=1} \frac{ f\big( \V{z}^{(m)} \rmv \big |\ist \V{x}^{(i,h,m)} \big)} { f_{\mathrm{fp}}\big( \V{z}^{(m)} \big) \omega^{(i,h,m)} }\vspace{1mm}$\; }
    \vspace{.8mm}
  Output: $\Big \{  \Big \{ \big\{  \V{x}^{(i,h,a)}, \omega^{(i,h,a)} \big\}^{N_{\mathrm{p}}}_{i=1}, \theta^{(h,a)} \Big\}^{N_{\mathrm{k}}}_{h=1}, \beta^{(a)}  \Big\}^{M}_{a=0} $
  \vspace{.3mm}
  \caption{\ac{gmm} Importance Sampling with Invertible Flow and DA -- Measurement Evaluation}
\vspace{-3.5mm}
\end{algorithm}
Note that for later use in Section \ref{sec:proposedMethod}, Algorithm \ref{alg:Evaluation} performs measurements evaluation by also taking a probability of existence, $p$, into account. Similarly, Algorithm \ref{alg:Update} also updates $p$. By using $p \rmv=\rmv 1$ as input for Algorithm \ref{alg:Evaluation} (see  Algorithm \ref{alg:PFDA}, line 2), Algorithm \ref{alg:PFDA} is equivalent to the estimation method discussed in Sections \ref{eq:MOU1} and \ref{eq:MOU2}. Furthermore, note that for consistency with Section \ref{sec:proposedMethod}, we introduced the notation $\beta^{(a)} \propto  \tilde{p}(a | \V{z}) $, $a \in \{0,\dots,M\}$ in Algorithm~\ref{alg:PFDA}, Algorithm~\ref{alg:Evaluation}, and \vspace{-.5mm}Algorithm~\ref{alg:Update}.

\section{Review of Graph-based Multisensor \ac{mot}}
\label{sec:review} \vspace{0mm}

We will first discuss the concept of \ac{po} states and then review the \ac{spa} messages that will later be calculated based on \ac{pf}. A summary of the system model and corresponding factor graph can be found in the supplementary material \cite{ZhaMey:J23a}. Graph-based \ac{mot} will be combined with a Gaussian mixture representation and \ac{pf}-based processing in Section\vspace{0mm}~\ref{sec:PFMessages}.

\begin{algorithm}
  \label{alg:Update}
 
  \scriptsize
    $\Big[ \big\{\boldsymbol{x}_{+}^{\ast(h)}\rmv\rmv, \boldsymbol{P}_{+}^{(h)}\big\}^{N_\mathrm{k}}_{h=1}, p_{+} \Big] \vspace{.7mm} = \hspace*{2mm} \text{Update}\Big( \Big \{  \Big \{ \big\{  \V{x}^{(i,h,a)}, \omega^{(i,h,a)} \big\}^{N_{\mathrm{p}}}_{i=1}, \theta^{(h,a)} \Big\}^{N_{\mathrm{k}}}_{h=1}\rmv, \beta^{(a)},  \kappa^{(a)}   \Big\}^{M}_{a=0} \hspace{0mm}, p  \Big) $\\[1.3mm]
    
   Compute association probabilities $\tilde{p}(a|\V{z}) = \frac{\beta^{(a)} \kappa^{(a)}}{\sum^{M}_{a\rmv=\rmv0} \beta^{(a)} \kappa^{(a)}}, a \rmv\rmv=\rmv\rmv 0,\dots,M$
      
      \For{$h = 1 : N_\mathrm{k}$}{
      \vspace{1mm}  
    
     \For{$a = 0 : M$}{
    
  \vspace{1.5mm}
    Update proposal weights (cf.~\eqref{eq:proposal1}--\eqref{eq:weightsFinal}), i.e., \\[.6mm]
     $\omega'^{(i,h,a)} \rmv=\rmv \tilde{p}(a | \V{z}) \omega^{(i,h,a)}$,
      $i \rmv\rmv=\rmv\rmv 1,\dots,N_\mathrm{p}$\vspace{0mm}\;
    }
    \vspace{1.5mm}
    Reindex particles and proposal weights using $l = i (a+1)\vspace*{.3mm}$ to obtain $\hspace*{-1.8mm}\big\{ \big(\ist \omega^{(\hspace{.2mm}  l,h)}\rmv\rmv, \V{x}^{(\hspace{.2mm}  l,h)}\big) \big\}^{L}_{l =1}$ from\vspace{.3mm} $\big\{  \big(\ist \omega'^{(i,h,a)}\rmv\rmv,$ $\V{x}^{(i,h,a)}\big) \big\}^{N_{\mathrm{p}}}_{i=1}\vspace*{-.3mm}$, $\hspace*{-1.8mm} a\rmv\in\rmv\{0,\dots,M\}$\;
    \vspace{2mm}
        
    Compute final particle weights according to (cf. \eqref{eq:weightsFinal})
    \vspace{1.5mm}
  
  \For{$l = 1 : L$}{
      $w^{( \hspace{.2mm} l,h)}  = \frac{ \Big ( \ist \kappa^{(0)}   (1 - p_{\mathrm{d}}) \ist + \ist \sum^{M}_{m = 1}  \rmv \frac{\kappa^{\rmv(m)} p_{\mathrm{d}} f\big( \V{z}^{(m)} \rmv \big |\ist \V{x}^{(\hspace{.2mm}  l,h )} \big )} { \mu_{\mathrm{fp}} f_{\mathrm{fp}}\big( \V{z}^{(m)} \big)}  \ist \Big) }{\omega^{( \hspace{.2mm} l,h )} }$\;
    }
    \vspace{.8mm}
    $\Big[\sim, \boldsymbol{P}^{(h)} \Big] = \text{GaussianRepresentation}\Big(\big\{{\boldsymbol{x}}^{(l,h)}, w^{(l,h)}\big\}_{l=1}^{L_\mathrm{s}} \Big) $\\[.8mm]
    \hspace{52mm} \tcp{see \textbf{Alg.~\ref{alg:GaussianApprox}}}
    }
\vspace{1mm}

 $w = \frac{p}{N_\mathrm{k} L_\mathrm{s}} \sum_{h=1}^{N_{\mathrm{k}}} \sum_{l=1}^{L_\mathrm{s}} w^{(l,h)}$
\vspace{1.3mm}

 $p_{+} = \frac{w}{(1-p) \tilde{p}(a = 0 | \V{z}) + w}$

\vspace{2mm}

$\Big[\big\{\boldsymbol{x}_{+}^{\ast(h)}\rmv\rmv, \boldsymbol{P}_{+}^{(h)}\big\}^{N_\mathrm{k}}_{h=1}\Big]\vspace{.3mm}\hspace{-.3mm}=\hspace{-.3mm} \text{Resampling}\Big(\Big\{ \boldsymbol{P}^{(h)}\rmv\rmv\rmv, \big\{ \boldsymbol{x}^{(l,h)}\rmv\rmv,w^{(l,h)}\big\}_{l=1}^{L_\mathrm{s}}  \Big\}^{N_\mathrm{k}}_{h=1} \ist \Big)\vspace{.3mm}$\\[.8mm]
  \hspace{57.5mm} \tcp{see \textbf{Alg.~\ref{alg:resampling}}}
  \vspace{.3mm}
  
  Output: $\big\{\boldsymbol{x}_{+}^{\ast(h)}\rmv\rmv, \boldsymbol{P}_{+}^{(h)}\big\}^{N_\mathrm{k}}_{h=1}$, $p_{+}$
  \caption{\ac{gmm} Importance Sampling with Invertible Flow and DA -- Measurement Update}
  \vspace{-5.5mm}
\end{algorithm}

\subsection{\ac{po} States} 
\vspace{-.5mm}
\label{sec:stateTransition}

As in \cite{MeyBraWilHla:J17,MeyKroWilLauHlaBraWin:J18}, we consider \ac{mot} for an unknown, time-varying number of objects by introducing \ac{po} states. 
The number of \acp{po} $\rv{J}_{k-1}$ at discrete time $k-1 \rmv\geq\rmv 0$ is the maximum possible number of objects that have generated a measurement up to time $k-1$. At time $k$, a new \ac{po} is introduced for each of the $\rv{M}_k$ observed measurements, and the total number of \acp{po} is updated as $\rv{J}_k = \rv{J}_{k-1} + \rv{M}_k$. All \acp{po} that have been introduced at previous time steps are referred to as legacy \acp{po}, i.e., at time $k$, there are $\rv{J}_{k-1}$ legacy \acp{po} and $\rv{M}_k$ new \acp{po}.

The augmented state of \ac{po} $j \in \{1,\dots, J_k\}$ is given by $\RV{y}_{k}^{(j)} \rmv\triangleq\rmv \big[\RV{x}^{(j)\T}_{k} \ist \rv{r}^{(j)}_{k} \big]^\T\rmv\rmv\rmv$, where the state $\RV{x}^{(j)}_{k}$ of \ac{po} $j$ consists of the position and possibly further parameters of the object represented by the \ac{po}. Furthermore, the existence variable $\rv{r}^{(j)}_{k} \rmv\in\rmv \{0,1\}$ models the existence/nonexistence of \ac{po} $j$ in the sense that \ac{po} $j$ exists at time $k$ if and only if $\rv{r}^{(j)}_{k} \rmv\rmv=\rmv\rmv 1$. For nonexistent \acp{po}, i.e., $\rv{r}^{(j)}_{k} \!=\! 0$, the state $\RV{x}^{(j)}_{k}$ is obviously irrelevant. Thus, all \acp{pdf} of augmented \ac{po} states $f\big(\V{y}^{(j)}_{k}\big) =\rmv f\big(\V{x}^{(j)}_{k}\rmv, r^{(j)}_{k}\big)$ can be expressed as $f\big(\V{x}^{(j)}_{k}\rmv, 0 \big) = f^{(j)}_{k} f_{\text{D}}\big(\V{x}^{(j)}_{k}\big)$, where $f_{\text{D}}\big(\V{x}^{(j)}_{k}\big)$ is an arbitrary ``dummy \ac{pdf}'' and $f^{(j)}_{k} \!\rmv\in [0,1]$ is a constant. To distinguish between legacy and new \acp{po}, we denote\vspace{.5mm} by $\underline{\RV{y}}^{(j)}_{k}$ and by $\overline{\RV{y}}^{(m)}_{k}$ the augmented state of a legacy \ac{po} and a new \ac{po} states, respectively.

The concept of legacy and new \acp{po} can be extended to scenarios with $S$ sensors as follows. Let  $\rv{M}_{k,s}$ be the number of measurements at time $k$ and sensor $s \rmv\in\rmv (1,\dots, S)$, where $(1,\dots, S)$ is an arbitrary processing order of the sensors. The maximum possible number of objects that generated a measurement up to time $k$ and sensor $s$ is $J_{k,s} = J_{k,s-1} + M_{k,s}$, with $J_{k,0} \triangleq J_{k-1}\vspace{-3mm}$.

\subsection{Problem Formulation and Selected Messages of the \ac{spa}}

At each time step $k \rmv\geq\rmv 1$, we consider the tracking of an unknown number of objects based on measurements $\V{z}_{1:k}$. Object detection is performed by comparing the existence probability $p\big(r_{k}^{(j)}\! \rmv= 1 \big| \V{z}_{1:k} \big)$ with a threshold $P_{\text{th}}$, i.e., \ac{po} $j \in \{1,\dots,J_k\}$ is declared to exist if $p\big(r_{k}^{(j)}\! \rmv=\rmv 1 \big| \V{z}_{1:k} \big) \rmv>\rmv P_{\text{th}}$. Note that $p\big(r_{k}^{(j)}\! \rmv=\rmv 1 \big| \V{z}_{1:k} \big) \rmv= \int f\big(\V{x}_k^{(j)}, r_k^{(j)}\! \rmv=\rmv 1 \big| \V{z}_{1:k}\big) \ist\mathrm{d}\V{x}_k^{(j)}$.
For existent \acp{po}, state estimation is performed by calculating the \ac{mmse} estimate \cite{Poo:B94}  as $\hat{\V{x}}_k^{(j)}
\ist\triangleq \int \V{x}_k^{(j)} f\big(\V{x}_k^{(j)} \big| r_k^{(j)} \rmv=\rmv 1, \V{z}_{1:k}\big) \ist\mathrm{d}\V{x}_k^{(j)}$, where $f\big(\V{x}_k^{(j)} \big| r_k^{(j)} \rmv=\rmv 1, \V{z}_{1:k}\big) \rmv=\rmv f\big(\V{x}_k^{(j)}, r_k^{(j)} \rmv=\rmv 1 \big| \V{z}_{1:k}\big)/ p\big(r_{k}^{(j)}$ $=\rmv 1\big| \V{z}_{1:k} \big)$.

Both object detection and estimation require the marginal posterior \acp{pdf} $f\big(\V{x}_k^{(j)}, r_k^{(j)} \big| \V{z}_{1:k})\rmv \triangleq f\big(\V{y}_k^{(j)} \big| \V{z}_{1:k})$, $j \in \{1,\dots,J_k\}$. However, calculating $f\big(\V{x}_k^{(j)}, r_k^{(j)} \big| \V{z}_{1:k}\big)$ by direct marginalization is infeasible due to the large number parameters in the joint posterior distribution in \cite[Eq.~(1)]{ZhaMey:J23a}.

As in \cite{MeyBraWilHla:J17,MeyKroWilLauHlaBraWin:J18}, we consider approximate calculation by performing the loopy \ac{spa} on the factor graph in Fig.~1 of  \cite{ZhaMey:J23a} and passing messages only forward in time. This makes it possible to efficiently calculate so-called beliefs $\tilde{f}\big(\V{x}_k^{(j)}\!, r_k^{(j)}\big) \triangleq \tilde{f}\big(\V{y}_k^{(j)}\big), \ist j \in \{1,\dots,J_{k} \}$ which accurately approximate the marginal posterior \acp{pdf} $f\big(\V{x}_k^{(j)}\!, r_k^{(j)} \big| \V{z}_{1:k}\big), \ist j \in \{1,\dots,J_{k} \}$ needed for object detection and estimation. To keep computational complexity feasible, at the end of each time $k$ with all sensors processed, a suboptimal pruning step has to be performed. Here, \acp{po} with probability of existence $p^{(j)}_{k} \!\triangleq\! \tilde{p}(r^{(j)}_{k} \!=\! 1|\V{z}_{1:n})$ below a threshold $P_{\text{pr}}$ are removed from the state space.

Next, we review the \ac{spa} messages that will later be calculated based on \ac{pf}. We will limit our discussion to messages and beliefs related to legacy \ac{po} states. Messages and beliefs related to new \ac{po} states are obtained by performing similar steps. (A complete description of message passing for \ac{mot} is provided in \cite[Section~IX-A]{MeyKroWilLauHlaBraWin:J18}.)

We consider sequential sensor processing, where the \ac{spa} incorporates sensor measurements sequentially and in an arbitrary processing order at each time step. The part of the factor graph that represents the processing of the measurements of one sensor at one time step is shown in Fig.~1 of  \cite{ZhaMey:J23a}. At the processing step related to time $k$ and sensor $s$, the ``prior messages'' of legacy \ac{po} states are denoted as $\alpha_{k,s}^{(j)}\big(\underline{\V{x}}_{k,s}^{(j)},\underline{r}_{k,s}^{(j)}\big) \rmv=\rmv \alpha_{k,s}^{(j)}\big(\underline{\V{y}}_{k,s}^{(j)} \big)$. At time $k$ and sensor $s \rmv=\rmv 1$, these message are computed by a prediction step \cite[Section~IX-A1]{MeyKroWilLauHlaBraWin:J18},\vspace{1mm} i.e.,
\begin{align}
&\alpha_{k,1}^{(j)}\big(\underline{\V{x}}_{k,1}^{(j)},\underline{r}_{k,1}^{(j)}\big) = \hspace{-3mm} \sum_{r_{k-1}^{(j)} \in \{0,1\}} \int f\big(\underline{\V{x}}_{k,1}^{(j)}, \underline{r}_{k,1}^{(j)}\big|\V{x}_{k-1}^{(j)}\!, r_{k-1}^{(j)}\big)  \nn\\[1mm]
&\hspace{39.3mm}\times \tilde{f}\big(\V{x}_{k-1}^{(j)}, r_{k-1}^{(j)}\big) \mathrm{d} \V{x}_{k-1}^{(j)} \label{eq:prediction}\\[-9.8mm]
\nn \\
\nn
\end{align}
that makes use of the state-transition function $f\big(\underline{\V{x}}_{k,1}^{(j)}, \underline{r}_{k,1}^{(j)}\big|$ $\V{x}_{k-1}^{(j)}, r_{k-1}^{(j)}\big) \triangleq f\big(\underline{\V{y}}_{k,1}^{(j)},\big|\V{y}_{k-1}^{(j)}\big)$. At the time $k$ and sensor $s > 1$, the ``prior message'' is the same as the belief after the previous sensor update, i.e., $\alpha_{k,s}^{(j)}\big(\underline{\V{x}}_{k,s}^{(j)},\underline{r}_{k,s}^{(j)}\big)\triangleq \tilde{f}\big(\V{x}_{k,s-1}^{(j)}\!, r_{k,s-1}^{(j)}\big)$.
For future reference, we also introduce $\underline{p}^{(j)}_{k,s} \rmv=\rmv \int \alpha_{k,s}^{(j)}\big(\underline{\V{x}}_{k,s}^{(j)},\underline{r}_{k,s}^{(j)} \rmv=\rmv1\big) \mathrm{d} \underline{\V{x}}_{k,s}^{(j)}$ as the predicted probability of existence for each legacy \ac{po}. Note that $\underline{p}^{(j)}_{k,s} + \int \alpha_{k,s}^{(j)}\big(\underline{\V{x}}_{k,s}^{(j)}, \underline{r}_{k,s}^{(j)} \rmv=\rmv0\big) \mathrm{d} \underline{\V{x}}_{k,s}^{(j)}  = 1$.

After the prior messages have been computed, a ``measurement evaluation'' step for each legacy and each new \ac{po} is performed. Here, we denote by $a^{(j)}_{k,s}$ the association variable related to \ac{po} $j$ at the update step related to sensor $s$ at time $k$ and by $b^{(m)}_{k,s}$ the association variable related to measurement $m$ at the update step related to sensor $s$ at time $k$ (see \cite[Sec.~1.2]{ZhaMey:J23a} for details). The \ac{spa} messages that are passed from the factor nodes $q( \underline{\V{x}}^{(j)}_{k,s}, \underline{r}^{(j)}_{k,s}, a^{(j)}_{k,s}\rmv; \V{z}_{k,s})$ in \cite[Eq.~(2)]{ZhaMey:J23a} and $v\big( \overline{\V{x}}^{(m)}_{k,s}\!, \overline{r}^{(m)}_{k,s}, b^{(m)}_{k,s}\rmv; \V{z}_{k,s}^{(m)} \big)$ in \cite[Eq.~(3)]{ZhaMey:J23a} to the adjacent variables nodes $a^{(j)}_{k,s}$ and $b^{(m)}_{k,s}$, respectively, are computed. For legacy \acp{po}, these messages are given by (see \cite[Section~IX]{MeyKroWilLauHlaBraWin:J18})
\begin{align}
  &\beta_{k,s}^{(j)}\big(a_{k,s}^{(j)}\big) = \rmv\int\rmv q\big( \underline{\V{x}}^{(j)}_{k,s}, 1, a^{(j)}_{k,s}\rmv; \V{z}_{k,s} \big) \alpha_{k,s}^{(j)}\!\big(\underline{\V{x}}_{k,s}^{(j)},1\big) \mathrm{d}\underline{\V{x}}_{k,s}^{(j)} \nn\\[.5mm]
  &\hspace{38mm}+ 1(a_{k,s}^{(j)}) \ist \big(1-\underline{p}^{(j)}_{k,s}\big) \ist.\label{eq:messageBeta}\\[-5mm]
  \nn
\end{align}
For new \acp{po}, the corresponding messages are denoted as $\xi_{k,s}^{(m)}\big(b_{k,s}^{(m)}\big)$, $m \in \{1,\dots,M_{k,s}\}$ and are calculated similarly (see \cite[Section~IX]{MeyKroWilLauHlaBraWin:J18}).

Next, probabilistic \ac{da} is performed by means of iterative \ac{spa} message passing with input messages $\beta_{k,s}^{(j)}\big(a_{k,s}^{(j)}\big)$, $j \in \{1,\dots,J_{k,s-1}\}$ and $\xi_{k,s}^{(m)}\big(b_{k,s}^{(m)}\big)$, $m \in \{1,\dots,M_{k,s}\}$ (see \cite[Section~IX-A3]{MeyKroWilLauHlaBraWin:J18} for details). After convergence, corresponding output messages $\kappa_{k,s}^{(j)}\big(a_{k,s}^{(j)}\big)$, $j \in \{1,\dots,J_{k,s-1}\}$ and $\iota_{k,s}^{(m)}\big(b^{(m)}_{k,s} \big)$, $m \in \{1,\dots,M_{k,s}\}$ are available for legacy \acp{po} and new \acp{po}, respectively. Probabilistic \ac{da} is followed by a ``measurement update'' step. Here, for legacy \acp{po}, messages $\gamma_{k,s}^{(j)}\big(\underline{\V{x}}_{k,s}^{(j)}, \underline{r}_{k,s}^{(j)}\big)$ passed from $q\big( \underline{\V{x}}^{(j)}_{k,s}, \underline{r}^{(j)}_{k,s}, a^{(j)}_{k,s}\rmv; \V{z}_{k,s} \big)$ to $\underline{\V{y}}_{k,s}^{(j)}$ are calculated \vspace{0mm}as
\begin{equation}
  \gamma_{k,s}^{(j)}\big(\underline{\V{x}}_{k,s}^{(j)}, 1\big) = \rmv\sum_{a_{k,s}^{(j)}\rmv=0}^{M_{k,s}} q\big( \underline{\V{x}}^{(j)}_{k,s}, 1, a^{(j)}_{k,s}\rmv; \V{z}_{k,s} \big)  \kappa_{k,s}^{(j)}\big(a_{k,s}^{(j)}\big)
  \label{eq:messageGamma}
  \vspace{0mm}
\end{equation}
and as $\gamma_{k,s}^{(j)}\big(\underline{\V{x}}_{k,s}^{(j)}, 0\big) = \gamma_{k,s}^{(j)} =  \kappa_{k,s}^{(j)}\big(0\big)$. Measurement update for new POs is performed by following similar steps  \cite[Section~IX-A]{MeyKroWilLauHlaBraWin:J18}.

Finally, beliefs are calculated to approximate the posterior \acp{pdf} of \acp{po}. For legacy \acp{po}, beliefs $\tilde{f}\big(\underline{\V{x}}_{k,s}^{(j)}, \underline{r}_{k,s}^{(j)}\big)$ approximating $f\big(\underline{\V{x}}_{k,s}^{(j)}, \underline{r}_{k,s}^{(j)}\rmv\big|\rmv\V{z}_{1:k}\big)$ are obtained\vspace{.5mm} as 
\begin{equation}
  \tilde{f}\big(\underline{\V{x}}_{k,s}^{(j)}, 1\big) \rmv=\rmv \frac{1}{\underline{C}_{k,s}^{(j)}}\alpha_{k,s}^{(j)}\big(\underline{\V{x}}_{k,s}^{(j)}, 1\big) \gamma_{k,s}^{(j)}\big(\underline{\V{x}}_{k,s}^{(j)}, 1\big)
  \label{eq:beliefLegacy}
\end{equation}
and as $\tilde{f}\big(\underline{\V{x}}_{k,s}^{(j)}, 0\big)\rmv=\rmv \underline{f}_{k,s}^{(j)} \ist f_{\text{D}}\big(\underline{\V{x}}^{(j)}_{k,s}\big)$ with\vspace{0mm} $\underline{f}_{k,s}^{(j)} \rmv= \big(1-\underline{p}^{(j)}_{k,s}\big) \ist \gamma_{k,s}^{(j)} /\underline{C}_{k,s}^{(j)}$. The constant $\underline{C}_{k,s}^{(j)}$ is given by $\underline{C}_{k,s}^{(j)}\rmv\triangleq \int \alpha_{k,s}^{(j)}\big(\underline{\V{x}}_{k,s}^{(j)}, 1\big) \ist \gamma_{k,s}^{(j)}\big(\underline{\V{x}}_{k,s}^{(j)}, 1\big) \ist \mathrm{d}\underline{\V{x}}_{k,s}^{(j)} + \big(1-\underline{p}^{(j)}_{k,s}\big) \ist \gamma_{k,s}^{(j)}\rmv$. 

Calculating the beliefs $\tilde{f}\big(\overline{\V{x}}_{k,s}^{(m)},\overline{r}_{k,s}^{(m)}\big)$, $m \in \{1,\dots,M_{k,s}\}$ for new POs is performed by following similar steps  \cite[Section~IX-A]{MeyKroWilLauHlaBraWin:J18}. Note that this calculation of new POs involves the messages $\iota_{k,s}^{(m)}\big(b^{(m)}_{k,s} \big)$ and $\varsigma^{(m)}_{k,s}\big(\overline{\V{y}}^{(m)}_{k,s} \big)$, $m \in \{1,\dots,M_{k,s}\}$ also shown in Fig.~1 of  \cite{ZhaMey:J23a}. The resulting beliefs for legacy and new POs are used as the prior messages for measurement update of sensor $s+1$ as discussed above, i.e.,  $\alpha_{k,s+1}^{(j)}\big(\underline{\V{x}}_{k,s+1}^{(j)}, \underline{r}_{k,s+1}^{(j)}\big) \triangleq \tilde{f}\big(\V{x}_{k,s}^{(j)}, r_{k,s}^{(j)}\big)$, $j \in \{1,\dots, J_{k,s}\}$. 
 When the measurements of the last sensor in the sequence have been processed, i.e., $s = S$, the resulting beliefs are used in the prediction steps \eqref{eq:prediction} of the next\vspace{-2mm} time step $k\rmv+\rmv1$.

\section{Graph-Based Multisensor \ac{mot}\\ with Invertible \ac{pf}}
\vspace{-.5mm}
\label{sec:PFMessages}
In nonlinear \ac{mot} scenarios, calculation of $\beta_{k,s}^{(j)}\big(a_{k,s}^{(j)}\big)$ in \eqref{eq:messageBeta} and $\tilde{f}\big(\underline{\V{x}}_{k,s}^{(j)}, \underline{r}_{k,s}^{(j)}\big)$ in \eqref{eq:beliefLegacy} related to legacy \ac{po} states as well as their counterparts $\xi_{k,s}^{(m)}\big(b_{k,s}^{(m)}\big)$ and $\tilde{f}\big(\overline{\V{x}}_{k,s}^{(m)}, \overline{r}_{k,s}^{(m)}\big)$, related to new \ac{po} states cannot be performed in closed form. We propose a particle-based implementation where a proposal \ac{pdf} is established using invertible \ac{pf} as introduced in Section \ref{sec:numericalImplementation}. This makes it possible to implement multisensor \ac{mot} with high dimensional states and nonlinear measurement models. A single time step of the proposed particle-based implementation is discussed next.  At first, we assume a single Gaussian kernel as the prior knowledge for each \ac{po}.  An extension to \ac{gmm} is also presented. In what follows, we consider a single time step, remove the time index $k$, and use the index $-$ short for $k-1$.
\vspace{-4mm}

\subsection{Prediction}
\vspace{-.5mm}
\label{sec:prediction}

It is assumed that the beliefs of legacy POs at time $k\rmv-\rmv1$ are represented by a single Gaussian distribution, i.e., $\tilde{f}\big(\V{x}_{-}^{(j)}, r_{-}^{(j)} \rmv\rmv=\rmv\rmv 1\big) \rmv\rmv=\rmv\rmv p^{(j)}_{-} \ist\ist \mathcal{N}(\V{x}_{-}^{(j)}\rmv;\V{x}_{-}^{\ast(j)}\rmv,\M{P}_{-}^{(j)})$, $j\!\in\! \{1,$
$\dots,J_{-}\}$. In \ac{mot} problems, the state transition function underlying the state transition model $f\big( \underline{\V{x}}_{1}^{(j)} \big| \V{x}_{-}^{(j)} \big)$ discussed in \cite[Section VIII-C]{MeyKroWilLauHlaBraWin:J18}
 is typically linear with additive Gaussian noise,\vspace{.2mm} i.e., $\underline{\RV{x}}^{(j)}_{1} = \M{G} \RV{x}^{(j)}_{-} + \RV{u}^{(j)}$ where $\M{G}$ is the state transition matrix and $\RV{u}^{(j)}$ is an additive Gaussian noise vector with mean $\V{u}^{\ast}$ and covariance matrix $\M{P}_{\RV{u}}$. Consequently, the messages computed in the prediction step are also represented by a Gaussian distribution, i.e., $\alpha_{1}^{(j)}\big(\underline{\V{x}}_{1}^{(j)},\underline{r}_{1}^{(j)}\big) \rmv=\rmv  \underline{p}^{(j)}_{1} \ist \mathcal{N}(\underline{\V{x}}_{1}^{(j)};\underline{\V{x}}_{1}^{\ast(j)},\underline{\M{P}}_{1}^{(j)})$, $j\!\in\! \{1,\dots,J_{-}\}$ with mean, covariance matrix, and existence probability given\vspace{.4mm} by $\underline{\V{x}}_{1}^{\ast(j)} \rmv=\rmv \M{G} \V{x}^{\ast(j)}_{-} \rmv+\rmv \V{u}^{\ast}$, $\underline{\M{P}}_{1}^{(j)} = \M{G} \M{P}_{-}^{(j)} \M{G}^{\mathrm{T}} \rmv+\rmv \M{P}_{\RV{u}}$, and\vspace{-.5mm}  ${\underline{p}}^{(j)}_{1} \rmv=\rmv p_{\mathrm{su}} \ist p^{(j)}_{-} $, respectively. \rdd{Here, $p_{\mathrm{su}}$ is the survival probability, i.e., the probability that an object that exists at time step $k\rmv-\rmv1$, still exists at time step $k$.} Here If the state transition function is not linear with additive Gaussian noise, for each $j\!\in\! \{1,\dots,J_{-}\}$, $N_{\mathrm{p}}$ particles are drawn from $\mathcal{N}(\V{x}_{-}^{(j)}\rmv;\V{x}_{-}^{\ast(j)}\rmv,\M{P}_{-}^{(j)})$, the prediction step of a conventional particle filter is performed \cite{AruMasGorCla:02}, and a predicted Gaussian representation $\mathcal{N}(\underline{\V{x}}_{1}^{(j)};\underline{\V{x}}_{1}^{\ast(j)},\underline{\M{P}}_{1}^{(j)})$ is computed from the resulting particles using Algorithm\vspace{-1mm} \ref{alg:GaussianApprox}.

\subsection{Measurement Evaluation}
\label{sec:samplingFlowEvaluation}

The following steps are performed sequentially for each sensor $s \rmv=\rmv 1,\dots, S$. First, a particle representation $\big\{\big(\underline{\V{x}}^{(i,j)}_{0,s},\underline{\omega}^{(i,j)}_{0,s}\big)\big\}_{i=1}^{N_\mathrm{p}}$ is obtained by drawing particles $\underline{\V{x}}^{(i,j)}_{0,s}$, $i \rmv\in\rmv \{1,\dots,N_{\mathrm{p}} \}$ from $\mathcal{N}(\underline{\V{x}}_{s}^{(j)};\underline{\V{x}}_{s}^{(j)\ast}\rmv\rmv,\underline{\M{P}}_{s}^{(j)})$ and setting the corresponding weights to $\underline{\omega}^{(i,j)}_{0,s} \rmv=\rmv \underline{p}^{(j)}_{s} /N_{\mathrm{p}}$. Next, we compute an extended set $\big\{\big\{\big(\underline{\V{x}}^{(i,a,j)}_{s},\underline{\omega}^{(i,a,j)}_{s}\big)\big\}_{i=1}^{N_{\mathrm{p}}}\big\}_{a=0}^{M_{s}}$, that consists of $N_{\mathrm{p}}$ particles and weights for each value \vspace{.2mm} of $a_{s}^{(j)} \in \{0,\dots,M_{s}\}$ and $j\!\in\! \{1,\dots, J_{s-1}\}$. (Note that measurement gating \cite{BarWilTia:B11} can be employed to reduce the number of measurements used for \ac{pf}.) For $a_{s}^{(j)} \rmv=\rmv 0$, we perform no flow, i.e., we set $\big\{\big(\underline{\V{x}}^{(i,0,j)}_{s}\rmv\rmv,\underline{\omega}^{(i,0,j)}_{s}\big)\big\}_{i=1}^{N_{\mathrm{p}}}\! =\! \big\{\big(\underline{\V{x}}^{(i,j)}_{0,s}\rmv\rmv,\underline{\omega}^{(i,j)}_{0,s}\big)\big\}_{i=1}^{N_{\mathrm{p}}}$. For $a_{s}^{(j)} \rmv= m\!\in\! \{1,\dots,M_{s}\}$, the \ac{pf} $\underline{\RV{x}}^{(j)}_{0,s}\rmv\xrightarrow[]\rmv \V{z}_{s}^{(m)} \rmv\xrightarrow[]\rmv \underline{\RV{x}}_{s}^{(j,m)}$ is applied to obtain new particles $\big\{\underline{\V{x}}^{(i,m,j)}_{s}\big\}_{i=1}^{N_\mathrm{p}}$ by migrating the particles $\big\{\underline{\V{x}}^{(i,m,j)}_{0,s}\big\}_{i=1}^{N_\mathrm{p}}$. By making use of the invertible \ac{pf} principle (cf.~\eqref{eq:invertibleMapping}) \cite{LiCoates:17}, the weights $\underline{\omega}_{s}^{(i,m,j)}$ corresponding to the migrated particles $\underline{\V{x}}^{(i,m,j)}_{s}$ are obtained \vspace{.5mm} as
\begin{equation}
  \underline{\omega}_{s}^{(i,m,j)} \rmv= \rmv\frac{\mathcal{N}(\underline{\V{x}}_{s}^{(i,m,j)};\underline{\V{x}}_{s}^{(j)\ast}\rmv\rmv\rmv,\underline{\M{P}}_{s}^{(j)}) \ist\underline{\theta}^{(j)}_{m,s}\!}{\mathcal{N}(\underline{\V{x}}_{0,s}^{(i,j)};\underline{\V{x}}_{s}^{(j)\ast}\rmv\rmv\rmv,\underline{\M{P}}_{s}^{(j)})} \ist \underline{\omega}_{0,s}^{(i,j)}, i \in \{1,\dots,N_{\mathrm{p}}\} \nn
  \vspace{.5mm}
\end{equation}
with mapping factor $\underline{\theta}^{(j)}_{m,s}$ (cf.~\eqref{eq:mappingFactor}). Note that the sets of weighted particles\vspace{.5mm} $\big\{\underline{\V{x}}_{s}^{(i,m,j)}\rmv\rmv,\underline{\omega}_{s}^{(i,m,j)}\big\}_{i=1}^{N_\mathrm{p}}$, $m \in \{1,\dots,$ $M_{s}\}$, despite all being based on a different proposal \ac{pdf} $q_{\mathrm{PFL}}\big(\underline{\V{x}}_{s}^{(j)}|\V{z}_{s}^{(m)}\big)$, $m \in \{1,\dots,$ $M_{s}\}$, still represent $\alpha_{s}^{(j)}\rmv\big(\underline{\V{x}}_{s}^{(j)}\rmv\rmv,1\big)$.
The result of this particle migration along the flow defined by a measurement $\V{z}_{s}^{(m)}$, $m \in \{1,\dots, M_{s}\}$, is that the particles are now at locations where the evaluation of the corresponding likelihood function $f\big( \V{z}_{s}^{(m)} \rmv\big|\ist \underline{\V{x}}_{s}^{(j)} \big)$ will produce a significant particle weight. Particle degeneracy and thus approximate message-passing operations accurately even if the dimension of the state is high\vspace{0mm} \cite{MeyBraWilHla:J17}.

The measurement evaluation step can now be performed on weighted particles $\big\{\big\{\big(\underline{\V{x}}_{s}^{(i, a,j)},\underline{\omega}_{s}^{(i,a,j)}\big)\big\}_{i=1}^{N_\mathrm{p}}\big\}_{a=0}^{M_{s}}$ by calculating an approximation $\tilde{\beta}_{s}^{(j)}(a)$ of the messages $\beta_{s}^{(j)}(a)$ in \eqref{eq:messageBeta} for all $j \!\in\! \{1,\dots,J_{s-1}\}$, $a \!\in\! \{0,\dots,M_{s}\}$\vspace{-1mm} as
\begin{align}
  \tilde{\beta}_{s}^{(j)}\big(a_{s}^{(j)} \rmv=\rmv a\big) &=\rmv \rmv\sum_{i=1}^{N_\mathrm{p}}\rmv q\big( \underline{\V{x}}^{(i,a,j)}_{s}\!, 1, a; \V{z}_{s} \big)\ist \underline{\omega}_{s}^{(i,a,j)} \nn\\
  &\hspace{23mm}+ 1(a) \ist \big(1-\tilde{\underline{p}}^{(j)}_{s}\big). \nn
\end{align} 

For the computation of a particle representation of new \ac{po} states, we first draw particles $\overline{\V{x}}^{(i)}_{0,s}$, $i \rmv\in\rmv \{1,\dots,N_{\mathrm{p}} \}$ from $f_{\mathrm{b}}\big(\overline{\V{x}}\big)$ as introduced in [Sec.~1.1]\cite{ZhaMey:J23a}. Next, for each new \ac{po} $j = J_{s-1}+m$, $m\!\in\! \{1,\dots,M_{s}\}$, new particles and corresponding weights $\big\{\overline{\V{x}}^{(m,i)}_{s}, \overline{w}^{(m,i)}_{s} \big\}_{i=1}^{N_\mathrm{p}}$ are obtained from $\big\{\overline{\V{x}}^{(i)}_{0,s}\big\}_{i=1}^{N_{\mathrm{p}}}$ by performing the invertible \ac{pf} $\overline{\RV{x}}_{0,s}\rmv\rmv\xrightarrow[]\rmv \V{z}_{s}^{(m)} \rmv\rmv\xrightarrow[]\rmv \overline{\RV{x}}_{s}^{(m)}$. Note that this flow relies on the mean $\overline{\boldsymbol{x}}_{\text{b}}$ and covariance matrix $\overline{\boldsymbol{P}}_{\text{b}}$ of $f_{\mathrm{b}}\big(\overline{\V{x}}\big)$. Finally, for each $m\!\in\! \{1,\dots,M_{s}\}$ approximate messages $\tilde{\xi}_{s}^{(m)}\big(b_{s}^{(m)}\big)$ are calculated from $\big\{\overline{\V{x}}^{(m,i)}_{s}\rmv\rmv,\overline{w}^{(m,i)}_{s}\big\}_{i=1}^{N_\mathrm{p}}$ by performing the same steps as described above for the calculation of $\tilde{\beta}_{s}^{(j)}\big(a_{s}^{(j)}\big)$. These messages are used as an input for the iterative \ac{spa} for data association \cite{WilLau:J14,MeyBraWilHla:J17,MeyKroWilLauHlaBraWin:J18} performed next (see \cite[Sec.~VI]{MeyKroWilLauHlaBraWin:J18} for details)\vspace{-3mm}.

\subsection{Measurement Update and Belief Calculation}
\vspace{-.5mm}
After the iterative loopy \ac{spa} for data association has been converged, the messages $\tilde{\kappa}_{s}^{(j)}\big(a_{s}^{(j)}\big)$, $j \!\in\! \{1,\dots,J_{s-1}\}$ and $\tilde{\iota}_{s}^{(m)}\big(b_{s}^{(m)}\big)$, $m \!\in\! \{1,\dots,M_{s}\}$ are available. These messages are used to obtain an approximation $\tilde{\gamma}_{s}^{(j)}\big(\underline{\V{x}}_{s}^{(j)}, 1\big) $ of the messages $\gamma_{s}^{(j)}\big(\underline{\V{x}}_{s}^{(j)}\!, 1\big) $, $j \!\in\! \{1,\dots,J_{s-1}\}$ in \eqref{eq:messageGamma} as well as an approximation $\tilde{\varsigma}_{s}^{(m)}\big(\overline{\V{x}}_{s}^{(m)}\!, 1\big)$ of the messages $\varsigma_{s}^{(m)}\big(\overline{\V{x}}_{s}^{(m)}\!, 1\big)$, $m \!\in\! \{1,\dots,M_{s}\}$  in \cite[Section~IX]{MeyKroWilLauHlaBraWin:J18}.

Beliefs approximating the posterior \ac{pdf} of \acp{po} are now computed by means of importance sampling. As in Section \ref{eq:MOU2}, we use the marginal association probabilities $p(a^{(j)}_{s} | \V{z}_{s})$ to weight the proposal \acp{pdf} $q_{\mathrm{PFL}}\big(\underline{\V{x}}_{s}^{(j)}|\V{z}_{s}^{(m)}\big)$ related to \ac{pf} based on measurements $\V{z}_{s}^{(m)}\rmv$, $m \rmv\in\rmv \{1,\dots,M\}$. Note that in \ac{mot} scenarios, accurate approximations of $p(a^{(j)}_{s} | \V{z}_{s})$ can be obtained as $\tilde{p}(a^{(j)}_{s} | \V{z}_{s}) \propto  \tilde{\beta}_{s}^{(j)}\big(a_{s}^{(j)} \big)  \tilde{\kappa}_{s}^{(j)}\big(a_{s}^{(j)} \big)$ (see \cite[Section~IX]{MeyKroWilLauHlaBraWin:J18} for details). Consequently, we obtain a new set of reindex particles and\vspace{0mm} weights $\big\{ \big(\ist \underline{\V{x}}_{s}^{(l,j)}\rmv\rmv, \underline{\omega}_{s}^{(l,j)}\rmv\rmv,$ $\big) \big\}^{L_{s}}_{l=1}$ from $\big\{  \big(\ist \underline{\V{x}}^{(i,a,j)}\rmv\rmv,$ $\underline{\omega}^{(i,a,j)}\big) \big\}^{N_{\mathrm{p}}}_{i=1}$, $a\rmv\in\rmv\{0,\dots,M_{s}\}$ by using $l = i (a+1)$, $\underline{\omega}_{s}^{(l,j)} = \underline{\omega}^{(i,a,j)}\rmv/\tilde{p}(a^{(j)}_{s} \rmv=\rmv a | \V{z}_{s})$, and \vspace{0mm} $L_{s} = N_{\mathrm{p}} \ist  (M_{s} + 1)$.

\begin{algorithm}
  \label{alg:SingleSensorUpdate}
  \scriptsize
  $\Big[\Big\{\rmv\big\{{\boldsymbol{x}}^{\ast(h,j)}_{+}, \boldsymbol{P}_{+}^{(h,j)}\big\}_{h=1}^{N_\mathrm{k}}, p_{+}^{(j)}\rmv\Big\}_{j=1}^{J_{s}} \Big] \rmv=\hspace*{5.4mm} \text{SingleSensorUpdate}\Big(\rmv\Big\{\rmv\big\{\underline{\boldsymbol{x}}^{\ast(h,j)}\rmv\rmv, \underline{\boldsymbol{P}}^{(h,j)}\big\}_{h=1}^{N_\mathrm{k}}, \underline{p}^{(j)}\Big\}_{j=1}^{J_{s-1}}\rmv\rmv,\V{z}_{s} \rmv\Big) $\\
  \vspace{.8mm}
  
  $\Big[ \Big \{ \big\{\overline{\boldsymbol{x}}_{+}^{\ast(h,m)}\rmv\rmv, \overline{\boldsymbol{P}}_{+}^{(h,m)}\big\}^{N_\mathrm{k}}_{h=1}, \xi^{(m)} \Big\}^{M_s}_{m=1}  \Big] = \text{NewObjects}\big(\V{z}_{s}\rmv\big)$ \\[.6mm]
  \hspace{50.4mm} \tcp{see \textbf{Alg. \ref{alg:newObjects}}}
  \vspace{1.2mm}
  
  \For{$j = 1 : J_{s-1}$}{
  \vspace{.8mm}

$\Big[ \Big \{  \Big \{ \big\{  \underline{\V{x}}^{(i,h,a,j)},  \underline{w}^{(i,h,a,j)} \big\}^{N_{\mathrm{p}}}_{i=1}, \theta^{(h,a,j)} \Big\}^{N_{\mathrm{k}}}_{h=1}, \beta^{(a,j)}  \Big\}^{M_s}_{a=0} \hspace{.2mm} \Big] \vspace{1mm} = \hspace*{0mm} \text{Evaluation}\Big( \big\{\underline{\boldsymbol{x}}^{\ast(h,j)}\rmv, \underline{\boldsymbol{P}}
^{(h,j)}\big\}^{N_\mathrm{k}}_{h=1}, \underline{p}^{(j)}\rmv\rmv, \V{z}_s \Big)$ \\[.5mm]
\hspace{45.0mm} \tcp{see \textbf{Alg. \ref{alg:Evaluation}}}}
\vspace{1.5mm}

  $\Big[ \big\{\big\{\kappa^{(a,j)} \big\}^{M_s}_{a = 0}\big\}^{J_{s-1}}_{j=1}\rmv, \big\{\iota^{(a)}\big\}^{M_s}_{a = 0}  \vspace{0mm}\Big] =\vspace{.5mm}$\\ $\hspace{8mm}\text{DataAssociation}\Big(\big\{\big\{\beta^{(a,j)} \big\}^{M_s}_{a = 0}\big\}^{J_{s-1}}_{j=1}\rmv\rmv, \big\{\xi^{(a)}\big\}^{M_s}_{a = 0}  \Big)$\\[1mm]
  \hspace{45mm} \tcp{see \textbf{\cite[Sec.~VI]{MeyKroWilLauHlaBraWin:J18}}}
\vspace{1.5mm}
  
\text{Compute}  $\overline{p}_{+}^{(m)} = \frac{(\xi^{(m)}-1)\iota^{(m)}}{(\xi^{(m)}-1)\iota^{(m)} + 1}$, $m \rmv\in\rmv \{1,\dots,M_s\} $ \\[2mm]
  
  \For{$j = 1 : J_{s-1}$}{
  \vspace{2mm}
  $\Big[ \big\{\underline{\boldsymbol{x}}_{+}^{\ast(h,j)}\rmv\rmv, \underline{\boldsymbol{P}}_{+}^{(h,j)}\big\}^{N_\mathrm{k}}_{h=1}, \underline{p_{+}}^{\rmv\rmv(j)} \Big] \vspace{.7mm} = \text{Update} \Big( \Big \{  \Big \{ \big\{  \underline{\V{x}}^{(i,h,a,j)},$\\
   \hspace*{3mm} $\underline{w}^{(i,h,a,j)} \big\}^{N_{\mathrm{p}}}_{i=1}, \theta^{(h,a,j)} \Big\}^{N_{\mathrm{k}}}_{h=1}\rmv, \beta^{(a,j)}, \kappa^{(a,j)}   \Big\}^{M}_{a=0}\rmv,  \underline{p}^{(j)}  \Big)$ \\[2mm]
\hspace{45mm} \tcp{see \textbf{Alg. \ref{alg:Update}}}
\vspace{1.5mm}
   }
   \vspace{1.5mm}
       For $\mathpzc{j} \rmv\rmv=\rmv\rmv 1,\dots,J_{s-1}$,\vspace{.3mm} reindex legacy object state information\vspace{.2mm} according to
       \hspace*{0mm}$\big\{{\boldsymbol{x}}^{\ast(h,\mathpzc{j})}_{+}\rmv, \boldsymbol{P}_{+}^{(h,\mathpzc{j})}\big\}_{h=1}^{N_\mathrm{k}} = \big\{{\underline{\boldsymbol{x}}}^{\ast(h,\mathpzc{j})}_{+}\rmv, \underline{\boldsymbol{P}}_{+}^{(h,\mathpzc{j})}\big\}_{h=1}^{N_\mathrm{k}}$ and $p_{+}^{(\hspace{.15mm} \mathpzc{j})} = \underline{p}_{+}^{(\mathpzc{j})}\rmv$.\\
    \vspace{2mm}
    
    For $\mathpzc{j} \rmv\rmv=\rmv\rmv J_{s-1} \rmv+\rmv 1, \dots J_s$, reindex new object state information\vspace{.2mm} following
       \hspace*{0mm}$\big\{{\boldsymbol{x}}^{\ast(h,\mathpzc{j})}_{+}\rmv, \boldsymbol{P}_{+}^{(h,\mathpzc{j})}\big\}_{h=1}^{N_\mathrm{k}} \rmv=\rmv \big\{{\overline{\boldsymbol{x}}}^{\ast(h,m)}_{+}\rmv, \overline{\boldsymbol{P}}_{+}^{(h,m)}\big\}_{h=1}^{N_\mathrm{k}}$ and\vspace{.2mm} $p_{+}^{(\hspace{.15mm} \mathpzc{j})} = \overline{p}_{+}^{(m)}\rmv$ using $m = \mathpzc{j} - J_{s-1}$ and $J_s = J_{s-1} + M_s$.\vspace{.3mm} \\
    \vspace{1.4mm}

  Output: $\Big\{\rmv\big\{{\boldsymbol{x}}^{\ast(h,\mathpzc{j})}_{+}, \boldsymbol{P}_{+}^{(h,\mathpzc{j})}\big\}_{h=1}^{N_\mathrm{k}}, p_{+}^{(\hspace{.15mm} \mathpzc{j})}\rmv\Big\}_{\mathpzc{j}=1}^{J_{s}}$
  \caption{Gaussian Mixture Implementation of Multisensor \ac{mot} -- Single Sensor Update Step}
  \vspace{-6mm}
\end{algorithm}

Next, based on \eqref{eq:beliefLegacy}, we update the particle weights of the legacy \acp{po} $j \!\in\! \{1,\dots,J_{s-1}\}$\vspace{1mm} by first computing
\begin{equation}
  \underline{w}_{s}'^{(l,j)} = \tilde{\gamma}_{s}^{(j)}\!\big(\underline{\V{x}}_{s}^{(l,j)}\rmv\rmv, 1\big) \ist \underline{\omega}_{s}^{(l,j)}, \quad l \in \{1,\dots,L_{s}\} \nn
  \vspace{.5mm}
\end{equation}
and then calculating normalized weights as 
\begin{align}
  \underline{w}_{s}^{(l,j)} = \frac{\underline{w}_{s}'^{(l,j)}}{\sum_{l=1}^{L_{s}}\underline{w}_{s}'^{(l,j)}+\big(1-\tilde{\underline{p}}^{(j)}_{s}\big) \tilde{\gamma}_{s}^{(j)}}, \quad l \in \{1,\dots,L_{s}\}. \nn \\[-4mm]
  \label{eq:weightNormLegacy} \\[-7mm]
  \nn
\end{align}
Note that denominator of \eqref{eq:weightNormLegacy} is a particle-based approximation of $\underline{C}_{s}^{(j)}$ in \eqref{eq:beliefLegacy}. The resulting particles and weights $\big\{\big(\underline{\V{x}}_{s}^{(l,j)}\rmv\rmv,\underline{w}_{s}^{(l,j)}\big)\big\}_{l=1}^{L_{s}}$ represent\vspace{.3mm} the belief $\tilde{f}\big(\underline{\V{x}}_{s}^{(j)}\rmv\rmv, 1 \big)$ of legacy \ac{po} $j \rmv\in\rmv \{1,\dots,J_{s-1}\}$. These particles can be used to calculate an approximation of the existence probability as $\tilde{p}_{s}^{(j)} = \sum_{l=1}^{L_{s}}\underline{w}_{s}^{(l,j)}$. A Gaussian representation of the belief $\tilde{f}\big(\V{x}_{s}^{(j)}\rmv\rmv, r_{s}^{(j)} \big) \rmv\triangleq\rmv \tilde{f}\big(\underline{\V{x}}_{s}^{(j)}\rmv\rmv, \underline{r}_{s}^{(j)} \big)$ is furthermore obtained by applying Algorithm \ref{alg:GaussianApprox}, which calculates a mean $\V{x}_{s}^{(j)\ast}$ and a covariance matrix $\M{P}_{s}^{(j)}$ from $\big\{\big(\underline{\V{x}}_{s}^{(l,j)}\rmv\rmv,\underline{w}_{s}^{(l,j)}/\tilde{p}_{s}^{(j)}\big)\big\}_{l=1}^{L_{s}}$\vspace{.5mm}.

For new \acp{po} $j \rmv\in\rmv \{J_{s-1} \rmv+\rmv1,\dots, J_{s}\}$\vspace{-.2mm}, approximate existence probabilities $\tilde{p}_{s}^{(j)}$ and a\vspace{0mm} Gaussian representation of the beliefs $\tilde{f}\big(\V{x}_{s}^{(j)}\rmv\rmv, r_{s}^{(j)} \big) \rmv\triangleq\rmv \tilde{f}\big(\overline{\V{x}}_{s}^{(m)}\rmv\rmv, \overline{r}_{s}^{(m)} \big)$,  for $m = j - J_{s-1}$ are calculated by performing similar steps as discussed above for legacy \acp{po}. Existence probabilities and Gaussian representations of the beliefs related to \ac{po} that have not been pruned are then used as input for processing measurements of the next sensor $s+1$ or, in case $s = S$, for processing at the next time step.  

For \acp{po} that have been declared to exist after the last sensor update, i.e., $\tilde{p}_{S}^{(j)} \rmv>\rmv P_{\text{th}}$, an approximate \ac{mmse} state estimate is directly given by the mean of the Gaussian representation, i.e., $\hat{\V{x}}^{(j)} \rmv\approx\rmv \V{x}_{S}^{\ast(j)}$.
\vspace{-3mm}

\begin{algorithm}
  \label{alg:newObjects}
  \scriptsize
  $\Big[ \Big \{ \big\{\overline{\boldsymbol{x}}_{+}^{\ast(h,m)}\rmv\rmv, \overline{\boldsymbol{P}}_{+}^{(h,m)}\big\}^{N_\mathrm{k}}_{h=1}, \xi^{(m)} \Big\}^{M}_{m=1}  \Big] \rmv=\rmv \text{NewObjects}\big(\V{z}_{s}\rmv\big) $\\[.8mm]
  
  Draw $\overline{\boldsymbol{x}}^{\ast(h)}_{\text{b}}\rmv\rmv, h \rmv\rmv=\rmv\rmv 1,\dots,N_{\mathrm{k}}$ from $f_{\mathrm{b}}(\overline{\V{x}})$\;
  \vspace{.8mm}
  
  Compute covariance matrix $\overline{\boldsymbol{P}}_{\text{b}}$ from $f_{\mathrm{b}}(\overline{\V{x}})$\;
  \vspace{1.2mm}
  
  \For{$m = 1 : M_{s}$}{
  \vspace{1.5mm}
    \For{$h = 1 : N_\mathrm{k}$} {
    \vspace{.8mm}
      $\Big[\big\{{\overline{\boldsymbol{x}}}^{(i,h,m)}, \overline{w}^{(i,h,m)}\big\}_{i=1}^{N_\mathrm{p}}\Big] =$\\[.5mm] $\hspace*{5.5mm} \text{InvertibleFlow}\Big(\ist\overline{\boldsymbol{x}}^{\ast(h)}_{\mathrm{b}}\rmv\rmv, \overline{\boldsymbol{P}}_{\mathrm{b}}, \V{z}^{(m)}_{s} \Big) $\;
     \hspace{48mm} \tcp{see \textbf{Alg. \ref{alg:invertibleFlow}}}
    \vspace{0mm}
      $\Big[\sim,\overline{\boldsymbol{P}}^{(h,m)} \Big]\vspace{.5mm} = \hspace*{0mm} \text{GaussianRepresentation}\Big(\big\{{\overline{\boldsymbol{x}}}^{(i,h,m)}, \overline{w}^{(i,h,m)}\big\}_{i=1}^{N_\mathrm{p}} \Big) \vspace{1.5mm}$\;
      \hspace{48mm} \tcp{see \textbf{Alg. \ref{alg:GaussianApprox}}}
    }
    \vspace{1mm}
    $\xi^{(m)}=1+\frac{\mu_\mathrm{b}}{N_{\mathrm{k}} N_{\mathrm{p}} \mu_\mathrm{fp} f_\mathrm{fp}\big(\boldsymbol{z}_{s}^{(m)}\big)}\sum_{h=1}^{N_{\mathrm{k}}} \sum_{i=1}^{N_{\mathrm{p}}}\overline{w}^{(i,h,m)}$ \;
\vspace{1.5mm}
$\Big[\big\{\overline{\boldsymbol{x}}_{+}^{\ast(h,m)}\rmv\rmv, \overline{\boldsymbol{P}}_{+}^{(h,m)}\big\}^{N_\mathrm{k}}_{h=1}\Big]\vspace{1mm}=$ \\[-.5mm]
 $\hspace{	11mm}\text{Resampling}\Big(\Big\{\overline{\boldsymbol{P}}^{(h,m)}\rmv\rmv\rmv, \big\{ \overline{\boldsymbol{x}}^{(i,h,m)}\rmv\rmv,\overline{w}^{(i,h,m)}\big\}_{i=1}^{N_\mathrm{p}}  \Big\}^{N_\mathrm{k}}_{h=1} \ist \Big)$\vspace{1mm}\;
 \hspace{53.5mm} \tcp{see \textbf{Alg. \ref{alg:resampling}}}
  }
    Output: $\Big \{ \big\{\overline{\boldsymbol{x}}_{+}^{\ast(h,m)}\rmv\rmv, \overline{\boldsymbol{P}}_{+}^{(h,m)}\big\}^{N_\mathrm{k}}_{h=1}, \xi^{(m)} \Big\}^{M}_{m=1}$
    \vspace{.5mm}
  \caption{Gaussian Mixture Implementation of Multisensor \ac{mot} -- Generation of New POs}
\vspace{-1mm}
\end{algorithm}

\subsection{The Proposed Multisensor \ac{mot} Method}
\vspace{0mm}
\label{sec:proposedMethod}

In multisensor \ac{mot} problems with nonlinear measurement models, object beliefs are non-Gaussian and potentially multimodal. Here, graph-based multisensor \ac{mot} with invertible \ac{pf} has to be combined with a \ac{gmm} discussed in Section \ref{sec:PFwithInvFlow}. Pseudocode for a single time step of the resulting multisensor \ac{mot} method is provided in \textbf{Algorithm  \ref{alg:PFDAPF-MM-GMM}}. Algorithm \ref{alg:PFDAPF-MM-GMM} relies on the single sensor update step provided in \textbf{Algorithm  \ref{alg:SingleSensorUpdate}} which, in turn, relies on the introduction of new \acp{po} presented in \textbf{Algorithm \ref{alg:newObjects}},  measurement evaluation presented in Algorithm \ref{alg:Evaluation} and measurement update presented in Algorithm \ref{alg:Update}. Note that at time $k=0$, Algorithm \ref{alg:PFDAPF-MM-GMM} is typically initialized by setting $J_{-} = 0$. However, if prior information is available, it can be incorporated in the form of the set $\Big\{\rmv\big\{\boldsymbol{x}_{-}^{\ast(h,j)}\rmv\rmv, \boldsymbol{P}_{-}^{(h,j)}\big\}_{h=1}^{N_\mathrm{k}}, p_{-}^{(j)}\Big\}_{\rmv j=1}^{\rmv J_{-}}$. Note that in a \ac{gmm}, an approximate MMSE estimate\vspace{.5mm} can be obtained as $\hat{\V{x}}^{(j)} \rmv\approx\rmv \frac{1}{N_{\mathrm{k}}} \sum^{N_{\mathrm{k}}}_{h\rmv=\rmv1} \V{x}_{S}^{\ast(h,j)}\vspace{0mm}$.
{\begin{algorithm}
  \label{alg:PFDAPF-MM-GMM}
 \vspace{.5mm}
  \scriptsize
  $\Big[ \ist \Big\{\rmv\big\{{\boldsymbol{x}}^{\ast(h,j)}\rmv\rmv, \boldsymbol{P}^{(h,j)}\big\}_{h=1}^{N_\mathrm{k}}, p^{(j)}\rmv\Big\}_{j=1}^{J} \ist \Big] \rmv=\rmv \text{MultisensorMOT}\Big[\rmv\Big\{\rmv\big\{\boldsymbol{x}_{-}^{\ast(h,j)}\rmv\rmv, \boldsymbol{P}_{-}^{(h,j)}\big\}_{h=1}^{N_\mathrm{k}}, p_{-}^{(j)}\Big\}_{\rmv j=1}^{\rmv J_{-}},\V{z} \Big] $\\[.5mm]
  
  Perform prediction\vspace{.5mm} step according to
  
    \For{$j = 1 : J_{-}$}{
    \vspace{.5mm}
    \For{$h = 1 : N_{\mathrm{k}}$}{
  \vspace{.8mm}
  
  $\underline{\V{x}}_{1}^{\ast(h,j)} \rmv=\rmv \M{G} \ist \V{x}^{\ast (h,j)}_{-} \rmv+\rmv \V{u}^{\ast}$\\[1.5mm]
  $\underline{\M{P}}_{1}^{(h,j)} = \M{G} \hspace{.1mm} \M{P}_{-}^{(h,j)} \hspace{-.1mm} \M{G}^{\mathrm{T}} \rmv+\rmv \M{P}_{\RV{u}}$\\[1.5mm]
  }
  \vspace{1mm}
  $\underline{p}^{(j)}_{1} = p_{\mathrm{su}} \ist p^{(j)}_{-} $
  \vspace{.5mm}
  }
  \vspace{1.5mm}

  Perform single sensor update steps\vspace{.5mm} sequentially, i.e.,
  
  \For{$s = 1 : S$}{
  \vspace{.8mm}
    $\Big[\rmv\Big\{\rmv\big\{{\boldsymbol{x}}_s^{\ast(h,j)}, \boldsymbol{P}_s^{(h,j)}\big\}_{h=1}^{N_\mathrm{k}}, p_{s}^{(j)}\rmv\Big\}_{j=1}^{J_{s}} \Big] \rmv\vspace{.8mm}=
     \hspace*{2mm}\text{SingleSensorUpdate}\Big[\rmv\Big\{\rmv\big\{\underline{\boldsymbol{x}}_{s}^{\ast(h,j)}\rmv, \underline{\boldsymbol{P}}_{s}^{(h,j)}\big\}_{h=1}^{N_\mathrm{k}}, \underline{p}_{s}^{(j)}\Big\}_{\rmv j=1}^{\rmv J_{s-1}}\rmv\rmv,\V{z}_{s} \rmv\Big]$ \\[1mm]
\hspace{52.5mm} \tcp{see \textbf{Alg. \ref{alg:SingleSensorUpdate}}}
\vspace{.8mm}


  }
  \vspace{.8mm}
  
  \mbox{Set $\rmv\Big\{\rmv\big\{{\boldsymbol{x}}^{\ast(h,j)}, \boldsymbol{P}^{(h,j)}\big\}_{h=1}^{N_\mathrm{k}}\rmv, p^{(j)}\rmv\Big\}_{\rmv j=1}^{\rmv J}\rmv\rmv = \Big\{\rmv\big\{{\boldsymbol{x}_S}^{\rmv\rmv\rmv\rmv\rmv\rmv\ast(h,j)}\rmv\rmv, \boldsymbol{P}_S^{(h,j)}\big\}_{h=1}^{N_\mathrm{k}}\rmv, p_{S}^{(j)}\rmv\Big\}_{\rmv j=1}^{\rmv J_{S}}$}
  where $J = J_{S}$
  \vspace{1mm}
  
  Output: $\Big\{\rmv\big\{{\boldsymbol{x}}^{\ast(h,j)}, \boldsymbol{P}^{(h,j)}\big\}_{h=1}^{N_\mathrm{k}}\rmv, p^{(j)}\rmv\Big\}_{\rmv j=1}^{\rmv J}$
   \vspace{.5mm}
  \caption{Single Time Step of Multisensor \ac{mot} with \ac{gmm} and Invertible Flow}
  \vspace{0mm}
\end{algorithm}}

\rd{Note that for the execution of the proposed method, for each time step and each object, we need to update each particle per kernel, pseudo-time, measurement, and sensor. The asymptotic complexity with respect to these parameters, per object and time step, thus reads $\mathcal{O}\big(N_{\lambda} N_{\mathrm{k}} N_{\mathrm{p}} \big(\sum^{S}_{s = 1} \rmv M_s\big) \big)$. The complexity of a conventional bootstrap implementation per object and time step is $\mathcal{O}\big(N_{\mathrm{b}} \big(\sum^{S}_{s = 1} \rmv M_s\big) \big)$, where $N_{\mathrm{b}}$ is the number of particles. The improved runtime-complexity tradeoff of the proposed method is due to the fact that, in challenging problems as the one considered in Sections \ref{sec:simResults}, for $N_{\lambda} N_{\mathrm{k}} N_{\mathrm{p}} = N_{\mathrm{b}}$, the proposed method strongly outperforms graph-based MOT that relies on conventional particle filtering. Since memory requirements per time step and object are $N_{\mathrm{k}} N_{\mathrm{p}} $ for the proposed method, and $N_{\mathrm{b}}$ for bootstrap particle filtering, and since $N_{\lambda} \gg 1$, for $N_{\lambda} N_{\mathrm{k}} N_{\mathrm{p}} = N_{\mathrm{b}}$, the memory requirements of the proposed method are significantly lower compared to conventional particle\vspace{-1.5mm} filtering.}

\section{Numerical Results}
\label{sec:simResults}

Next, we report simulation results assessing the performance of our method and comparing it with that of two reference methods for multisensor-multiobject tracking\vspace{-3mm} .

\subsection{Tracking Scenario and Reference Methods}\label{sec:trackingScenario}

We consider an underwater 3-D surveillance scenario where eight objects are tracked by two static sonar hydrophone arrays. $200$ time steps are considered. The hydrophones are deployed about 1300 m below sea level. The object states at time $k$ consist of 3-D position and velocity, i.e., $\RV{x}^{(j)}_{k} = [\rv{x}^{(j)}_{1,k}\iist\rv{x}^{(j)}_{2,k}\iist\rv{x}^{(j)}_{3,k}\iist\rv{\dot{x}}^{(j)}_{1,k}\iist\rv{\dot{x}}^{(j)}_{2,k}\iist\rv{\dot{x}}^{(j)}_{3,k}]^{\text{T}}$, $j=1,\dots,8$ and evolve according to a constant-velocity model \cite[Sec.~6.3.2]{BarRonKir:01}, where the dynamic noise has the physical interpretation as an acceleration with variance $\sigma^2_{\RV{w}}$. The \ac{roi} is $[-1000\ist\text{m}, \ist 1000\ist\text{m} ]  \times [-1000\ist\text{m}, \ist 1000\ist\text{m}] \times [-1500\ist\text{m}, \ist -500\ist\text{m}]$. 
Objects appear at $k \rmv\in\rmv \{1,10,20,30,40,50,60,70\}$ and disappear at $k \rmv\in\rmv \{130,140,150,160,170,180,190,200\}$. To simulate a tracking scenario with challenging data association, we generate the initial state of the objects as follows. The initial state of the first object is randomly generated by setting its position at a circle centered at the origin with a radius of 50 m and a depth of 1000 m, i.e., for its position, we have $\sqrt{{\rv{x}^{(1)2}_{1,0}}+{\rv{x}^{(1)2}_{2,0}}}\rmv=\rmv50$ m and $\rv{x}^{(1)}_{3,0}\rmv=\rmv-1000$ m. The velocity is obtained by setting the vertical speed to zero and the horizontal velocity vector pointing to the circle center with $\sqrt{\rv{\dot{x}}^{(1)2}_{1,0}+\rv{\dot{x}}^{(1)2}_{2,0}}\rmv=\rmv0.3\ist$m/s. For the appearance of each further object $j=2,\dots,8$, an initial position is randomly generated near the initial position of the previously appeared object, i.e., around a circle with radius 50 m centered at $[x^{(j\rmv-\rmv1)}_{1,k}\iist x^{(j\rmv-\rmv1)}_{2,k}\iist x^{(j\rmv-\rmv1)}_{3,k}]$. The initial velocity vector is set with respect to the center circle, as discussed above. As a result of this initialization procedure, tracks start in close proximity in time and space. This makes it challenging to perform data association and declare the existence of newborn objects. \rd{The times when objects appear and disappear, as well as the states of appearing objects, are unknown to all simulated tracking methods. All methods aim to detect the presence of a new object and sequentially estimate its state across time merely based on TDOA measurements and the statistical model. The prior intensity of object birth is modeled at each timestep by a Poisson point process with mean $\mu_\text{b} = 0.05$. The prior \ac{pdf} for newborn object, $f_{\mathrm{b}}\big( \overline{\V{x}}_{k} \big)$, is uniform on the \ac{roi} for the 3-D position and Gaussian with zero mean and covariance matrix $5$m$^2$/s$^2 \M{I}_3$ for the 3-D velocity. The survival probability is $p_\mathrm{su}=0.95$. The object declaration threshold is set to $P_{\text{th}} \rmv=\rmv 0.5$ and the pruning threshold\vspace{1mm} to $P_{\text{pr}} \rmv=\rmv 10^{-4}$.} 

The two hydrophone arrays have the geometry of the array described in \cite{WigHil:C07} and are located at $[519 \ist \text{m} \ist\ist\ist\ist\ist 137 \ist \text{m} \ist\ist -\rmv\rmv\rmv\rmv\rmv1300 \ist \text{m}]^{\T}$ and $[-519 \ist \text{m} \ist\ist\ist -\rmv\rmv137 \ist \text{m} \ist\ist -\rmv\rmv1300 \ist \text{m}]^{\T}\rmv\rmv$, respectively. Each hydrophone array consists of 4 receivers. Hence, there are six receiver pairs at each array. Each receiver pair generates \ac{tdoa} measurements and is considered a sensor for \ac{mot}. This means that multisensor measurements $\RV{z}^{(m)}_{k,s} \rmv$, $m \rmv\in\rmv \big\{ 1,\dots,\rv{M}_{k,s} \big\}$ and $s \rmv\in\rmv \big\{ 1,\dots,12 \big\}$ are obtained by the two arrays at time $k$. At each sensor $s$, a random number of $\rv{M}_{k,s}$ measurements are generated. In particular, the \ac{tdoa} measurement $\rv{z}^{(m)}_{k,s}$ of a detected object with state $\RV{x}^{(j)}_{k}$ is modeled \vspace{1.3mm} as $\rv{z}^{(m)}_{k,s} = \frac{1}{c} \Big( \big\| \big[\rv{x}^{(j)}_{1,k}\ist\ist\ist \rv{x}^{(j)}_{2,k}\ist\ist\ist \rv{x}^{(j)}_{3,k}\big]^{\text{T}} - \V{p}_{s_\text{L}}\big\| - \big\| \big[\rv{x}^{(j)}_{1,k}\ist\ist\ist \rv{x}^{(j)}_{2,k}\ist\ist\ist \rv{x}^{(j)}_{3,k}\big]^{\text{T}} - \V{p}_{s_\text{R}} \big\|\Big)+\rv{v}_{k,s}^{(m)}$ where $\V{p}_{s_\text{L}}$ and $\V{p}_{s_\text{R}}\rmv$ are the paired receiver positions of sensor $s$, $c=1500\ist$m/s is the propagation speed, and $\rv{v}_{k,s}^{(m)}$ is additive zero-mean Gaussian noise with standard deviation $\sigma_{\rv{v}}$ that is assumed statistically independent across $s$, $k$, and $m$. The \ac{pdf} of \ac{fp} measurements, $f_{\mathrm{fp}}\big( z^{(m)}_{k,s} \big)$, is uniform on $\frac{1}{c} \big[-\| \V{p}_{s_\text{L}} - \V{p}_{s_\text{R}}\|,\| \V{p}_{s_\text{L}} - \V{p}_{s_\text{R}}\|\big] $.

We compare the proposed \ac{spa}-based \ac{mot} with the embedded particle flow sampling strategy (``SPA-PF")  with two reference sampling strategies. The first (``SPA-PM'') follows the sampling strategy of the bootstrap particle filter \cite{GorSalSmi:93,AruMasGorCla:02} and uses predicted beliefs as proposal \ac{pdf} \cite{MeyBraWilHla:J17}. The second (yet unpublished) (``SPA-UT'') follows the sampling strategy of the unscented particle filter \cite{MerDouFre:00}, i.e., it uses a Gaussian mixture representation and the unscented transformation to calculate an informative proposal \acp{pdf}. For SPA-PM, we use $N_{\mathrm{b}}=10^6$ particles for newborn \acp{po} and $N_{\mathrm{b}}=6 \cdot 10^4$ particles for legacy \acp{po}. For all other simulated methods, we use $N_{\mathrm{k}}=100$ kernels. For each kernel representing a newborn \ac{po}, we set  $N_{\mathrm{p}} = 500$ for SPA-UT and SPA-PF and for each kernel representing a legacy \ac{po}, we se $N_{\mathrm{p}} = 30$ for SPA-UT and SPA-PF. \rd{Since $N_{\mathrm{b}} / N_{\mathrm{p}} N_{\mathrm{k}}  = 20$ the memory requirements of SPA-PM are 20 time higher compared to SPA-PF and SPA-UT.}

 We also simulate two variants of the second reference method to obtain a similar runtime for SPA-PF. In particular, for ``SPA-UT-1'', we set $N_{\mathrm{p}} = 4000$ for kernels representing newborn \acp{po} and $N_{\mathrm{p}} = 250$ for kernels representing legacy \acp{po}. Furthermore, for ``SPA-UT-2'', we set  $N_{\mathrm{p}} = 6000$ for kernels representing newborn \ac{po} and $N_{\mathrm{p}} = 30$ for kernels representing legacy \ac{po}. Note that the memory requirements of SPA-UT-1 and SPA-UT-2 are higher than the ones of SPA-UT and SPA-PF due to their higher values of $N_{\mathrm{p}}$. Finally, we also simulate a method (``SPA-PF-H'') that uses the sampling strategy of SPA-PF for kernels representing newborn \acp{po} and the sampling strategy of SPA-PM for kernels representing legacy \acp{po}. Using fewer samples for the kernels representing legacy \ac{po}, or even using the strategy of SPA-PF, is motivated by the fact that the beliefs of legacy \acp{po} are typically unimodal and quite informative. We set $N_\lambda=20$ for SPA-PF. For all considered methods, 100 simulation runs are performed. All methods are implemented in MATLAB, and each simulation run is processed on a single core of a 2.6GHz Intel Xeon Gold 6240  processor.

The performance of the six \ac{mot} methods is evaluated w.r.t. to changes in four system parameters (i) object driving noise standard deviation $\sigma_{\RV{w}}$, (ii) mean number of \acp{fp} $\mu_{\mathrm{fp}}$, (iii) detection probability $p_{\text{d}}$ and (iv) measurement noise standard deviation $\sigma_{\RV{v}}$. Note that for the setting $\sigma_{\RV{v}}=5\!\times\!10^{-7}\!$ s, the number of samples $N_{\mathrm{p}}$ was doubled for all methods to yield high tracking performance. The tracking accuracy of the various methods is measured by the Euclidean distance-based \ac{ospa} metric with cutoff parameter $C\!=\!50$ \cite{SchVoVo:J08}. 
\vspace{-2mm}

\subsection{Performance Comparison}\label{subsec:performanceComparison}
\begin{figure}[ht!]
\centering
    \psfrag{s01}[b][b][0.8]{\color[rgb]{0.15,0.15,0.15}\setlength{\tabcolsep}{0pt}\begin{tabular}{c}\raisebox{.3mm}{OSPA}\end{tabular}}%
    \psfrag{s04}[t][t][0.8]{\color[rgb]{0.15,0.15,0.15}\setlength{\tabcolsep}{0pt}\begin{tabular}{c}\raisebox{-1mm}{time steps $k$}\end{tabular}}%
    \color[rgb]{0.15,0.15,0.15}%
    \psfrag{x01}[t][t][0.8]{0}%
    \psfrag{x02}[t][t][0.8]{50}%
    \psfrag{x03}[t][t][0.8]{100}%
    \psfrag{x04}[t][t][0.8]{150}%
    \psfrag{x05}[t][t][0.8]{200}%
    \psfrag{v01}[r][r][0.8]{0}%
    \psfrag{v02}[r][r][0.8]{10}%
    \psfrag{v03}[r][r][0.8]{20}%
    \psfrag{v04}[r][r][0.8]{30}%
    \psfrag{v05}[r][r][0.8]{40}%
    \psfrag{v06}[r][r][0.8]{50}%
    \psfrag{SPA-PM}[l][l][.7]{SPA-PM}%
    \psfrag{SPA-UT}[l][l][.7]{SPA-UT}%
    \psfrag{SPA-UT-1}[l][l][.7]{SPA-UT-1}%
    \psfrag{SPA-UT-2}[l][l][.7]{SPA-UT-2}%
    \psfrag{SPA-PF (Proposed)}[l][l][.7]{SPA-PF (Proposed)}%
    \psfrag{SPA-PF-H (Proposed)}[l][l][.7]{SPA-PF-H (Proposed)}%
    \raisebox{2.5mm}{\includegraphics[height=50mm, width=72mm]{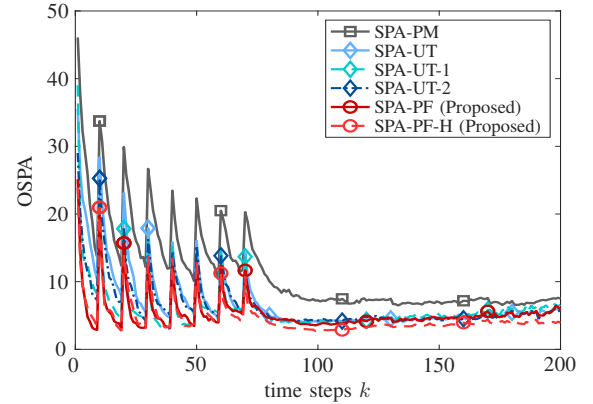}\label{subfig:dstd}}%
    \vspace{-1mm}   
\caption{\small OSPA performance for high uncertainty of prior with $\sigma_{\RV{w}} = 1$ m/s$^2$. Other parameters are set as $\mu_{\mathrm{fp}}=5$, $\sigma_{\RV{v}}=10^{-6}\!$ s, $p_{\text{d}}=0.9$. }
\vspace{-2mm}
\label{fig:trackingPerformanceUncertainPrior}
\end{figure}

\begin{figure}[ht!]
\centering
    \psfrag{s04}[t][t][0.8]{\color[rgb]{0.15,0.15,0.15}\setlength{\tabcolsep}{0pt}\begin{tabular}{c}\raisebox{-1mm}{time steps $k$}\end{tabular}}%
    \psfrag{s03}[b][b][0.8]{\color[rgb]{0.15,0.15,0.15}\setlength{\tabcolsep}{0pt}\begin{tabular}{c}\raisebox{.3mm}{OSPA}\end{tabular}}%
    \color[rgb]{0.15,0.15,0.15}%
    \psfrag{x01}[t][t][0.8]{0}%
    \psfrag{x02}[t][t][0.8]{50}%
    \psfrag{x03}[t][t][0.8]{100}%
    \psfrag{x04}[t][t][0.8]{150}%
    \psfrag{x05}[t][t][0.8]{200}%
    \psfrag{v01}[r][r][0.8]{0}%
    \psfrag{v02}[r][r][0.8]{10}%
    \psfrag{v03}[r][r][0.8]{20}%
    \psfrag{v04}[r][r][0.8]{30}%
    \psfrag{v05}[r][r][0.8]{40}%
    \psfrag{v06}[r][r][0.8]{50}%
    \psfrag{SPA-PM}[l][l][.7]{SPA-PM}%
    \psfrag{SPA-UT}[l][l][.7]{SPA-UT}%
    \psfrag{SPA-UT-1}[l][l][.7]{SPA-UT-1}%
    \psfrag{SPA-UT-2}[l][l][.7]{SPA-UT-2}%
    \psfrag{SPA-PF (Proposed)}[l][l][.7]{SPA-PF (Proposed)}%
    \psfrag{SPA-PF-H (Proposed)}[l][l][.7]{SPA-PF-H (Proposed)}%
    \raisebox{2.5mm}{\includegraphics[height=50mm, width=72mm]{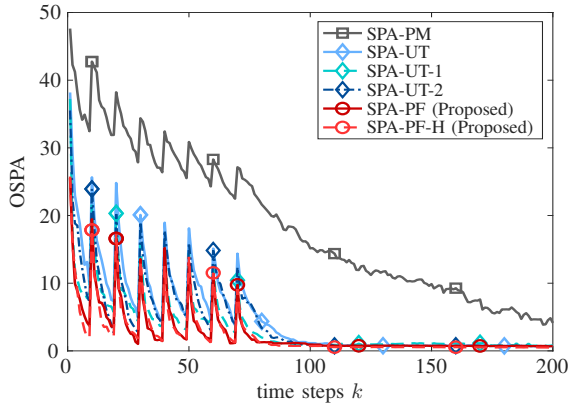}\label{subfig:mstsd}}%
\vspace{-2mm}
\caption{\small OSPA performance for informative measurement model with $\sigma_{\RV{v}}=5\!\times\!10^{-7}\!$ s. Other parameters are set as $\mu_{\mathrm{fp}}=5$, $\sigma_{\RV{w}}=0.1$ m/s$^2$, $p_{\text{d}}=0.9$. }
\vspace{-2mm}
\label{fig:trackingPerformanceInformativeMeasurement}
\end{figure}

\vspace{0mm}
In what follows, we discuss two scenarios where particle degeneracy is particularly pronounced. Particle degeneracy can be caused by uninformative prior information or a very informative likelihood function. To obtain a scenario with uninformative prior information, we consider the case $\sigma_{\RV{w}} = 1$ m/s$^2$. In addition, to obtain a scenario with an informative likelihood function, we consider the case  $\sigma_{\RV{v}}=5\!\times\!10^{-7}\!$ s. Fig.~\ref{fig:trackingPerformanceUncertainPrior} and Fig.~\ref{fig:trackingPerformanceInformativeMeasurement} show the \ac{mospa} error---averaged over 100 simulation runs---of all methods versus time $k$ for these two scenarios. It can be seen that all methods yield error peaks at time steps where the objects appear. This is because the MOT methods \rd{do not know when a new object appears} and often need few time steps after the appearance of an object to declare its existence \rd{based on TDOA measurements and the statistical model.} However, the proposed methods, i.e., SPA-PF and SPA-PF-H, have lower error peaks at the time steps when objects appear, i.e., SPA-PF and SPA-PF-H can often declare the existence of an object faster. In addition, SPA-PF and SPA-PF-H can outperform the other reference methods at almost all time steps. SPA-PM, in particular performs very poorly due to particle degeneracy. It can be noted that SPA-UT-1 and SPA-UT-2 have improved performance compared to SPA-UT but are still outperformed by SPA-PF and SPA-PF-H despite their more extensive memory\vspace{-2mm} requirements.

Furthermore, the \ac{mospa} error and the runtime per time step of the six methods for different system parameter values are shown in Table \ref{fig:table1}. The default value of the four parameters are $\sigma_{\RV{w}}=0.1$ m/s$^2$, $\sigma_{\RV{v}}=1\!\times\!10^{-6}\!$ s, $\mu_{\mathrm{fp}}=5$ and $p_{\text{d}}=0.9$. For each row corresponding to a particular system parameter value, the value of the other three parameters is set to the default value. For each system parameter value corresponding to one row, the \ac{mospa} and runtime are averaged over 100 simulation runs and 200 time steps. The best and the second best \ac{mospa} value corresponding to each system parameter value is marked by an underline and dashed underline, respectively. \rd{As can be noted, the performance of SPA-PM significantly degrades when measurement noise standard deviation is reduced to $\sigma_{\RV{v}}=5\!\times\!10^{-7}\!$ s. This unwanted and counterintuitive behavior is a clear indicator of particle degeneracy in SPA-PM. Only with the proposed method is it possible to yield improved tracking performance as measurement noise variance is reduced.}

It can be seen that for almost all system parameter values, the proposed SPA-PF and SPA-PF-H outperform the reference methods. At the same time, their runtime is comparable with SPA-UT-1 and SPA-UT-2, which yield higher memory requirements. The only scenario where SPA-PF and SPA-PF-H do not result in the lowest MOSPA is when $\sigma_{\RV{v}}=2\!\times\!10^{-6}\!$ s. In this case, since measurements are not very informative, SPA-PM does not suffer from particle degeneracy. It can also be seen that SPA-PF-H outperforms SPA-PF both in terms of \ac{mospa} and runtime, while both SPA-PF and SPA-PF-H typically outperform SPA-UT. We can thus conclude that the main challenge for the sampling method is the initial time step after a new object appears in the scene. Here, beliefs of new \acp{po} are highly uninformative and have complicated shapes. At later time steps, beliefs become informative and unimodal and can thus be computed accurately with fewer samples and the sampling strategy of the bootstrap particle\vspace{0mm} filter.  \rd{For real-time processing, it is expected that an adaptation of the proposed method for execution on graphical processing units (GPUs), can strongly reduce runtimes by exploiting the highly parallelizable nature of \ac{pf}. }

\begin{figure}[ht!]
\centering
    \psfrag{s03}[t][t][0.8]{\color[rgb]{0.15,0.15,0.15}\setlength{\tabcolsep}{0pt}\begin{tabular}{c}\raisebox{-1mm}{time steps $k$}\end{tabular}}%
    \psfrag{s04}[b][b][0.8]{\color[rgb]{0.15,0.15,0.15}\setlength{\tabcolsep}{0pt}\begin{tabular}{c}\raisebox{.3mm}{OSPA}\end{tabular}}%
    \color[rgb]{0.15,0.15,0.15}%
    \psfrag{x01}[t][t][0.8]{0}%
    \psfrag{x02}[t][t][0.8]{50}%
    \psfrag{x03}[t][t][0.8]{100}%
    \psfrag{x04}[t][t][0.8]{150}%
    \psfrag{x05}[t][t][0.8]{200}%
    \psfrag{v01}[r][r][0.8]{0}%
    \psfrag{v02}[r][r][0.8]{10}%
    \psfrag{v03}[r][r][0.8]{20}%
    \psfrag{v04}[r][r][0.8]{30}%
    \psfrag{v05}[r][r][0.8]{40}%
    \psfrag{v06}[r][r][0.8]{50}%
    \psfrag{SPA-PF-H (Nk = 1)}[l][l][.7]{SPA-PF-H ($N_{\mathrm{k}}=1$)}%
    \psfrag{SPA-PF-H (Nk = 5)}[l][l][.7]{SPA-PF-H ($N_{\mathrm{k}}=5$)}%
    \psfrag{SPA-PF-H (Nk = 10)}[l][l][.7]{SPA-PF-H ($N_{\mathrm{k}}=10$)}%
    \psfrag{SPA-PF-H (Nk = 50)}[l][l][.7]{SPA-PF-H ($N_{\mathrm{k}}=50$)}%
    \psfrag{SPA-PF-H (Nk = 100)}[l][l][.7]{SPA-PF-H ($N_{\mathrm{k}}=100$)}%
    \psfrag{SPA-PM}[l][l][.7]{SPA-PM}%
    \raisebox{2.5mm}{\includegraphics[height=50mm, width=72mm]{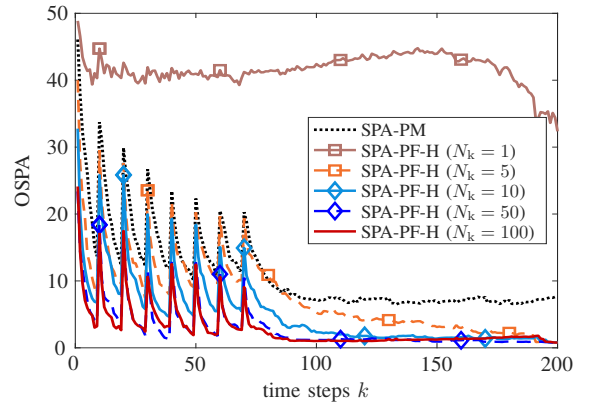}\label{subfig:kernel}}%
\vspace{-2mm}
\caption{\small OSPA performance of SPA-PM and SPA-PF-H for different number number of kernels, $N_k$. }
\vspace{-3mm}
\label{fig:trackingPerformanceKernel}
\end{figure}

\rd{In what follows, we numerically investigate the effect of the number of kernels $N_k$ on system performance. In particular, we set the number of kernels as $N_{\mathrm{k}} \in \{1,5,10,50,100\}$ in SPA-PF-H and compare SPA-PF-H  for these different values of $N_{\mathrm{k}}$ with SPA-PM.  The parameters $\sigma_{\RV{v}},\mu_{\mathrm{fp}},\sigma_{\RV{w}}$, and $p_{\text{d}}$ are set as dicussed above. For comparable runtimes of all SPA-PF-H variants, we set $N_{\mathrm{p}} \in \{100000, 20000, 8000, 1300, 500\}$ for newborn object and $N_{\mathrm{p}} \in \{15000, 1000, 400, 70, 30\}$  for legacy objects. As expected, it can be seen in Fig.~\ref{fig:trackingPerformanceKernel} that the accuracy of SPA-PF-H improves with increasing $N_{\mathrm{k}}$. While SPA-PF-H performs very poorly for $N_{\mathrm{k}} = 1$, notably, it can already outperform SPA-PM for $N_{\mathrm{k}} = 5$. The average runtime per time step of SPA-PF-H is $27.14$s, $15.06$s, $10.85$s, $11.35$s, and $12.72$s for $N_{\mathrm{k}}=1$,$N_{\mathrm{k}}=5$, $N_{\mathrm{k}}=10$, $N_{\mathrm{k}}=50$, and $N_{\mathrm{k}}=100$, respectively, as well as $22.66$s for SPA-PM\vspace{-1mm}. } 


\begin{table*}[ht!]
\centering
\begin{footnotesize}
\hspace{-.5mm}
\begin{tabular}{  C{1cm} | C{1.25cm} | C{0.85cm} C{0.85cm} |  C{0.85cm} C{0.85cm} | C{0.85cm} C{0.85cm} | C{0.85cm} C{0.85cm} | C{0.85cm} C{0.85cm} | C{0.85cm} C{0.85cm} }
\hline
\multirow{2}{*}{\rule{-1.5mm}{6mm} Parameter}&\multirow{2}{*}{\rule{0mm}{6mm} Value}&\multicolumn{2}{c|}{\rule{0mm}{3mm}SPA-PM}&\multicolumn{2}{c|}{SPA-UT}&\multicolumn{2}{c|}{SPA-UT-1}&\multicolumn{2}{c|}{SPA-UT-2}&\multicolumn{2}{c|}{SPA-PF}&\multicolumn{2}{c}{SPA-PF-H}\\[1mm]
    & & {OSPA} & {Runtime (s)}  & {OSPA} & {Runtime (s)}  & {OSPA} & {Runtime (s)}  & {OSPA} & {Runtime (s)}  & {OSPA} & {Runtime (s)}  & {OSPA} & {Runtime (s)}  \\
   
\hline

\rowcolor{black!15!white}\cellcolor[gray]{1} & $1$ m/s$^2$ & $11.38$ & $20.94$ & $7.02$ &  $3.22$ & $6.06$ & $19.98$ & $6.45$ & $25.84$ & \dashuline{$5.31$} & $21.66$ & \ul{$4.64$} & $13.19$\rule{0mm}{3.3mm} \\[.5mm]
\cellcolor[gray]{1} & $0.5$ m/s$^2$ & $9.13$ & $23.54$ & $5.84$ &  $3.35$ & $4.50$ & $21.42$ & $5.11$ & $28.10$ & \dashuline{$3.80$} & $22.54$ & \ul{$3.45$} & $13.90$\rule{0mm}{3.3mm} \\[.5mm]
\rowcolor{black!15!white}\multirow{-4.5}{*}{\cellcolor[gray]{1}$\sigma_{\RV{w}}$} & $0.1$ m/s$^2$ & $7.61$ & $23.99$ & $4.82$ & $3.35$ & $3.45$ & $21.59$ & $4.35$ & $29.10$ & \dashuline{$2.96$} & $22.48$ & \ul{$2.67$} & $13.68$\rule{0mm}{3.3mm} \\[.5mm]
\hline
\cellcolor[gray]{1} & $2\!\times\!10^{-6}\!$ s & \ul{$4.51$} & $21.69$  & $5.25$ & $3.19$ & \dashuline{$4.78$} & $19.85$ & $5.31$ & $25.64$ & $5.37$ & $22.53$ & $5.13$ & $14.43$\rule{0mm}{3.3mm} \\[.5mm]
\rowcolor{black!15!white}\cellcolor[gray]{1} & $1\!\times\!10^{-6}\!$ s & $7.61$ & $20.77$  & $4.69$ & $3.05$ & $3.44$& $19.33$ & $4.45$ & $25.95$ & \dashuline{$3.02$} & $20.64$ & \ul{$2.69$} & $12.62$\rule{0mm}{3.3mm} \\[.5mm]
\multirow{-4.5}{*}{\cellcolor[gray]{1}$\sigma_{\RV{v}}$} & $5\!\times\!10^{-7}\!$ s & $18.60$ & $42.83$  & $5.10$ & $5.06$ & $3.59$& $37.94$ & $4.28$ & $51.82$ & \dashuline{$2.61$} & $31.72$ & \ul{$2.40$} & $21.26$\rule{0mm}{3.3mm} \\[.5mm]
\hline
\rowcolor{black!15!white}\cellcolor[gray]{1} & $10$ & $7.79$ & $33.88$  & $5.85$ & $5.29$ & $4.89$& $33.13$ & $6.74$ & $41.76$ & \dashuline{$4.30$} & $41.21$ & \ul{$4.10$} & $29.48$\rule{0mm}{3.3mm} \\[.5mm]
\cellcolor[gray]{1} & $5$ & $7.61$ & $22.66$  & $4.78$ & $3.15$ & $3.48$& $20.20$ & $4.38$ & $27.02$ & \dashuline{$3.28$} & $21.58$ & \ul{$2.82$} & $13.21$\rule{0mm}{3.3mm} \\[.5mm]
\rowcolor{black!15!white}\multirow{-4.5}{*}{\cellcolor[gray]{1}$\mu_{\mathrm{fp}}$} & $2$ & $9.70$ & $15.66$  & $5.03$ & $2.40$ & \dashuline{$2.36$} & $14.20$ & $4.23$ & $19.22$ & $2.67$ & $13.42$ & \ul{$2.33$} & $7.27$\rule{0mm}{3.3mm} \\[.5mm]
\hline
\cellcolor[gray]{1} & $0.85$ & $11.78$ & $20.72$  & $8.52$ & $3.27$ & \dashuline{$4.89$} & $20.15$ & $8.00$ & $26.36$ & \ul{$4.32$} & $22.31$ & $5.02$ & $14.89$\rule{0mm}{3.3mm} \\[.5mm]
\rowcolor{black!15!white}\cellcolor[gray]{1} & $0.9$ & $7.61$ & $22.15$  & $5.01$ & $3.02$ & $3.29$& $18.98$ & $4.65$ & $24.97$ & \dashuline{$3.09$} & $21.15$ & \ul{$2.66$} & $12.50$\rule{0mm}{3.3mm} \\[.5mm]
\multirow{-4.5}{*}{\cellcolor[gray]{1}$p_{\text{d}}$} & $0.95$ & $8.06$ & $21.80$  & $3.90$ & $3.26$ & $2.58$& $20.77$ & $3.01$ & $27.83$ & \dashuline{$2.56$} & $21.10$ & \ul{$2.35$} & $12.40$\rule{0mm}{3.3mm} \\[.5mm]
\hline
\end{tabular}
\end{footnotesize}
\vspace{2.5mm}

\caption{\small \ac{mospa} and runtime per time step of different algorithms for the considered tracking scenario w.r.t related parameters.} 
\label{fig:table1}
\vspace{-5mm}
\end{table*}

\subsection{Real Data Experiment: Echolocation in Oceanography}\label{subsec:realDataExperiment}
\begin{figure}[ht!]
  \centering
\psfrag{s04}[][][0.7]{\color[rgb]{0.15,0.15,0.15}\setlength{\tabcolsep}{0pt}\begin{tabular}{c}\raisebox{6mm}{$x_1$ (m)}\end{tabular}}%
\psfrag{s02}[][][0.7]{\color[rgb]{0.15,0.15,0.15}\setlength{\tabcolsep}{0pt}\begin{tabular}{c}\raisebox{6mm}{$x_2$ (m)}\end{tabular}}%
\psfrag{s01}[][][0.7]{\color[rgb]{0.15,0.15,0.15}\setlength{\tabcolsep}{0pt}\begin{tabular}{c}\raisebox{6mm}{$x_3$ (m)}\end{tabular}}%
%
\color[rgb]{0.15,0.15,0.15}%
%
\psfrag{x01}[t][t][0.7]{-600}%
\psfrag{x02}[t][t][0.7]{0}%
\psfrag{x03}[t][t][0.7]{600}%
%
\psfrag{v01}[r][r][0.7]{-500}%
\psfrag{v02}[r][r][0.7]{0}%
\psfrag{v03}[r][r][0.7]{500}%
%
\psfrag{z01}[r][r][0.7]{-1300}%
\psfrag{z02}[r][r][0.7]{-800}%
\psfrag{z03}[r][r][0.7]{-300}%
\psfrag{SPA-PF-H}[l][l][0.6]{SPA-PF-H}%
\psfrag{Ref}[l][l][0.6]{Ref}%
\psfrag{Sensor}[l][l][0.6]{HARP}%
%
\hspace{-2mm}\raisebox{0.5mm}{\includegraphics[scale=0.3]{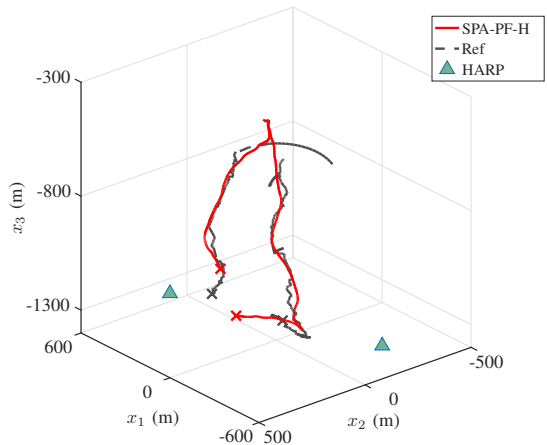}}%
    \vspace{-1mm}   
\caption{\small  Underwater acoustic tracking tracking scenario with two Cuvier's beaked whales.}
\vspace{-1mm}
\label{fig:whale1}
\end{figure}

\rd{We further validate our proposed \ac{mot} method in an underwater acoustic tracking scenario. The acoustic signals ``clicks'' emitted by two Cuvier's beaked whales are recorded by two \ac{harp}\cite{Wig:07}, each of which is equipped with four hydrophones.  \acp{harp} are deployed at a depth of 1330 m and approximately 1 km apart. Preprocessing of acoustic data is described in \cite{JanMeySnyWigBauHil:A22}. Each pair of hydrophones on each \ac{harp} acts as a sensor that provides TDOA measurements every $7$s. Since there are 6 pairs of hydrophones on each \ac{harp}, there are a total of $S\rmv=\rmv12$ sensors providing TDOA measurements. We use a dataset that consists of 172 time steps and has a total duration of roughly 20 minutes. It was recorded on July 1st, 2018, in Southern California. Tracking results are shown in Fig.~\ref{fig:whale1}. The red solid lines show the estimated tracks of two whales provided by SPA-PF-H. The two whales were initially detected at depths of about 450 m and then kept diving until a depth of 1300 m. For obvious reasons, no ground truth information exists for this scenario. However, we have added reference tracks of the two whales that are the result of a trained operator hand-annotating preprocessed acoustic\vspace{0mm} data.}

\begin{figure}[ht!]
  \centering
\psfrag{s12}[][][0.7]{\color[rgb]{0.15,0.15,0.15}\setlength{\tabcolsep}{0pt}\begin{tabular}{c}time (s)\end{tabular}}%
\psfrag{s11}[][][0.7]{\color[rgb]{0.15,0.15,0.15}\setlength{\tabcolsep}{0pt}\begin{tabular}{c}\raisebox{6mm}{$x_3$ (m)}\end{tabular}}%
\psfrag{s04}[][][0.7]{\color[rgb]{0.15,0.15,0.15}\setlength{\tabcolsep}{0pt}\begin{tabular}{c}\raisebox{6mm}{$x_1$ (m)}\end{tabular}}%
\psfrag{s10}[][][0.7]{\color[rgb]{0.15,0.15,0.15}\setlength{\tabcolsep}{0pt}\begin{tabular}{c}time (s)\end{tabular}}%
\psfrag{s08}[][][0.7]{\color[rgb]{0.15,0.15,0.15}\setlength{\tabcolsep}{0pt}\begin{tabular}{c}time (s)\end{tabular}}%
\psfrag{s01}[][][0.7]{\color[rgb]{0.15,0.15,0.15}\setlength{\tabcolsep}{0pt}\begin{tabular}{c}\raisebox{6mm}{$x_2$ (m)}\end{tabular}}%
%
\color[rgb]{0.15,0.15,0.15}%
%
\psfrag{x01}[t][t][0.7]{0}%
\psfrag{x02}[t][t][0.7]{200}%
\psfrag{x03}[t][t][0.7]{400}%
\psfrag{x04}[t][t][0.7]{600}%
\psfrag{x05}[t][t][0.7]{800}%
\psfrag{x06}[t][t][0.7]{1000}%
\psfrag{x07}[t][t][0.7]{1200}%
\psfrag{x08}[t][t][0.7]{0}%
\psfrag{x09}[t][t][0.7]{200}%
\psfrag{x10}[t][t][0.7]{400}%
\psfrag{x11}[t][t][0.7]{600}%
\psfrag{x12}[t][t][0.7]{800}%
\psfrag{x13}[t][t][0.7]{1000}%
\psfrag{x14}[t][t][0.7]{1200}%
\psfrag{x15}[t][t][0.7]{0}%
\psfrag{x16}[t][t][0.7]{200}%
\psfrag{x17}[t][t][0.7]{400}%
\psfrag{x18}[t][t][0.7]{600}%
\psfrag{x19}[t][t][0.7]{800}%
\psfrag{x20}[t][t][0.7]{1000}%
\psfrag{x21}[t][t][0.7]{1200}%
%
\psfrag{v01}[r][r][0.7]{-1500}%
\psfrag{v02}[r][r][0.7]{-1000}%
\psfrag{v03}[r][r][0.7]{-500}%
\psfrag{v04}[r][r][0.7]{-200}%
\psfrag{v05}[r][r][0.7]{0}%
\psfrag{v06}[r][r][0.7]{200}%
\psfrag{v07}[r][r][0.7]{400}%
\psfrag{v08}[r][r][0.7]{-400}%
\psfrag{v09}[r][r][0.7]{-200}%
\psfrag{v10}[r][r][0.7]{0}%
\psfrag{v11}[r][r][0.7]{200}%
\psfrag{SPA-PM}[l][l][0.6]{SPA-PM}%
\psfrag{SPA-PF-H}[l][l][0.6]{SPA-PF-H}%
\psfrag{Ref}[l][l][0.6]{Ref}%
%
\hspace{-2mm}\raisebox{0.5mm}{\includegraphics[height=70mm, width=82mm]{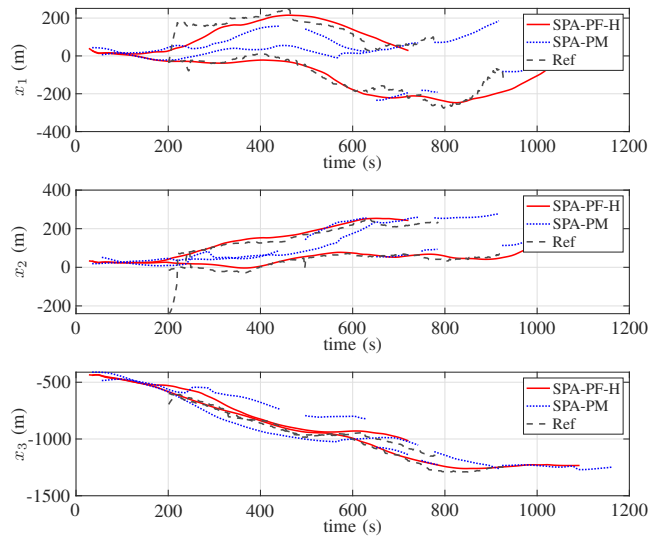}}%
    \vspace{-1mm}   
\caption{\small A comparison between the estimated tracks provided by SPA-PF-H and SPA-PM. The hand-annotated tracks of the two whales are marked for reference (gray dashed line). Each axis of the 3-D domain is shown individually.}
\vspace{-3mm}
\label{fig:whale2}
\end{figure}

\rd{In Figure \ref{fig:whale2}, we show the estimate tracks of SPA-PF-H compared with SPA-PM in 2-D. SPA-PM does break and merge tracks, which makes it very difficult to determine how many whales are actually there. Most importantly SPA-PF-H can potentially replace the human operator, while SPA-PM cannot. The overall number of particles of SPA-PM is 20 times that of SPA-PF-H. The runtime per time step of SPA-PM is 4.65 s while 2.35 s of SPA-PF-H, i.e., SPA-PF-H is faster than SPA-PM. Since the measurement interval is 7 s, both SPA-PF-H  and SPA-PM can be used in real-time. For larger scenarios with more than two whales, a GPU implementation is required for real-time\vspace{2mm} processing.}

\section{Conclusion}
\label{sec:conclusion}

We presented a graph-based Bayesian method for multisensor MOT with high-dimensional object states. Particle degeneracy is avoided by performing operations on the graph using \ac{pf}. Our numerical results indicate that the main challenge for sampling is representing the posterior distribution at the initial time step after a new object appears in the scene. Compared to state-of-the-art reference methods, we show favorable tracking performance in a 3-D \ac{mot} scenario. The introduced approach is expected to be particularly appealing for passive surveillance problems \cite{FerMunTesBraMeyPelPetAlvStrLeP:J17}. Future research avenues include graph-based processing with embedded stochastic \ac{pf} \cite{DaiDau:J22,LiyDau:J22} and applications including extended object tracking \cite{MeyWin:J20,MeyWil:J21}, simultaneous localization and object tracking \cite{MeyHliHla:J16}, and information-seeking\vspace{0mm} control \cite{MeyWymFroHla:J15}. \rd{We also aim to demonstrate real-time processing capabilities of the proposed approach by execution on graphical processing units (GPUs), exploiting the highly parallelizable nature of \ac{pf}\vspace{-1mm}.}

\section*{Acknowledgement}

The authors would like to thank Dr.~Thomas Kropfreiter and Mr.~Junsu Jang for carefully reading the\vspace{4mm} manuscript.

DISTRIBUTION STATEMENT A: Approved for public release. Distribution is unlimited. This material is based upon work supported by the Under Secretary of Defense for Research and Engineering under Air Force Contract No. FA8702-15-D-0001. Any opinions, findings, conclusions, or recommendations expressed in this material are those of the author(s) and do not necessarily reflect the views of the Under Secretary of Defense for Research and Engineering\

\renewcommand{\baselinestretch}{.97}
\selectfont
\bibliographystyle{IEEEtran}
\bibliography{StringDefinitions,IEEEabrv,Papers,Books,Temp}

\end{document}